\documentclass[longauth]{aa}  

\usepackage{graphicx}

\usepackage{txfonts}
\usepackage{hyperref}
\usepackage{xcolor}
\usepackage[version=4]{mhchem}

\usepackage{array}
\newcolumntype{L}[1]{>{\raggedright\let\newline\\\arraybackslash\hspace{0pt}}m{#1}}
\newcolumntype{C}[1]{>{\centering\let\newline\\\arraybackslash\hspace{0pt}}m{#1}}
\newcolumntype{R}[1]{>{\raggedleft\let\newline\\\arraybackslash\hspace{0pt}}m{#1}}

\usepackage{multirow}
\makeatletter   
\renewcommand*\aa@pageof{, page \thepage{} of \pageref*{LastPage}}
\makeatother

\newcommand{\Teq}{T_{\textrm{eq}}}
\newcommand{\Teff}{T_{\textrm{eff}}}
\newcommand{\Tint}{T_{\textrm{int}}}

\newcommand*\chem[1]{\ensuremath{\mathrm{#1}}}
\newcommand{\kzz}{K_{zz}}

\newcommand{\degrees}{^\circ}
\newcommand{\partfrac}[2]{\frac{\partial #1}{\partial #2}}  

\definecolor{HydraGreen}{HTML}{029E73}
\definecolor{PyratRed}{HTML}{D55E00}
\definecolor{YangBlue}{HTML}{0173B2}

\begin{document}
    
  \title{Phase-dependent chemistry of WASP-43~b revealed with a suite of one-, two-, and three-dimensional models}


   \author{Robin Baeyens \inst{1}
   \and
   Julianne I. Moses \inst{2}
    \and
    Jasmina Blecic \inst{3, 4}
    \and 
    Elspeth K.~H. Lee \inst{5}
    \and 
    Lucas Teinturier \inst{6, 7}
    \and 
    Shang-Min Tsai \inst{8}
    \and 
    Jeehyun Yang \inst{9,15}
    \and 
    Jingxuan Yang \inst{10}
    \and 
    Ludmila Carone \inst{11}
    \and 
    Renyu Hu \inst{12,13,14,15} 
    \and
    Sven Kiefer \inst{16,17,11,18} 
    \and 
    Anjali A.~A. Piette \inst{19}
    \and
    Taylor J. Bell \inst{20}
    \and
    Nicolas Crouzet \inst{21}
    \and
    Ian Dobbs-Dixon \inst{3}
    \and 
    Christiane Helling \inst{11}
    \and 
    Nicolas Iro \inst{22}
    \and
    Dominic Samra \inst{9}
    \and 
    Olivia Venot \inst{23, 24}
    \and 
    Jean-Michel Désert \inst{25, 1} 
    }

   \institute{Anton Pannekoek Institute for Astronomy, University of Amsterdam, Science Park 904, 1098 XH Amsterdam, The Netherlands
        \email{r.l.l.baeyens@uva.nl}
        \and Space Science Institute, 4765 Walnut St, Suite B, Boulder, CO 80301, USA
        \and Department of Physics, New York University Abu Dhabi, Abu Dhabi, United Arab Emirates
        \and Center for Astrophysics and Space Science (CASS), New York University Abu Dhabi, Abu Dhabi, United Arab Emirates
        \and Center for Space and Habitability, University of Bern, Gesellschaftsstrasse 6, CH-3012 Bern, Switzerland
        \and LIRA, Observatoire de Paris, Université PSL, CNRS, Sorbonne Université, Université Paris Cité, 5 Place Jules Janssen, Meudon, 92195, France
        \and Laboratoire de Météorologie Dynamique, IPSL, CNRS, Sorbonne Université, Ecole Normale Supérieure, Université PSL, Ecole Polytechnique, Institut Polytechnique de Paris, 4 Place Jussieu, Paris, 75005, France
        \and Institute of Astronomy \& Astrophysics, Academia Sinica, Taipei 10617, Taiwan
        \and Department of Astronomy and Astrophysics, The University of Chicago, Chicago, IL 60637, USA
        \and Atmospheric, Oceanic and Planetary Physics, Department of Physics, University of Oxford, Oxford OX1 3PU, UK
        \and Space Research Institute, Austrian Academy of Sciences, Schmiedlstrasse 6, A-8042 Graz, Austria
        \and Department of Astronomy \& Astrophysics, The Pennsylvania State University, University Park, PA 16802, USA
        \and Center for Exoplanets and Habitable Worlds, The Pennsylvania State University, University Park, PA 16802, USA
        \and Institute for Computational and Data Science, The Pennsylvania State University, University Park, PA 16802, USA
        \and Jet Propulsion Laboratory, California Institute of Technology, Pasadena, CA 91109, USA
        \and University of Texas at Austin, Department of Astronomy, 2515 Speedway C1400, Austin, TX 78712, USA
        \and Institute of Astronomy, KU Leuven, Celestijnenlaan 200D, 3001 Leuven, Belgium
        \and Institute for Theoretical Physics and Computational Physics, Graz University of Technology, Petersgasse 16, 8010 Graz, Austria
        \and School of Physics and Astronomy, University of Birmingham, Edgbaston, Birmingham, B15 2TT, UK
        \and AURA for the European Space Agency (ESA), Space Telescope Science Institute, 3700 San Martin Drive, Baltimore, MD 21218, USA
        \and Kapteyn Astronomical Institute, University of Groningen, P.O. Box 800, 9700 AV Groningen, The Netherlands
        \and Institute of Space Research, German Aerospace Center (DLR), Germany
        \and Université Paris Cité and Univ Paris Est Creteil, CNRS, LISA, F-75013 Paris, France
        \and Univ. Grenoble Alpes, CNRS, IPAG, 38000 Grenoble, France
        \and Leibniz-Institut für Astrophysik Potsdam (AIP), An der Sternwarte 16, 14482 Potsdam, Germany
        }

   \date{Received XXX; accepted YYY}

 
  \abstract
   {Spectroscopic phase curves of tidally locked giant exoplanets provide us with a unique global picture of their atmospheric chemistry, as different regions of the planet rotate in and out of view. However, the spatially changing composition of tidally locked exoplanets -- governed by stellar irradiation and atmospheric circulation -- make phase-resolved observations exceptionally hard to interpret. 
   }
   {Our goal is to investigate the chemistry of the hot Jupiter WASP-43~b in detail using theoretical models, considering the constraints of the James Webb Space Telescope's MIRI phase curve.}
   {With a suite of pseudo-two-dimensional and three-dimensional photochemical models, we simulate the composition of WASP-43~b in various configurations (different metallicities and wind speeds), and compare them with atmospheric retrieval models. Additionally, we probe the robustness of photochemical predictions by comparing the results of four one-dimensional models that have equivalent setups.}
   {We confirm that disequilibrium chemistry in our theoretical models reduces the methane concentration on the planet's night side for wind jet speeds $\gtrsim \, 500$~m~s$^{-1}$. Varying the metallicity in the models induces large changes in the \ce{CO2} and \ce{SO2} concentrations, with \ce{SO2} producing mid-infrared absorption features in synthetic emission spectra of the night side at atmospheric metallicities $\gtrsim 10 \times$ solar. Our models provide evidence for pole-to-equator circulation enhancing the \ce{CH4}, \ce{NH3}, and HCN abundances, which is nonetheless insufficient for detectable spectral features. Finally, we show that \ce{H2O}, CO, and \ce{CO2} are robustly modeled, but species affected by photochemistry (\ce{CH4}, \ce{NH3}, \ce{SO2}, HCN) are more sensitive to model-specific assumptions and pathways.}
   {We conclude that horizontal quenching is the prime mechanism that explains the non-detection of methane in the MIRI phase-curve of WASP-43~b. This mechanism requires only moderate wind speeds and is operative at various thermal structures and atmospheric metallicities. Furthermore, coupled carbon-sulfur chemistry leads to an additional decrease in methane compared to previous models in the literature that did not contain sulfur chemistry. We do not favor a high metallicity as it would have led to observable \ce{SO2} features in the MIRI spectra. 
   Our study shows that phase-dependent photochemistry models are essential tools in the interpretation of hot-Jupiter phase curves, but benchmarking is needed to improve the accuracy of photochemical models in the future.
   }



\titlerunning{Phase-dependent chemistry of WASP-43~b}
\authorrunning{R.~Baeyens et al.}

   \maketitle
   \nolinenumbers

%
\section{Introduction}

The atmospheres of tidally locked exoplanets are complex systems. Viewed globally, their chemical composition is largely determined by their bulk elemental composition, the amount of stellar irradiation they receive, and their evolutionary history. But viewed locally, the planet's temperature and pressure structure, atmospheric circulation, and radiative transfer dictate the chemical composition of the gas and the molecular signatures that can be observed. Marrying the global and the local views is one of the key challenges for exoplanet atmosphere science.

Spectroscopic phase curves provide arguably the richest possible dataset that can be obtained for transiting exoplanets \citep{parmentier_exoplanet_2018}. During such observation, the planetary emission spectrum is being measured as a function of phase, for the full duration of its orbit, yielding three-dimensional (3D) information of the planet's thermal structure and any spectrally active molecules in its atmosphere \citep{stevenson_thermal_2014, kreidberg_global_2018, arcangeli_climate_2019, kempton_reflective_2023}. 
Not only does this enable probing the three-dimensional processes occurring in exoplanet atmospheres -- such as atmospheric circulation \citep{komacek_atmospheric_2017, may_new_2022, splinter_precise_2025}, frictional drag and magnetism \citep{beltz_exploring_2022}, molecular dissociation and recombination \citep{mansfield_evidence_2020}, inhomogeneous chemistry \citep{moses21pseudo2d, baeyens_photodissociation_2024, evans-soma_SiO_2025}, and cloud coverage \citep{parmentier_cloudy_2021, coulombe_highly_2025}, but spectroscopic phase curves also provide valuable constraints to atmospheric retrievals. By making reasonable assumptions, models, or parametrizations of the 3D thermal structure and underlying bulk composition, phase curve retrievals help break degeneracies that plague 1D models \citep{feng_impact_2016, blecic_implications_2017, taylor_understanding_2020, irwin_2.5D_2020, changeat_exploration_2021, chubb_exoplanet_2022, Yang_simultaneous_2024}.

With its short orbital period ($P_\textrm{orb} = 0.81$~days) and favorable planet-to-star flux contrast, WASP-43~b \citep[$M= 2.0$~$M_\textrm{Jup}$, $R = 1.0$~$R_\textrm{Jup}$, $\Teq \approx 1450$~K, ][]{hellier2011wasp, gillon_TRAPPIST_2012} has historically been a favorable target for phase curve observations. \textit{Hubble Space Telescope} (HST) WFC3 data and \textit{Spitzer Space Telescope} photometry at 3.6~$\mu$m and 4.5~$\mu$m have revealed an unusually large brightness contrast between the day- and night-side hemispheres \citep{stevenson_thermal_2014, stevenson_spitzer_2017}. Several theoretical explanations have been brought forth to explain this large day-night contrast, including night-side clouds \citep{mendonca_revisiting_2018, Venot_global_2020, murphy_lack_2023}, disequilibrium chemistry \citep{mendonca_three-dimensional_2018, Venot_global_2020}, disruption of the equatorial jet stream \citep{carone_equatorial_2020}, or an elevated atmospheric metallicity \citep{kataria_atmospheric_2015}. Additionally, studies have reanalyzed the infrared \textit{Spitzer} data, yielding higher night-side temperatures and bringing them more in line with theoretical predictions of heat redistribution \citep{keating_revisiting_2017, mendonca_revisiting_2018, morello_independent_2019, bell_comprehensive_2021}.
One puzzle piece to WASP-43~b's peculiar phase curve -- especially compared to hot Jupiters in the same temperature range \citep{dang_comprehensive_2025} -- is its short orbital period of 19.5~hours and high surface gravity, affecting the atmospheric circulation \citep{Showman_3D_2015, carone_equatorial_2020, baeyens_grid_2021, Roth_hot_2024} and cloud distribution \citep{helling_cloud_2021, roman_clouds_2021, parmentier_cloudy_2021}. Nonetheless, recent climate models have yielded robust predictions of equatorial superrotation \citep{Schneider_exploring_2022, teinturier_radiative_2024}, corroborated by a measured jet wind speed of $5.4^{+2.8}_{-2.1}$~km~s$^{-1}$ from high-resolution Doppler spectroscopy \citep{lesjak_retrieval_2023}. 

Several molecules have been detected in WASP-43~b's atmosphere, including \ce{H2O} \citep{kreidberg_precise_2014}, \ce{CO} and/or \ce{CO2} \citep{stevenson_spitzer_2017}, and tentative evidence for AlO \citep{chubb_aluminium_2020}.
Under the assumption that the measured molecular abundances are representative of a gas mixture in chemical equilibrium, observations with HST have constrained the heavy element composition of WASP-43~b to 0.4--3.5 times the solar metallicity \citep{kreidberg_precise_2014}, or 0.3--1.7 times solar when \textit{Spitzer} data is included \citep{stevenson_spitzer_2017}, at $1\sigma$ confidence. 
These constraints are in agreement with 
high-resolution spectroscopy of the planet's day side \citep{lesjak_retrieval_2023}.
Classical one-dimensional retrievals, however, may bias abundance determinations based on planetary emission spectra \citep{feng_impact_2016, taylor_understanding_2020}. While mitigating such biases, retrievals of multiple phases simultaneously have yielded water abundances and metallicities that are broadly consistent with previous work \citep{feng_2D_2020, irwin_2.5D_2020}. Still, important caveats remain regarding model degeneracies \citep{changeat_exploration_2021}, as well as the data reduction that is used \citep{may_introducing_2020, chubb_exoplanet_2022}.

As part of the Transiting Exoplanets Community Early Release Science program, \citet{Bell_nightside_2024} observed a phase curve of WASP-43~b with the \textit{James Webb Space Telescope}'s (JWST) Mid-Infrared Instrument (MIRI). Their main findings were a confirmation of the large day-night brightness temperature contrast (1524 $\pm$ 35~K and 863 $\pm$ 23~K respectively), most likely caused by night-side clouds at 100~mbar, while the day side remains cloud-free. They also report a precisely measured phase offset of $7^\circ$ east of the substellar point, in agreement with the hot-spot offset measured with subsequent eclipse mapping studies on those same data \citep{hammond_two-dimensional_2024, Challener2024}, but smaller than what is predicted by 3D general circulation models (GCMs). Phase-dependent 1D retrieval models detect a uniform water abundance at each phase, but are inconclusive about additional molecules \citep{Bell_nightside_2024}. Interesting is also the non-detection of methane on the night-side hemisphere of WASP-43~b, despite the low temperature. The lack of \ce{CH4} could be explained through disequilibrium chemistry, in which gas on the night side is efficiently replenished by methane-depleted gas from the hot day side. This type of horizontal quenching was already predicted by \citet{Venot_global_2020}. A simultaneous phase curve retrieval of the MIRI/LRS data confirms the presence of \ce{H2O} and lack of \ce{CH4}, and in addition reports evidence for \ce{NH3} \citep{Yang_simultaneous_2024}. Their best-fit model, however, would require a two-orders-of-magnitude enhancement in elemental nitrogen, leaving the true nature of the \ce{NH3} spectral detection ambiguous.
Notably, neither the photochemical models of \citet{Venot_global_2020} or the phase curve retrieval of \citet{Yang_simultaneous_2024} include sulfur chemistry, which has now been shown to be important in the atmospheres of hot gas giants \citep{tsai23wasp39b}.

Despite being a major step-up in data quality from HST, the \textit{JWST}/MIRI spectroscopic phase curve showed only water absorption. Nonetheless, a stringent upper limit on \ce{CH4} \citep{Bell_nightside_2024}, tentative evidence for \ce{NH3} (\citealt{Yang_simultaneous_2024}, but also \citealt{feng_2D_2020}), and non-detection of the mid-infrared \ce{SO2} absorption that is observed in other hot giants \citep{dyrek_SO2_2024, powell_sulfur_2024}, provide a rich context for investigations into the three-dimensional chemical structure of the atmosphere. Additionally, another spectroscopic phase curve of WASP-43~b has been observed with \textit{JWST}'s NIRSpec/G395H (GTO 1224, PI: Stephan Birkmann), potentially probing spectral features of \ce{H2O}, CO, \ce{CO2}, and \ce{SO2} (Crouzet et al., in prep.). Hence, we aim to answer the questions: 
\begin{itemize}
    \item What are the requirements for methane depletion on the night side of WASP-43~b?
    \item At what metallicities are photochemical models consistent with the phase-dependent spectra emerging from the \textit{JWST}/MIRI and NIRSpec phase curves?
    \item How does sulfur chemistry affect the models and conclusions of \citet{Venot_global_2020}? 
    \item What are the main sources of uncertainty in photochemical modelling?
\end{itemize}

In this paper, we investigate the gas-phase chemistry of WASP-43~b using a suite of chemical kinetics models. We focus on describing the chemical composition of the planet self-consistently, highlighting model differences, and comparing our results with existing constraints from \citet{Bell_nightside_2024}, including additional, unpublished retrieval models (see Appendix~\ref{sec_retrievals} for details). The photochemical models that were used are described in Sec.~\ref{sec_models}. We discuss the chemistry of WASP-43~b in detail in Secs.~\ref{sec_CO}-\ref{sec_S}, and investigate the effects of the metallicity (Sec.~\ref{sec_metallicity}), jet speed (Sec.~\ref{sec_jetspeed}), and vertical mixing strength (Sec.~\ref{sec_kzz}). A self-consistent 3D climate-chemistry model is presented in Sec.~\ref{sec_3Dchemistry}. To illustrate the observational impact, we computed synthetic emission spectra at different phases, which are presented in Sec.~\ref{sec_spectra}. Additionally, we test the robustness of photochemical model predictions with a model intercomparison presented in Sec.~\ref{sec_comparison}. Finally, we discuss the observed methane depletion, cloud-chemistry feedback, and metallicity determination (Sec.~\ref{sec_discussion}), and present our conclusions (Sec.~\ref{sec_conclusions}).

\section{Forward models: codes and assumptions}\label{sec_models}

Throughout this paper, we mainly employ three different photochemical models to study the phase-dependent chemical composition of WASP-43~b: \textit{ACE-PAC}, \textit{KINETICS}, and \textit{Exo-FMS} with \textit{mini-chem}. Each of the models is set up differently, with a specific geometry, underlying thermal structure, and chemical reaction network in order to address different science questions. 

\begin{table*}
    \caption{Overview of the longitude-dependent photochemical models and their scientific application in this paper.}
    \label{tab_models}
    \centering
    \setlength\tabcolsep{0pt}
    \begin{tabular*}{\linewidth}{@{\extracolsep{\fill}}L{0.14\linewidth}C{0.28\linewidth}C{0.28\linewidth}C{0.28\linewidth}}
\hline\hline
\noalign{\smallskip}
Model  &  \textit{ACE-PAC} (Sec.~\ref{sec_ace-pac})  &  \textit{KINETICS} (Sec.~\ref{sec_kinetics})  &  \textit{Exo-FMS + mini-chem} (\ref{sec_exo-FMS})  \\
\noalign{\smallskip}
\hline
\noalign{\smallskip}
Science case & Impact of metallicity and wind~speed & Impact of sulfur chemistry, \qquad cf.~\citet{Venot_global_2020} & Self-consistent thermal, chemical, and dynamical coupling \\
\noalign{\smallskip}
\hline
\noalign{\smallskip}
Geometry  &  pseudo-2D   &  pseudo-2D  &  3D   \\
Thermal structure  &  \textit{Generic PCM}   &  \textit{ATMO}  &  self-consistent   \\
Elements  &  C, H, O, N, S   &  C, H, O, N, S, Cl  &  C, H, O, N  \\
Chemical network  &  96 species, 1140 reactions   &  150 species, 2350 reactions  &  13 species, 20 reactions \\
Photochemistry & Yes & Yes & No \\
Jet speed (km~s$^{-1}$) & 3.4 & 4.6 & self-consistent ($\approx3.3$) \\
Eddy diffusion, \\\quad $\kzz$ $(\textrm{cm}^2~\textrm{s}^{-1})$  &  $8.3\cdot10^7 \left(p_\textrm{bar}\right)^{-0.50}$  &  $10^7 \left(p_\textrm{bar}\right)^{-0.65}$  &  self-consistent \\
Stellar flux$^a$ \\\quad (erg~s$^{-1}$~cm$^{-2}$)   &  $\sim20$   &   $\sim30$    &   N/A  \\
\noalign{\smallskip}
\hline
\noalign{\smallskip}
Molar fractions  & Substellar / Antistellar  & Substellar / Antistellar & Substellar / Antistellar \\
\noalign{\smallskip}
$\log\left(\ce{H2O}\right)_\textrm{1~mbar}$  &  -3.5 \quad/\quad -3.5  &  -3.5 \quad/\quad -3.5  &  -3.5 \quad/\quad -3.5  \\
$\log\left(\ce{NH3}\right)_\textrm{1~mbar}$  &  -6.7 \quad/\quad -7.0  &  -7.2 \quad/\quad -7.5  &  -6.7 \quad/\quad -6.7  \\
$\log\left(\ce{CH4}\right)_\textrm{1~mbar}$  &  -8.2 \quad/\quad -8.2  &  -12.0 \quad/\quad -11.3  &  -7.3 \quad/\quad -7.3  \\
$\log\left(\ce{CO}\right)_\textrm{1~mbar}$  &  -3.4 \quad/\quad -3.4  &  -3.3 \quad/\quad -3.3  &  -3.3 \quad/\quad -3.3  \\
$\log\left(\ce{CO2}\right)_\textrm{1~mbar}$  &  -7.1 \quad/\quad -6.5  &  -7.1 \quad/\quad -6.4  &  -7.0 \quad/\quad -6.7  \\
$\log\left(\ce{HCN}\right)_\textrm{1~mbar}$  &  -8.1 \quad/\quad -8.0  &  -6.4 \quad/\quad -6.4  &  -7.8 \quad/\quad -7.8  \\
$\log\left(\ce{SO2}\right)_\textrm{1~mbar}$  &  -11.8 \quad/\quad -8.2  &  -11.4 \quad/\quad -7.9  &  N/A  \\
\noalign{\smallskip}
\noalign{\smallskip}
$\log\left(\ce{H2O}\right)_\textrm{100~mbar}$  &  -3.5 \quad/\quad -3.5  &  -3.5 \quad/\quad -3.5  &  -3.5 \quad/\quad -3.5  \\
$\log\left(\ce{NH3}\right)_\textrm{100~mbar}$  &  -6.9 \quad/\quad -6.9  &  -7.5 \quad/\quad -7.5  &  -6.8 \quad/\quad -6.8  \\
$\log\left(\ce{CH4}\right)_\textrm{100~mbar}$  &  -7.7 \quad/\quad -7.9  &  -8.6 \quad/\quad -8.0  &  -8.0 \quad/\quad -7.9  \\
$\log\left(\ce{CO}\right)_\textrm{100~mbar}$  &  -3.4 \quad/\quad -3.4  &  -3.3 \quad/\quad -3.3  &  -3.3 \quad/\quad -3.3  \\
$\log\left(\ce{CO2}\right)_\textrm{100~mbar}$  &  -7.2 \quad/\quad -6.9  &  -7.3 \quad/\quad -6.9  &  -7.3 \quad/\quad -6.9  \\
$\log\left(\ce{HCN}\right)_\textrm{100~mbar}$  &  -8.1 \quad/\quad -8.1  &  -8.4 \quad/\quad -8.5  &  -7.9 \quad/\quad -7.9  \\
$\log\left(\ce{SO2}\right)_\textrm{100~mbar}$  &  -13.3 \quad/\quad -15.9  &  -12.3 \quad/\quad -15.4  &  N/A  \\
\noalign{\smallskip}
\hline
\noalign{\smallskip}
\end{tabular*}
\begin{minipage}{\linewidth} 
    {\footnotesize $^a$ Integrated high-energy flux between 10~nm and 200~nm at 1~au. The full stellar spectra are shown in Fig.~\ref{fig_stellarspectra}. 
    }
\end{minipage}
\end{table*}

The specific setups enable us to investigate several aspects of WASP-43~b's chemistry. Primarily, the pseudo-2D model \textit{ACE-PAC} is used to simulate the longitude-dependent chemistry under various metallicities and wind speeds. Additionally, we employ \textit{KINETICS} with the exact same input parameters as used in \citet{Venot_global_2020} in order to study the impact of sulfur chemistry. Finally, the 3D model \textit{Exo-FMS + mini-chem} is complementary to the pseudo-2D models and enables self-consistent simulations with chemistry fully coupled to dynamics. Table~\ref{tab_models} provides a general overview of the different models, their science cases, setups, and result. 

While different in their implementation, these codes above have in common that they numerically calculate the chemical kinetics, namely the rate of change in chemical concentrations over time, including production and loss through chemical reactions and the transport of species. Specifically, the following system of partial differential equations is solved for each species $i$:
\begin{equation}\label{eq_chem}
    \partfrac{n_i}{t} + \vec{\nabla} \cdot (n_i \vec{v_i}) = P_i - L_i    
\end{equation} where $n_i$ is the number density of species $i$, $\vec{v_i}$ is the transport velocity vector, and $P_i$ and $L_i$ are the chemical production and loss rates. Transport may occur both via diffusion or advection. The pseudo-2D models consider vertical transport via eddy diffusion and molecular diffusion. In addition, they mimic zonal advection to first order, by modelling the rotation of the atmospheric column around the planet over time. The 3D climate-chemistry model \textit{Exo-FMS} likewise adopts vertical diffusive transport, in addition to chemical advection computed directly via the velocity vector. Photochemical kinetics has the advantage of taking into account compositional changes due to chemical reactions, but also due to potentially important photochemical dissociation or dynamical mixing processes. 

Next, we describe the specific implementations of the photochemical models used in this work. The models that are used to generate the underlying thermal structures are further detailed and compared in Appendix~\ref{sec_temperature}.

\subsection{ACE-PAC model}\label{sec_ace-pac}

The \textit{ACE-PAC} model \citep{Agundez_pseudo_2014} implements equation~\eqref{eq_chem} in a pseudo-2D framework. This means that the underlying grid is a one-dimensional, vertical column, but the assumed background temperature and irradiation angle are changed over time in order to represent different equatorial longitudes on the planet. As such, this setup calculates photochemical kinetics in the atmospheric column, as the entire column rotates around the planetary equator at a speed defined by the superrotating equatorial wind jet. Given the dominance of superrotation on hot Jupiter climates \citep[e.g.][]{Showman_3D_2015, baeyens_grid_2021, Roth_hot_2024}, a pseudo-2D model approximates chemical mixing in the equatorial region of irradiated exoplanet atmospheres to first order, without having to run computationally expensive 3D models (see also Sec.~\ref{sec_3Dchemistry}). As such, this code has been applied previously to investigate day-to-night transport of photochemical products \citep{Konings_impact_2022, baeyens_grid_2022, baeyens_photodissociation_2024}. In this paper, we leverage the relatively quick runtime of this framework and use \textit{ACE-PAC} to investigate the impact of various metallicities and wind speeds on the photochemistry of WASP-43~b.

The chemical reaction network used here is the C-H-O-N-S network of \textit{VULCAN} \citep{tsai_comparative_2021}, incorporating 96 species and 570 reversible reactions. Backward reaction rates are calculated from the forward rates based on their equilibrium constants. The chemical network has been applied previously to explain the presence of \chem{SO_2} on WASP-39~b \citep{tsai23wasp39b, powell_sulfur_2024}, and has been compared with other sulfur-bearing networks for hot Jupiters \citep{veillet_inclusion_2025}. Photochemical rates are computed based on an observed stellar ultraviolet (UV) spectrum of WASP-43~b \citep{behr_muscles_2023} assuming a stellar radius of 0.667~R$_\odot$, orbital distance of 0.01528~AU, and zero albedo \citep{bonomo_gaps_2017}. Additionally, we use photodissociation cross-sections and branching ratios from a mix of theory and experiments \citep{venot_chemical_2012, hebrard_photochemistry_2013, dobrijevic_coupling_2014, heays_photodissociation_2017, hrodmarsson_photodissociation_2023}. 

To setup our pseudo-2D photochemical model for the \textit{ACE-PAC} model, pressure- and longitude-dependent temperatures were extracted from the \textit{Generic PCM} 3D GCM discussed in Sec.~\ref{sec_gpcm} by taking a cosine-weighted average over latitude in the equatorial region ($\pm20^\circ$ latitude). The temperature grid uses 90 longitude samples (i.e.~$4^\circ$ spacing) and 120 pressures logarithmically spaced between 100~bar and $10^{-8}$~bar. Since the upper boundary of the GCM cannot be placed at such low pressures while maintaining numerical stability, temperatures are extrapolated isothermally in the upper atmosphere. 
The jet speed for the pseudo-2D model is analogously computed from the \textit{Generic PCM}, but is additionally averaged between 1~bar and 10$^{-4}$~bar and over all longitudes. This procedure yields a nominal value of 3.4~km~s$^{-1}$. The eddy diffusion coefficient $K_{zz}$ is computed using the parametrized equation (1) in \citet{moses21pseudo2d}, with $H_\textrm{1 mbar} = 103$~km and $\Teff= 1450$~K. 
Each model is initialized at thermochemical equilibrium at the substellar point and run for 50 planetary rotations ($\sim10^7$~s), which is sufficient to reach a periodic steady-state.

\subsection{KINETICS model}\label{sec_kinetics}

To compare our photochemical model results with earlier predictions from \citet{Venot_global_2020}, we also use the Caltech/JPL \textit{KINETICS} model \citep{allen81,yung84} within the above-described pseudo-2D framework to explore how sulfur photochemistry can affect the overall species distributions. Therefore, we adopt the exact same input parameters as \citet{Venot_global_2020}, aside from the inclusion of sulfur chemistry in this work. 

The \textit{KINETICS} model has been used previously in the pseudo-2D mode for WASP-43 b \citep[see Sections 3.3.2 and 4.2 of][]{Venot_global_2020} and other exoplanets \citep{moses21pseudo2d}, without the inclusion of sulfur species and reactions. Here we follow similar procedures as in those previous works, including adopting the same \textit{2D-ATMO} pressure-temperature profiles as a function of longitude for WASP-43~b that are described in \citep{Venot_global_2020} and Section~\ref{sec_atmo}.  However, instead of using the C-H-O-N chemical network from \citep{moses13gj436b}, we now adopt an updated chemical network that includes sulfur and chlorine species and reactions, as is described for \textit{KINETICS} in \citet{tsai23wasp39b}.  This fully reversed chemical network contains $\sim$150 total species and 2350 reactions. Although the addition of chlorine has little effect on the results other than adding a non-trivial amount of HCl to the atmosphere, the sulfur-based photochemistry can affect the C-H-O-N species abundances, as we discuss further below.

The eddy diffusion coefficient $K_{zz}$, which controls atmospheric transport in the vertical direction, is assumed to be longitude-independent and follows the expression described by \citet{Venot_global_2020}: $K_{zz}$ (cm$^2$ s$^{-1}$) = 10$^7 \, P_\textrm{bar} ^{-0.65}$ for pressures $P_\textrm{bar}$ $<$ 300 bar, and $K_{zz}$ = 10$^{10}$ cm$^2$ s$^{-1}$ for $P_\textrm{bar}$ $>$ 300 bar, with $K_{zz}$ also capped at a maximum of 10$^{10}$ cm$^2$ s$^{-1}$ in the upper atmosphere.  Note that this parameterization for $K_{zz}$ differs somewhat from that assumed  in the \textit{ACE-PAC} model described in Sec.~\ref{sec_ace-pac}, but the two descriptions are equal within one order of magnitude. The model extends from 64~bar to 10$^{-11}$~bar, with 198 vertical grid points spaced equally in log pressure. The stellar spectrum adopted for the model is described in Section 3.3.1 of \citet{Venot_global_2020}.

Also following \citet{Venot_global_2020}, the \textit{KINETICS} models assume an equatorial jet speed of 4.6 km s$^{-1}$, which defines the effective rate at which the atmospheric column rotates about the planet.  Both the thermal structure and incidence angle of the incoming stellar radiation change with longitude as the atmospheric column rotates, and those values remain fixed for an amount of time defined by the effective rotation rate and the longitude grid spacing \citep[36$^{\circ}$ for this model, see][]{Venot_global_2020}. From the residence time at one longitude, the volume mixing ratios as a function of pressure are passed as initial conditions to the next longitude.  We start the calculations on the day side at 36$^{\circ}$ longitude with the initial conditions set to be the fully converged 1D steady-state thermo/photochemical model solution. The calculations are run for 100 full rotations ($\sim10^7$~s) at which point the results for each longitude exhibit no further changes.


\subsection{3D Exo-FMS GCM + mini-chem}\label{sec_exo-FMS}

In addition to the pseudo-2D photochemical models, we also include a simulation of WASP-43~b using the 3D \textit{Exo-FMS} GCM \citep{Lee2021} coupled to the \textit{mini-chem} chemical kinetics scheme \citep{Tsai2022,Lee2023}.
This GCM output was also used previously as part of the NIRSpec/G395H white light curve and thermal mapping analysis in \citet{Challener2024}. The model is complementary to the aforementioned pseudo-2D models. It simplifies the chemical network and does not incorporate sulfur chemistry nor photodissociation, but it computes the full thermal-chemical-dynamical coupling self-consistently. As such, the model enables us to verify the pseudo-2D assumptions and investigate 3D effects such as meridional transport.

The \textit{Exo-FMS} GCM setup for hot Jupiters is primarily described in \citet{Lee2021}, including further physics module additions from \citet{Lee2024}.
For the current WASP-43~b simulations, we took planetary parameters from \citet{Bell_nightside_2024} and assumed solar metallicity.
In this study, we only use the GCM simulation output without the effects of cloud formation and their opacity from \citet{Challener2024}.
We include correlated-k radiative transfer with 11 wavelength bins \citep{kataria_three-dimensional_2013}, taking into account the effect of thermal feedback from the time-dependent changing chemical composition following the adaptive equivalent extinction method in \citet{Amundsen2017}.
Mixing length theory \citep[e.g.][]{Marley2015} is included, which enables emulating tracer transport from convective motions through estimating a vertical $K_{\rm zz}$ value coupled to a tracer diffusive scheme \citep{Lee2024}.
This convective mixing component is applied in addition to the regular 3D advective transport tracer scheme in the GCM. The zonal component of the wind is purely advective, and the simulation has a mean equatorial wind speed of 3.3~km~s$^{-1}$, similar to the \textit{Generic PCM}.
We assume a constant Na and K abundance at solar metallicity, assuming a number fraction of 10$^{-6}$ and 10$^{-7}$ for Na and K respectively.

The mini-chem scheme includes 10 forward reactions and 13  chemical species (OH, \ce{H2}, \ce{H2O}, H, CO, \ce{CO2}, O, \ce{CH4}, \ce{C2H2}, \ce{NH3}, \ce{N2}, HCN, He), utilizing net forward reaction rate tables to significantly reduce the size of the network, while retaining good accuracy when compared to a full network implementation \citep{Tsai2022}.
This makes the scheme more computationally amenable for inclusion in large scale 3D atmospheric modelling such as GCMs.
The current scheme does not include the effects of photochemistry or sulfur species, however.

\section{WASP-43~b chemistry}\label{sec_chemistry}


\begin{figure*}%
    \centering%
    \begin{minipage}{0.49\linewidth}%
    \includegraphics[width=\linewidth]{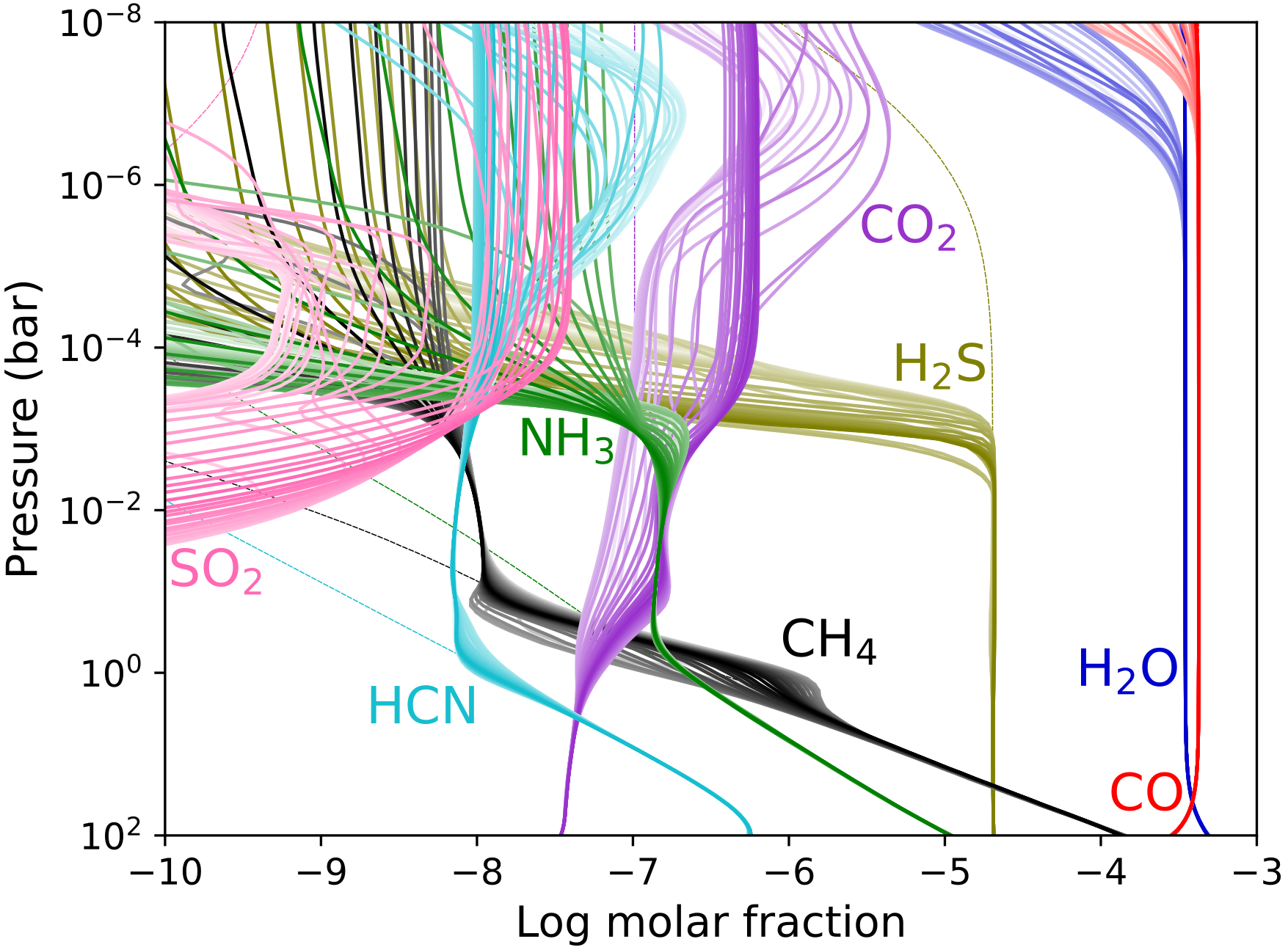}%
    \caption{The chemical abundances of major species in WASP-43~b, computed using the \textit{ACE-PAC} pseudo-2D kinetics model (\textit{solid}), assuming solar metallicity. Different lines represent different equatorial longitudes, and are lighter for longitudes closer to the substellar point. The chemical equilibrium composition at substellar point is indicated (\textit{dotted}).}%
    \label{fig_chem_acepac}%
    \end{minipage}\hfill%
    \begin{minipage}{0.49\linewidth}%
    \includegraphics[width=\linewidth]{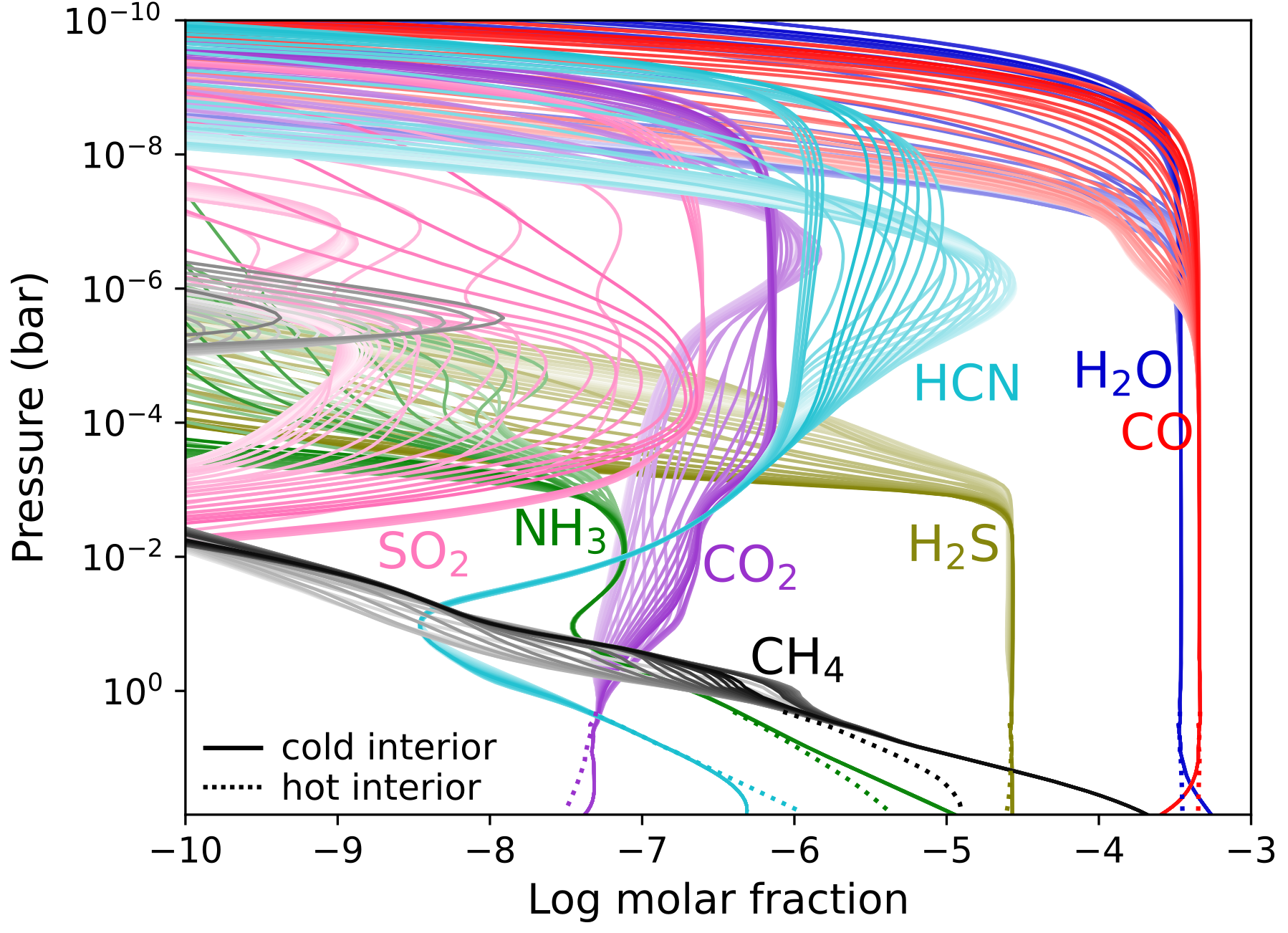}%
    \caption{The chemical abundances of major species in WASP-43~b, computed using the \textit{KINETICS} pseudo-2D photochemical model, assuming solar metallicity. The adopted temperature profiles were computed with the radiative transfer code \textit{2D-ATMO} \citep[as in][]{Venot_global_2020}. Two assumptions for interior heat transport are used, resulting in compositional differences for a cold (\textit{solid}) and hot interior (\textit{dotted}).}%
    \label{fig_chem_kinetics}%
    \end{minipage}%
\end{figure*}




The nominal \textit{ACE-PAC} and \textit{KINETICS} pseudo-2D photochemical models are shown in Figs.~\ref{fig_chem_acepac} and \ref{fig_chem_kinetics}. For dominant species, the results are consistent with each other and with \citet{Venot_global_2020}. At solar metallicity, H$_2$O and CO are the dominant species, whereas CH$_4$ and NH$_3$ remain below 1 ppm, except in the deepest parts of the atmosphere. Furthermore, N$_2$ (not shown) and H$_2$S are the largest reservoirs for nitrogen and sulfur throughout most of the pressure range. At low pressures below $\sim$1~mbar, however, H$_2$S is destroyed and sulfur is mostly present in its atomic form.

Chemical disequilibrium through transport-induced quenching can be seen in both WASP-43~b models. There are clear deviations from chemical equilibrium for \chem{CH_4}, \chem{NH_3}, and HCN from the $\sim$0.1~bar level upward (due to vertical mixing). The same species also show homogeneous, longitude-independent concentrations in the lower and middle atmosphere (due to horizontal mixing). The effects of photochemistry are likewise visible, as clear photodissociation of \chem{H_2S}, \chem{CH_4}, and \chem{NH_3} occurs at low pressures ($p \lesssim 1$~mbar), combined with an increase in \chem{SO_2}, \chem{CO_2}, and HCN. Overall, the 2D photochemical models show transitions in trace species from the deep to the upper atmosphere, but the dominant species remain fairly unaffected. 

\subsection{Carbon and oxygen chemistry}\label{sec_CO}

\begin{figure}
    \centering
    \includegraphics[width=\linewidth]{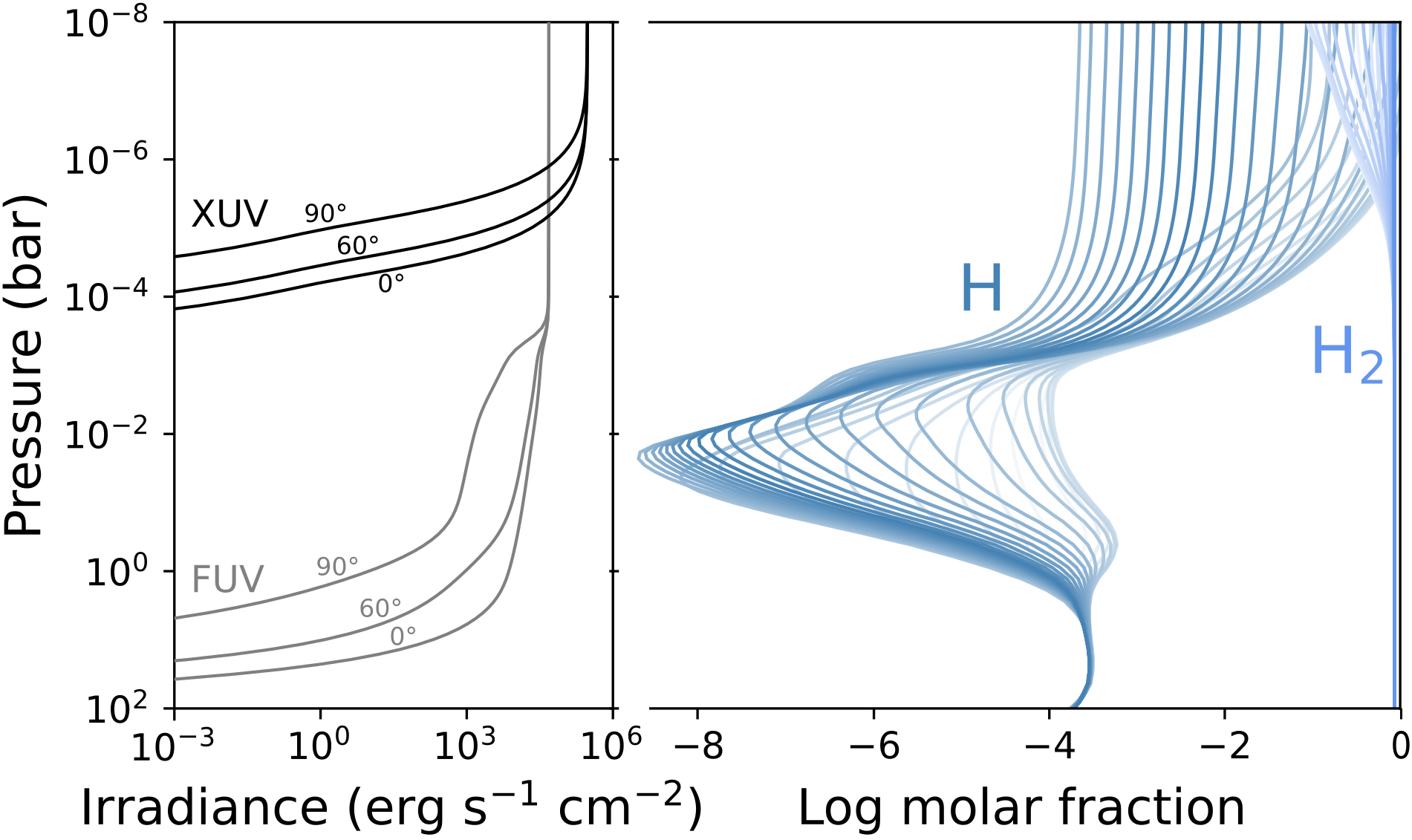}
    \caption{Stellar radiative flux and the abundances of H and H$_2$ as a function of atmospheric pressure for the \textit{ACE-PAC} solar metallicity model. Displayed are the wavelength-integrated XUV (10~nm - 121~nm) and FUV (121~nm - 200~nm) fluxes, showing different penetration depth in the atmosphere. Three zenith angles are shown: 0$\degrees$ (substellar point), 60$\degrees$, and 90$\degrees$ (limb). The different abundance profiles correspond to different longitudes of the planet, with lighter shades representing locations closer to the substellar point.
    }
    \label{fig_H}
\end{figure}

The coupled carbon and oxygen chemistry on WASP-43~b is dominated by \chem{H_2O} and \chem{CO}. Even at solar metallicity, the second-most abundant carbon species is \chem{CO_2}. It is marginally more abundant on the night side, and mostly produced via the direct combination reaction CO + O($^3P$)/O($^1D$) $\rightarrow$ \chem{CO_2}. Here, atomic oxygen originates from attacks of \chem{H_2O} by atomic hydrogen. The subsequently produced hydroxyl radical OH either reacts with a second hydrogen atom to produce O($^3P$) + H$_2$ or it is photodissociated to produce O($^1D$) + H. The net reaction is \chem{H_2O} + 2~H $\rightarrow$ 2~\chem{H_2} + O. The above mechanism is controlled by the abundance of atomic hydrogen, which is shown in Fig.~\ref{fig_H}. The concentration of atomic hydrogen peaks at pressures lower than 1~mbar, which is where the majority of the extreme ultraviolet (XUV) radiation gets absorbed. In the upper atmosphere, the day side contains about hundred to thousand times as much atomic H as the night side. Atomic H replaces molecular H$_2$ as the dominant species on the star-facing hemisphere at pressures $\lesssim 1~\mu$bar. 

The methane concentrations appear to be model sensitive. The \textit{ACE-PAC} model shows a more vertically quenched profile at $\sim$millibar pressures, leading to a molar fraction of 10$^{-8}$ (Fig.~\ref{fig_chem_acepac}). The \textit{KINETICS} model predicts less vertical quenching and an even lower fraction of $\lesssim 10^{-10}$ (Fig.~\ref{fig_chem_kinetics}). Here, on the other hand, the upper atmosphere shows a local peak in \chem{CH_4} due to CO photolysis on the morning limb and the subsequent reduction of atomic carbon into \ce{CH4}. This mechanism does not lead to a buildup of methane, however, as the gas crosses into the day-side hemisphere and gets attacked by atomic hydrogen. This behavior is not present in the \textit{ACE-PAC} model, due to differences in temperature (140~K colder) and horizontal advection timescale ($35\%$ longer), resulting in a different fate of atomic carbon at microbar pressures. These differences are also reflected in the coupled carbon-sulfur chemistry (Sec~\ref{sec_S}). 

Nonetheless, the general behavior for methane is consistent, namely a low concentration that decreases with altitude and is strongly zonally quenched. As a result, little to no day-night differences are expected, despite the low night-side temperatures favoring \chem{CH_4} over CO. The interconversion of CO to \chem{CH_4} is well-studied for extrasolar giant planets \citep{visscher_quenching_2011, moses_disequilibrium_2011, zahnle_methane_2014, tsai_toward_2018} and also for Jupiter \citep[e.g.,][]{yang2026coupled}. In our models, the pathway is bottlenecked by either the formation of methanol: \chem{CH_2OH} + H $\rightarrow$ \chem{CH_3OH}, or the subsequent dissociation of methanol: \chem{CH_3OH} $\rightarrow$ \chem{CH_3} + OH, depending on altitude. The characteristic timescale of this process is $10^4$~s -- $10^5$~s. This value is comparable to the dynamical timescale for zonal advection $\pi R_p / u$ for zonal wind speeds $u > 1$~km/s, which explains the inability of the atmosphere to produce substantial amounts of night-side methane. 

\subsection{Nitrogen chemistry}\label{sec_N}

Nitrogen chemistry in our $1\times$ solar metallicity models is mostly determined by the balance between \chem{N_2} and \chem{NH_3}, as well as the increase in HCN at higher altitudes. The dominant nitrogen-bearing species at all pressures is \chem{N_2} with a molar fraction of 60~ppm ($6\cdot10^{-5}$), except at the lowest pressures of the model where it is instead found in atomic form.

The dominant pathway from ammonia to nitrogen gas is different on the day and night side of the planet. The strong day-side irradiation enables photolysis of the triple-bonded N$_2$ into atomic nitrogen. Subsequent reactions with hydrogen enable further reduction: N($^4$S) + \chem{H_2} $\rightarrow$ \chem{NH_2} followed by \chem{NH_2} + H $\rightarrow$ \chem{NH_3}. The latter reaction is not thermodynamically favored, however, and runs faster in the opposite direction, explaining why nitrogen photolysis does not result in the formation of ammonia. In the deeper layers ($p \approx 10^{-2}$~bar), this pathway is throttled by the diminished photolysis of \chem{N_2}.

On the night side, on the other hand, photolysis is absent as the main mechanism to break the strong \chem{N_2} bonds. Instead, \ce{N2} destruction is facilitated by the presence of atomic oxygen forming nitrous oxide, \chem{N_2O}. Nitrogen is then freed up via the reaction \chem{N_2O} + H $\rightarrow$ NO + NH, which has a long chemical timescale of $\sim$$10^7$~s. The full scheme is given by
\begin{align*}
    &\chem{N_2} + \textrm{O(}^3\textrm{P)} \rightarrow \chem{N_2O} \\
    &\chem{N_2O} + \chem{H} \rightarrow \chem{NO} + \chem{NH} \\
    &\chem{NH} + \chem{H_2} \rightarrow \chem{NH_2} + \chem{H} \\
    &\chem{NO} + \chem{H} \rightarrow \chem{HNO} \\
    &\chem{HNO} + \chem{H_2} \rightarrow \chem{NH} + \chem{H_2O} \\
    &\chem{NH_2} + \chem{H} \rightarrow \chem{NH_3} \\
    \noalign{\smallskip}
    \hline
    \noalign{\smallskip}
    &\chem{N_2} + \textrm{O(}^3\textrm{P)} + 4~\chem{H_2} \rightarrow 2~\chem{NH_3} + \chem{H_2O} \quad \textrm{(net)}.
\end{align*} However, due to long chemical timescales compared to the wind jet speed, there is insufficient time to build up ammonia on the night side.

Another important nitrogen-bearing molecule is HCN, and its concentration differs between our two models. In the \textit{ACE-PAC} model of WASP-43~b, HCN is vertically quenched at low abundance (10$^{-8}$) at pressures lower than 1~bar. Efficient horizontal mixing eliminates day-night differences (Fig.~\ref{fig_chem_acepac}). However, at low pressures ($\sim$$10^{-6}$~bar), the HCN concentration on the day side of the planet increases by more than an order of magnitude. The production of high-altitude HCN is enabled by photochemical destruction of \chem{N_2} and \chem{NH_3}, leading to an increased amount of atomic nitrogen. Additionally, photodissociation of CO produces atomic carbon. Most of it is recycled back into CO, but 1:1~000~000 carbon atoms ends up in CN via N($^4$S) + C $\rightarrow$ CN. The reaction is partially catalyzed by the presence of atomic oxygen through the slightly faster scheme N($^4$S) + O($^3$P) $\rightarrow$ NO followed by NO + C $\rightarrow$ CN + O($^3$P). The final fate of most CN is to form HCN, whose main sinks are photodissociation and attacks by atomic oxygen leading to breaking the C-N bond into CO + NH. 

Notably, the \textit{KINETICS} model produces much more HCN, peaking at over 10~ppm and making it the second-most abundant carbon-bearing molecule (Fig.~\ref{fig_chem_kinetics}). The reason for the higher production rate is a higher abundance of nitrogen- and carbon-donor species, due to strong photolysis at low pressure. The dominant reaction scheme is:
\begin{align*}
    &\chem{CO} + \textrm{h}\nu  \rightarrow    \chem{C} + \textrm{O($^3P$)} \\
    &\chem{N_2} + \textrm{h}\nu  \rightarrow    2~\textrm{N($^4S$)} \\
    &\chem{H_2O} + \chem{H}     \rightarrow     \chem{OH} + \chem{H_2} \\
    &\textrm{N($^4S$)} + \chem{OH}       \rightarrow     \chem{NO} + \chem{H} \\
    &\chem{C} + \chem{NO}       \rightarrow     \chem{CN} + \textrm{O($^3P$)} \\
    &\chem{CN} + \chem{H_2}     \rightarrow     \chem{HCN} + \chem{H} 
\end{align*} Although these reactions are also represented in the nominal \textit{ACE-PAC} model, small differences in the stellar high-energy spectrum, the upper thermal structure of the model, molecular diffusion, and $\kzz$ will affect the photolysis rates of \ce{N_2} and CO and the subsequent fate of atomic radicals. Because of these reasons, the HCN abundance is not as robustly predicted as \ce{H2O}, \ce{CO}, or \ce{CO2} (see also Fig.~\ref{fig_compare}).

\subsection{Sulfur chemistry}\label{sec_S}

The variation of sulfur species with longitude in our 1$\times$ solar atmospheric metallicity pseudo-2D models extends deep to $\gtrsim$1 bar pressure levels in the atmosphere, indicating short photochemical lifetimes for many of the sulfur species.  The most abundant photochemical products on the day side are S, SH, CS, and to a lesser extent S$_2$ (not shown in Figs.~\ref{fig_chem_acepac} and \ref{fig_chem_kinetics}).  Atomic S is the dominant product in the 10$^{-3}$--10$^{-10}$ bar region at all longitudes, essentially completely replacing H$_2$S at high altitudes in our \textit{ACE-PAC} and \textit{KINETICS} models.  Atomic S is produced largely through SH + H $\rightarrow$ S + H$_2$, with the SH produced from a variety of sources, including the reverse of this reaction, or through H + H$_2$S = SH + H$_2$ or S($^1$D) + H$_2$ = SH + H, depending on altitude.  The large atomic H abundance on the planet's day side and at high altitudes favors the production of atomic S and inhibits the recycling back to H$_2$S.  The chemical lifetime of atomic S is extremely short at depth ($<1$~s at 1 bar) but increases with increasing altitude to exceed a year at 
$p$ $\lesssim$10$^{-8}$ bar.  Atomic S actually increases at night at very high altitudes but drops off significantly at night in the 1--10$^{-3}$ bar region due to the smaller abundance of atomic H (Fig.~\ref{fig_H}) and to a shift in the balance of thermochemical reactions back towards eventual reformation of H$_2$S at these pressures. The long lifetime at high altitudes allows horizontal transport to homogenize the zonal abundance of S through much of the stratosphere. 

The large H abundance on the day side --- due to both photolysis of hydrogen-bearing species and due to higher day-side temperatures --- also favors SH production on the day side, but the SH chemical lifetime is much shorter than that of atomic S at high altitudes, remaining at less than a second at 10$^{-8}$ bar during both day and night.  Therefore, SH drops in importance as a sulfur product at pressures less than 10$^{-5}$~bar, achieving a peak mixing ratio of 6 or 8~ppm in \textit{ACE-PAC} and \textit{KINETICS} respectively, each time at $\sim4 \cdot \, 10^{-4}$~bar on the day side, exceeding the abundance of H$_2$S at this and lower pressures.  Note that the H$_2$S mixing ratio itself drops off fairly precipitously at pressures less than $10^{-3}$~bar and shows longitudinal variation only for pressures less than $\sim 10^{-3}$ bar. Within the region where H$_2$S is beginning to rapidly fall off, S$_2$ becomes an important sulfur product.  The S$_2$ is produced largely through S + SH $\rightarrow$ S$_2$ + H, and is lost through the reverse of that reaction and through photolysis.  At night when photolysis does not occur, the S$_2$ survives longer and thus has an elevated mixing ratio through the night-side longitudes.  The S$_2$ lifetime is short enough that large variations with longitude are expected.

Notably, some coupled carbon-sulfur photochemistry does occur. CS becomes an important product, while CS$_2$ does not under WASP-43~b temperature conditions \citep[see also][]{veillet_inclusion_2025}.  The CS is produced via reactions such as SO + C $\rightarrow$ CS + O($^3P$) and SH + C $\rightarrow$ CS + H, where the C ultimately derives from CO photolysis. The CS is confined to high altitudes (e.g., $p$ $\lesssim$ 10$^{-3}$ bar) on WASP-43~b and has a narrower peak during the daytime, expanding vertically during the night when loss processes such as reactions with OH, O, and SH are less effective due to smaller night-side abundances of these short-lived destructive radicals.  Trace amounts of CS$_2$ are produced (peak molar fractions of $10^{-12}$ in \textit{ACE-PAC} and $10^{-10}$ in \textit{KINETICS}), but the dominant production reaction --- SH + CS $\rightarrow$ CS$_2$ +  H --- is ineffective due to the fact that the SH abundance drops off in abundance at the higher altitudes where CS is confined.

Coupled oxygen-sulfur photochemistry also occurs in our models.  SO and SO$_2$ achieve mixing ratios as large as $\sim$1 ppm and $\sim$0.25 ppm in the upper stratosphere in the 1$\times$ solar metallicity \textit{KINETICS} model, and about an order of magnitude lower in \textit{ACE-PAC}, both species more abundant and more vertically extended on the night side than the day side.  As explained by \citet{tsai_day-night_2023}, this apparent paradox of enhanced night-side abundances occurs because although day-side photochemistry is required for production of the precursor species for SO and SO$_2$ formation, these precursors such as OH and O can be transported by zonal winds and/or form on the night side, where SO and SO$_2$ can continue to be produced without the corresponding photochemical destruction. The main reaction for SO production is S + OH $\rightarrow$ SO + H  during both day and night.  The dominant SO$_2$ production reaction is SO + OH $\rightarrow$ SO$_2$ + H, again both during the day and at night.  The long chemical lifetime of atomic S has already been mentioned, and that supplies a key reactant to the night side.  The key radical species OH has a loss timescale on the order of minutes on the night side at 10$^{-8}$ bar, being lost via OH + H$_2$ $\rightarrow$ H$_2$O + H.  However, OH is produced on the night side at these high altitudes via O + H$_2$ $\rightarrow$ OH + H, so the net photochemical lifetime for OH is rather on the order of days, depending on longitude. As a result, SO and SO$_2$ continue to be produced on the night side. Water, too, has a very long chemical lifetime once it reaches the night side, and some OH is produced via H + H$_2$O $\rightarrow$ OH + H$_2$, leading to additional SO and SO$_2$.  Note that this large night-side abundance of SO and SO$_2$ is a consequence of the horizontal transport of oxygen species and, obviously, does not occur in 1D models of the night side \citep[see also][]{tsai_day-night_2023}.  

While not a dominant effect, we do find signs of sulfur chemistry affecting the atmospheric composition of non-sulfur species in our models. As an example, CH$_4$ is less abundant when sulfur species are present than when they are not. In a similar pseudo-2D model presented by \citet{Venot_global_2020} with the same thermal structure and other assumptions as our \textit{KINETICS} model, except without sulfur species in the model, the CH$_4$ that is lost during the day recovers at night to mixing ratios of $\sim 2\cdot 10^{-7}$ throughout the upper stratosphere, with an additional region of CH$_4$ production in the $\sim$10$^{-3}$--10$^{-6}$ bar region that persists at most longitudes. The current model \textit{with} sulfur photochemistry likewise shows this $\sim$microbar \chem{CH_4} peak (Fig.~\ref{fig_chem_kinetics}), but it is an order of magnitude smaller than in \citet{Venot_global_2020}. Additionally, the methane peak diminishes progressively during the day, and does not recover at night in the stratosphere, like it did in the model without sulfur chemistry \citep{Venot_global_2020}. The explanation lies in the fate of high-altitude atomic carbon. In the model of \citet{Venot_global_2020}, C-atoms are transported by horizontal winds from their production site to the night side, ending up in CH$_x$ hydrocarbons and eventually CH$_4$. In our model, conversely, that atomic C is being scavenged by sulfur species, eventually forming CS through the processes discussed above. Thus, more of the carbon is tied up in CS, with little extra production of CH$_4$.  With sulfur present, the abundance of CH$_4$ is suppressed at all longitudes, but especially at night, in comparison with the models without sulfur. We conclude that coupled carbon-sulfur chemistry is a secondary mechanism that acts to further deplete the night-side methane abundance, aside from the chemical disequilibrium induced by horizontal transport \citep{Venot_global_2020}.

\subsection{Compositional changes with metallicity}\label{sec_metallicity}
\label{sec:metallicity}

\begin{figure}
    \centering
    \includegraphics[width=\linewidth]{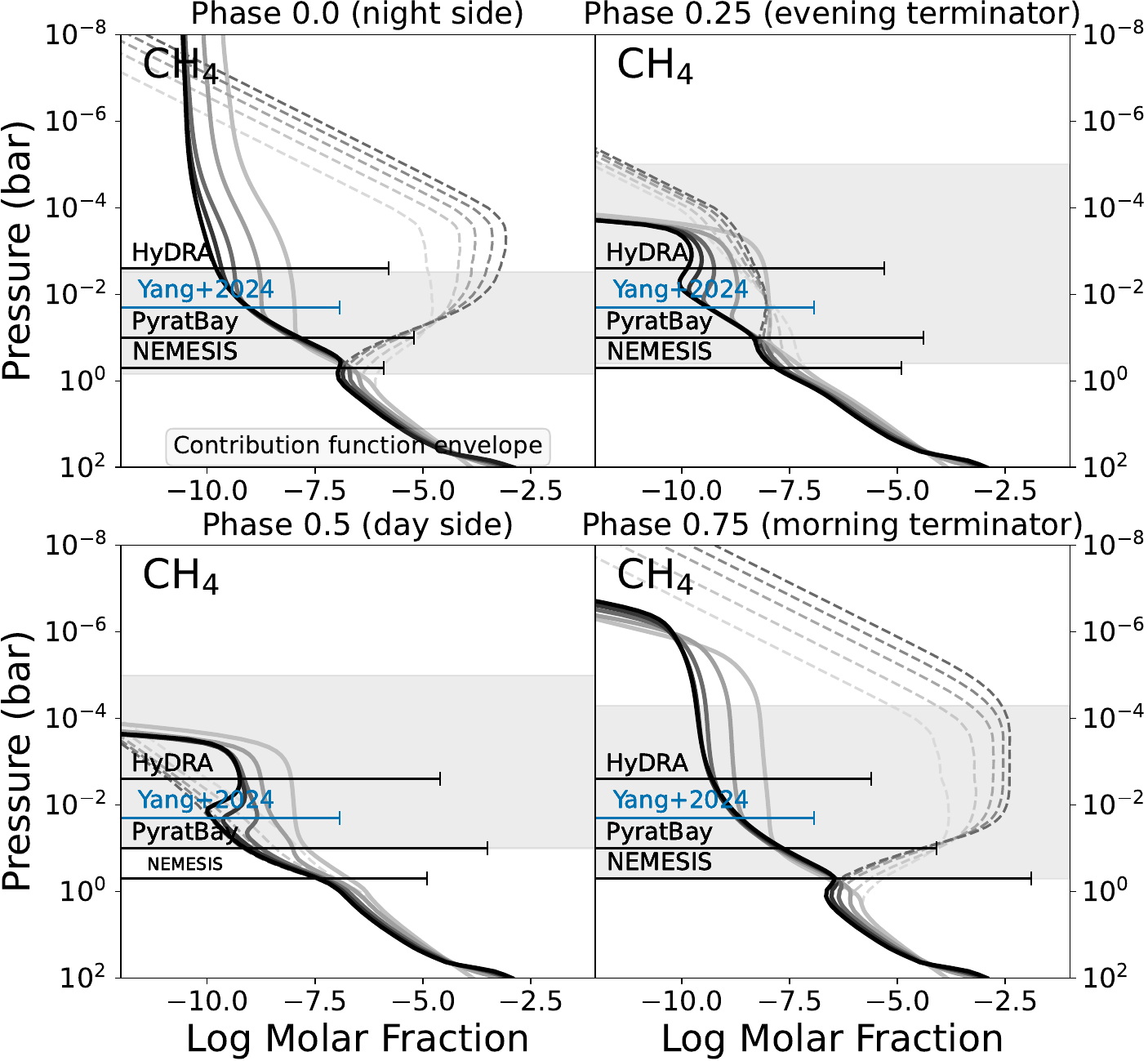}
    \caption{The methane abundance calculated using the \textit{ACE-PAC} pseudo-2D kinetics model (\textit{solid}) deviates strongly from chemical equilibrium (\textit{dashed}) on the planet's night side and cold morning terminator. The results agree with upper limits from atmospheric retrievals, in contrast to the chemical equilibrium predictions. Metallicities between 1 $\times$ solar to 10 $\times$ solar are shown, with darker colors corresponding to a higher metallicity. Retrieval constraints are from \citet{Bell_nightside_2024} (\textbf{black}), and \citet{Yang_simultaneous_2024} (\textbf{\color{YangBlue}{blue}}). The lines corresponding to different retrievals are vertically displaced for clarity. Outlined in gray is the contribution envelope as reported in \citet{Bell_nightside_2024}.}
    \label{fig_CH4_metallicities}
\end{figure}

\begin{figure}
    \centering
    \includegraphics[width=\linewidth]{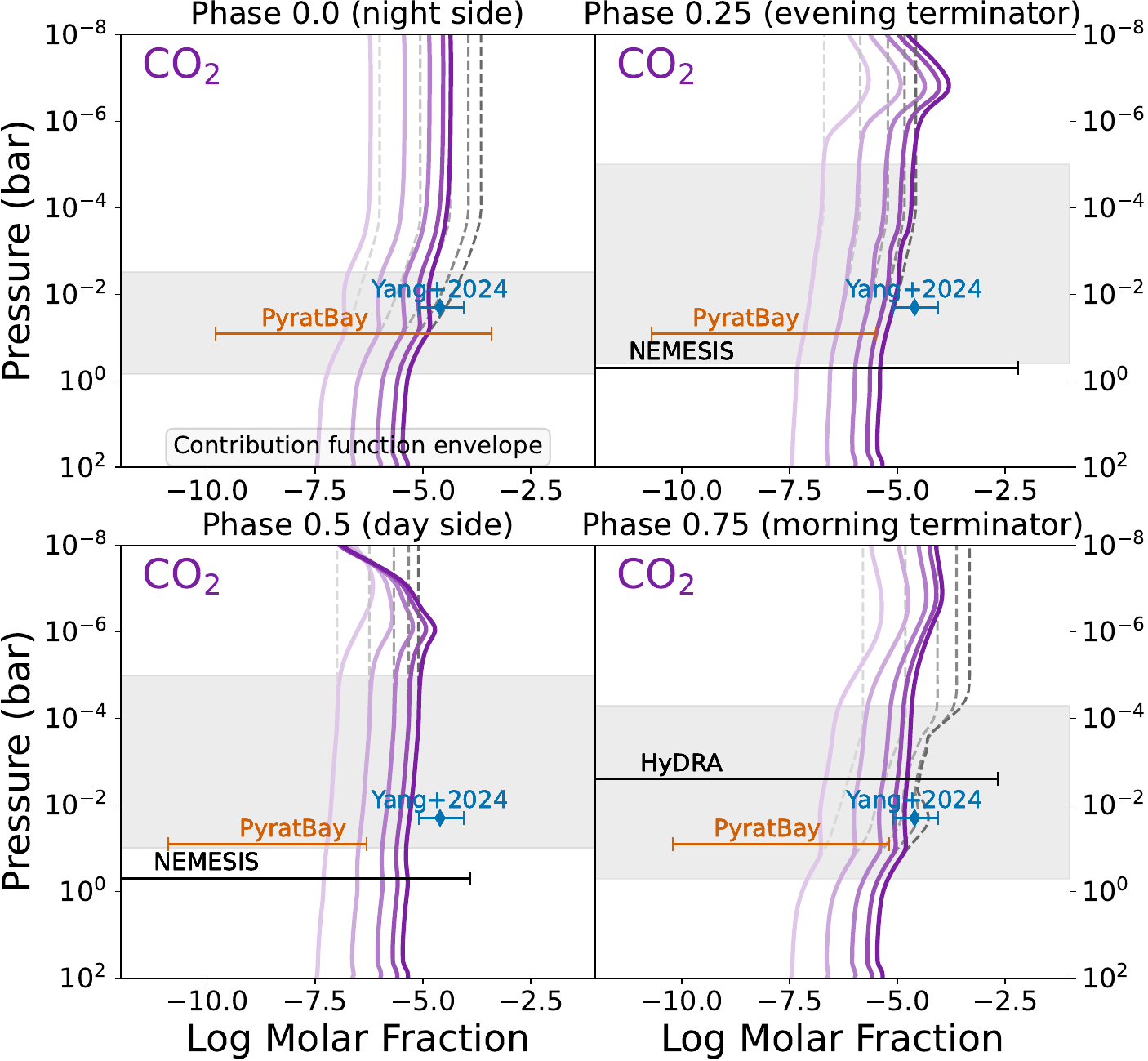}
    \caption{Analogous to Fig.~\ref{fig_CH4_metallicities}, but showing \ce{CO2}. The abundance is moderately dependent on metallicity. In addition to the retrievals from \citet{Bell_nightside_2024} (\textbf{black}) and \citet{Yang_simultaneous_2024} (\textbf{\color{YangBlue}{blue}}), we also show a more permissive retrieval model (\textbf{\color{PyratRed}{red}}, see Appendix~\ref{sec_retrievals} for details).}
    \label{fig_CO2_metallicities}
\end{figure}

\begin{figure}
    \centering
    \includegraphics[width=\linewidth]{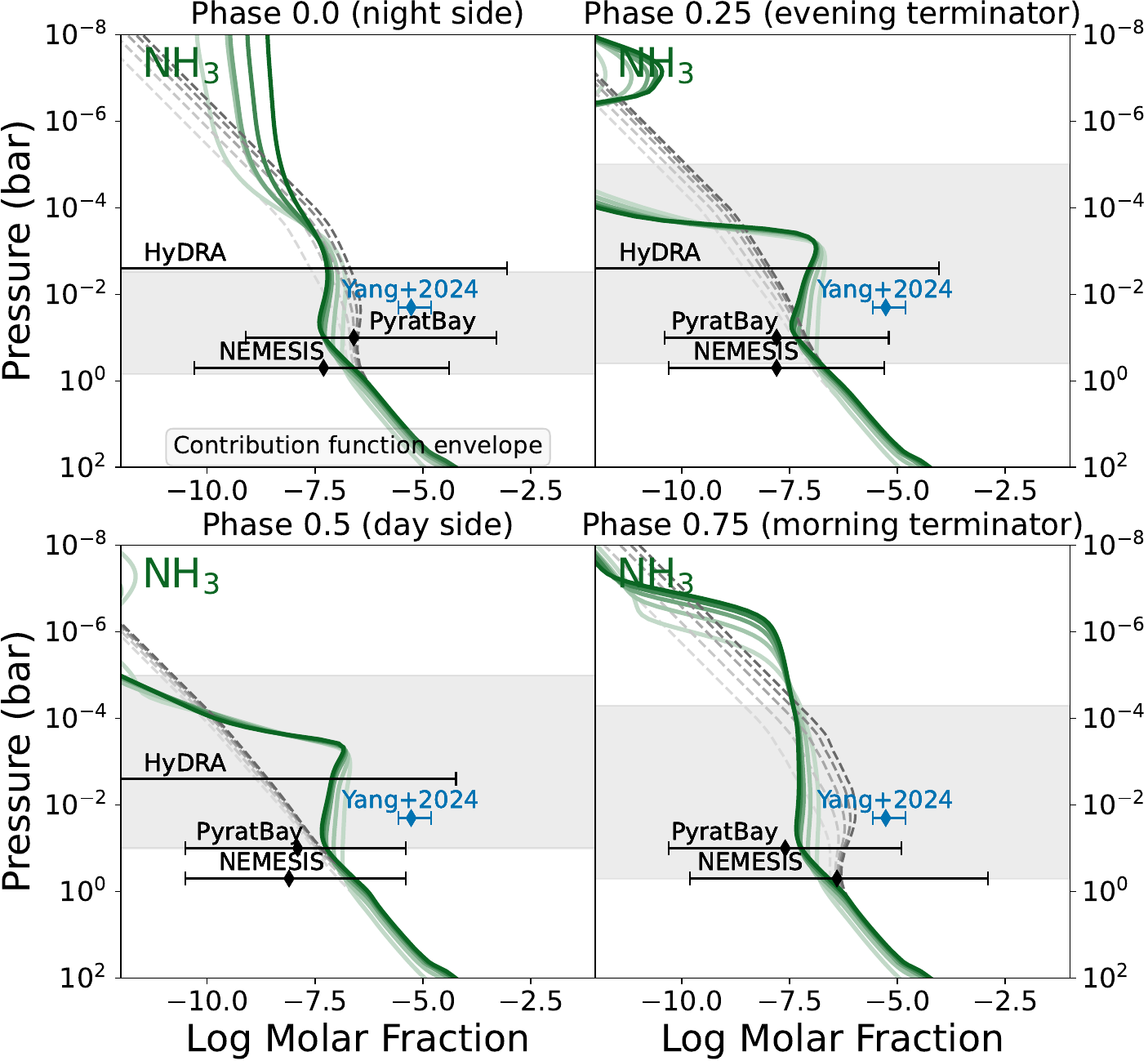}
    \caption{Analogous to Fig.~\ref{fig_CH4_metallicities}, but showing \ce{NH3}. Very little metallicity dependence can be seen. }
    \label{fig_NH3_metallicities}
\end{figure}

\begin{figure}
    \centering
    \includegraphics[width=\linewidth]{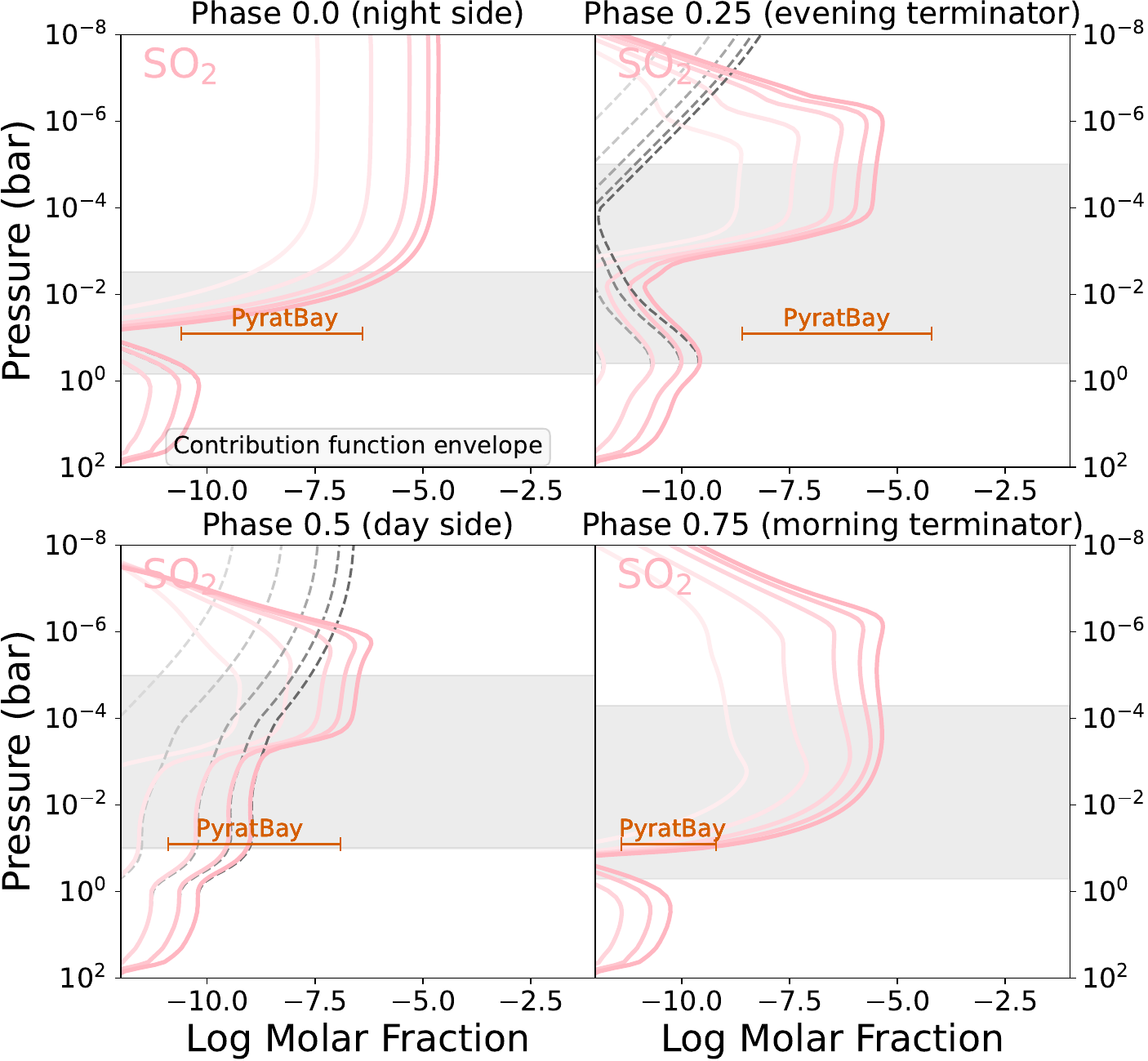}
    \caption{Analogous to Fig.~\ref{fig_CH4_metallicities}, but showing \ce{SO2}. Strong deviations from chemical equilibrium can be seen, as well as a strong sensitivity to the assumed atmospheric metallicity.}
    \label{fig_SO2_metallicities}
\end{figure}

The phase curve of \citet{Bell_nightside_2024} did not strongly constrain the atmospheric metallicity of WASP-43~b, with metallicities of 1-10 times the solar value agreeing with the data. Here, we analyze the phase-dependent chemical composition at metallicities of 1, 2.5, 5, 7.5, and 10 times solar \citep{lodders_solar_2019} to assess the observability of species at different levels of enrichment. We run \textit{ACE-PAC} pseudo-2D models using thermal profiles of the \textit{Generic PCM} with different metallicities, changing the elemental abundances, thermal structure, and wind jet speed correspondingly. Other model parameters are kept the same. Where possible, the chemical model predictions are compared to constraints from atmospheric retrievals of \citet{Bell_nightside_2024}. The abundances of methane, carbon dioxide, ammonia, and sulfur dioxide are shown in Figs.~\ref{fig_CH4_metallicities} -- \ref{fig_SO2_metallicities}. Indications from retrieval models are $1\sigma$-uncertainty intervals or 95\%-upper limits, except where indicated. We refer to Appendix~\ref{sec_retrievals} for further details about the specific retrieval models. 

Methane does not exhibit a strong metallicity dependence in our models. Across the range of metallicities considered, the \chem{CH_4} concentration varies by an order of magnitude, with the highest concentration found at $1\times$ solar metallicity ($10^{-8}$ molar fraction, vertically quenched at all phases). The methane concentration further drops as the metallicity increases. Interestingly, our chemical equilibrium calculations of the night-side methane abundance show the opposite trend (dashed lines in Fig.~\ref{fig_CH4_metallicities}), namely a maximal \chem{CH_4} abundance at $10\times$ solar metallicity. We can understand the observed decoupling of the methane abundance from the 
chemical equilibrium trends at the night side, since methane is horizontally quenched. As such, it traces the day side composition, where it is still a minor constituent in a CO-dominated atmosphere.  
Chemical equilibrium predictions of night-side and morning-terminator methane clearly do not agree with observed upper limits (Fig.~\ref{fig_CH4_metallicities}). The disagreement is worse at high metallicities. However, the inclusion of horizontal advection in our photochemical model produces results that are in agreement with observations, regardless of metallicity.

In contrast to methane, carbon dioxide does exhibit a moderate amount of metallicity dependence (Fig.~\ref{fig_CO2_metallicities}). This goes for both the actual photochemical kinetics models and the predictions from equilibrium chemistry, as \chem{CO_2} remains in equilibrium throughout most of the atmosphere (some vertical quenching can be observed on the night side and morning terminator). The \citet{Bell_nightside_2024}-data does not constrain the \chem{CO_2} abundance particularly well, and all metallicities at each phase appear to match a 99\%-upper limit from the \textit{HyDRA} and \textit{NEMESIS} retrievals. Constraints from the \textit{PyratBay} model are likewise uncertain, and able to match most metallicities within 1$\sigma$, with a slight preference for lower metallicities. Comparisons with results from a two-dimensional retrieval of all phases at the same time \citep{Yang_simultaneous_2024}, however, appear to hint at a high metallicity of 7.5 -- 10 times the solar value. Although this metallicity estimate stands in contrast to the metallicity quoted in \citet{Yang_simultaneous_2024} ($1.6^{+4.9}_{-1.0}\times$ solar), the existing tension with the retrieved \chem{CO_2} abundance ($-4.60^{+0.55}_{-0.50}$) can also be seen in their Figure~11.

Ammonia, analogous to methane, exhibits little dependence on the assumed metallicity in our WASP-43~b models (Fig.~\ref{fig_NH3_metallicities}). In addition, only minor departures from thermochemical equilibrium are seen, as the atmosphere of WASP-43~b is generally too hot for ammonia quenching to compete with very fast thermal dissociation reactions of \chem{NH_3} and \chem{NH_2}. The rate-limiting step and the eventual fate of the majority of nitrogen is the formation of \chem{N_2} via N($^4S$) + N($^4S$) $\rightarrow$ \chem{N_2} (low metallicity) or \chem{NH_2} + \chem{NO} $\rightarrow$ \chem{N_2} + \chem{H_2O} (high metallicity). The retrievals from \citet{Bell_nightside_2024} have large uncertainties and are in agreement with the photochemical predictions and chemical equilibrium at all metallicities. Noteworthy is that the \citet{Yang_simultaneous_2024}-retrieval model provides indications for ammonia abundances higher than either our equilibrium or kinetic model predictions. While the retrieved value may be explained simply by a strongly ($>200 \times$) enhanced N/H value \citep{Yang_simultaneous_2024}, invoking such ad-hoc hypothesis may be avoided by assuming a higher $\kzz$ value, dredging up more ammonia from the deeper layers. A hypothesis of strong vertical mixing is explored in Sec.~\ref{sec_kzz} and further discussed in Sec.~\ref{sec_discussion}.

Finally, sulfur dioxide displays very clear deviations from chemical equilibrium, as well as a strong positive correlation with metallicity (Fig.~\ref{fig_SO2_metallicities}). 
Despite very active sulfur photochemistry in our WASP-43~b models, low atmospheric metallicities prevent photochemically produced sulfur species like SO$_2$ from forming abundantly.  At 1$\times$ solar metallicity, for example, the thermochemical equilibrium H$_2$S mixing ratio is only $\sim 2.7 \, \times \, 10^{-5}$, so the mixing ratios of the sulfur photochemical products never exceed this amount.  The abundance of species such as SO$_2$ (and CO$_2$, see Fig.~\ref{fig_CO2_metallicities}) that have three heavy elements are particularly suppressed at low atmospheric metallicity. In contrast, the order-of-magnitude larger H$_2$S abundance at 10$\times$ solar metallicity would lead to more abundant sulfur-bearing photochemical products, including SO$_2$, which becomes increasingly a dominant product at higher metallicities for warmer H$_2$-dominated atmospheres such as that of WASP-43~b \citep[e.g.][]{tsai23wasp39b, polman_h2s_2023, veillet_inclusion_2025}. Indeed, our models predict a high SO$_2$ concentration of 1-10~ppm at all phases. Therefore, SO$_2$ joins CO$_2$ as an excellent probe of atmospheric metallicity, and the lack of notable SO$_2$ absorption in the WASP-43~b emission spectra in the 7--8 $\mu$m region \citep{Bell_nightside_2024} may help constrain the planet's atmospheric metallicity. The lack of SO$_2$ is corroborated by the precisely inferred low abundance posed by the \textit{PyratBay} model on the morning terminator, which can be interpreted as a fairly strict upper limit (Fig.~\ref{fig_SO2_metallicities}). An important consideration is the dependence of \ce{SO2} formation on both the stellar irradiation and planetary gravity \citep{dyrek_SO2_2024, konings_reliability_2025}, beyond the atmospheric metallicity. As such, planet-specific models are highly useful to predict the observability of \ce{SO2}. Finally, we note that equilibrium chemistry at high metallicity predicts a day-side SO$_2$ abundance that is not much lower than the photochemically produced maximum (Fig.~\ref{fig_SO2_metallicities}). Nonetheless, high amounts of sulfur dioxide at the planetary limbs and night side would be a telltale sign of photochemistry. 

\subsection{Compositional changes with jet speed}\label{sec_jetspeed}

\begin{figure}
    \centering
    \includegraphics[width=\linewidth]{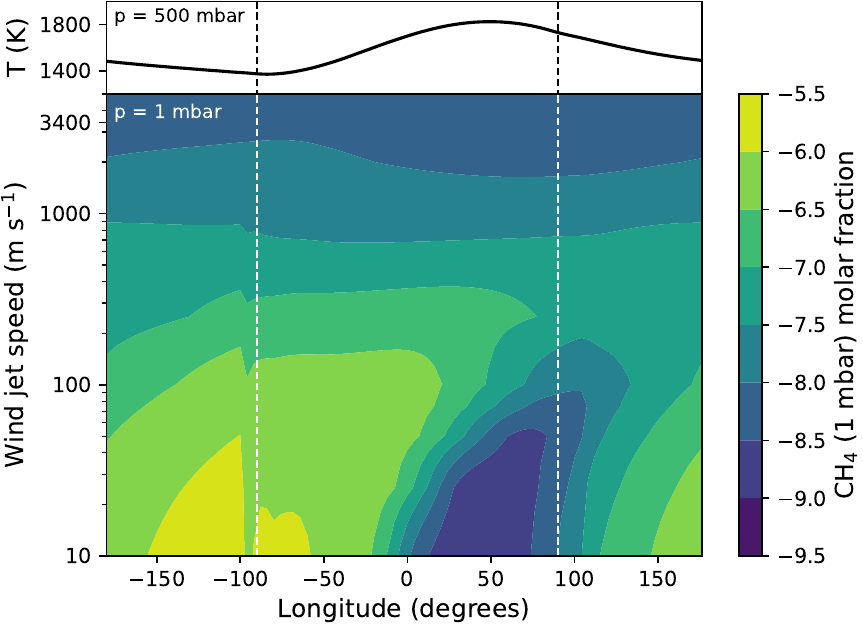}
    \caption{Map of the methane abundance at 1~mbar as a function of longitude  and wind jet speed, computed from a suite of pseudo-2D models. The nominal wind speed predicted by the GCM is 3400~m~s$^{-1}$. The plot on the top shows the temperature deeper in the atmosphere, near the vertical quench level ($\sim$500~mbar), which ultimately determines the disequilibrium \chem{CH_4} abundance if horizontal advection would be neglected. Dashed vertical lines highlight the morning and evening limbs, at $-90^\circ$ and $90^\circ$ respectively. Note that the underlying thermal structure remains the same for all photochemical kinetics calculations, namely a GCM with superrotation.}
    \label{fig_CH4_windspeeds}
\end{figure}

The parameter that controls horizontal mixing in the pseudo-2D models is the wind speed of the equatorial jet stream. Given the evidence for transport-induced quenching of night-side methane on WASP-43~b \citep[][and Fig.~\ref{fig_CH4_metallicities}]{Bell_nightside_2024}, we vary the wind jet speed in the solar-abundance \textit{ACE-PAC} model and investigate which values are compatible with the non-detection of night-side methane. All other parameters are kept fixed. This includes $\kzz$, although vertical mixing and horizontal advection are in principle correlated through atmospheric circulation \citep{komacek_atmospheric_2016, komacek_vertical_2019, baeyens_grid_2021}. 

Figure~\ref{fig_CH4_windspeeds} shows that the methane concentration becomes increasingly more homogeneous as the wind jet increases in speed. At low wind speeds on the order of 10--100~m~s$^{-1}$, \ce{CH4} tracks the temperature profile near the vertical quench point. As such, the hot day side and evening limb are depleted, while the morning limb -- corresponding to the lowest temperature -- is methane-rich. At the nominal value of $1\times$ solar metallicity, the methane concentration can peak up to several ppm, which is likely near its detection threshold. For faster jet speeds ($\gtrsim500$~m~s$^{-1}$), however, this morning-evening dichotomy disappears. At the nominal speed (3.4~km~s$^{-1}$), \ce{CH4} is strongly depleted ($< 10^{-8}$) and uniformly distributed across the day and night. 

\subsection{Compositional changes with vertical mixing strength}\label{sec_kzz}

\begin{figure*}
    \centering
    \includegraphics[width=\linewidth]{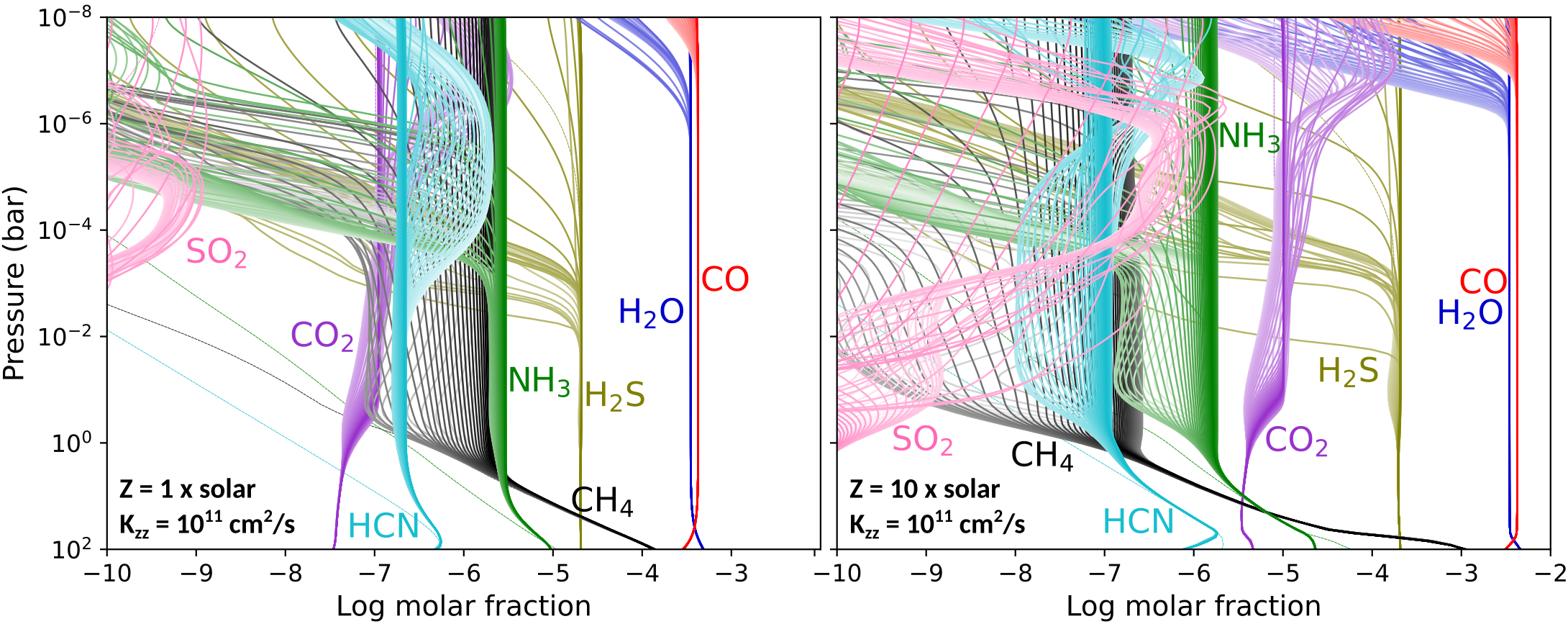}
    \caption{The chemical abundances of major species in WASP-43~b modelled using \textit{ACE-PAC} for strong vertical mixing ($\kzz = 10^{11}$~cm$^{2}$s$^{-1}$), assuming solar (\textit{left}) and $10 \times$ solar (\textit{right}) metallicity. Different lines represent different equatorial longitudes, with lighter shades being located closer to the substellar longitude.}
    \label{fig_kzz}
\end{figure*}

Motivated by the unexplained indications of a high ammonia abundance (see Fig.~\ref{fig_NH3_metallicities}), as well as a general lack of vertical mixing constraints in exoplanet atmospheres \citep[e.g.][]{komacek_vertical_2019, Baxter_2021, soni_signature_2024}, we perform a sensitivity test of the WASP-43~b atmosphere using a high constant value for the eddy diffusion coefficient $\kzz = 10^{11}$~cm$^{2}$s$^{-1}$. This value is still plausible, but on the high end of what we expect for non-inflated hot Jupiters \citep{baeyens_grid_2021} (see also Sec.~\ref{sec_disc_methane}). The results of this experiment, for solar and $10 \times$ solar metallicity, are shown in Fig.~\ref{fig_kzz}.

Upon comparing the high-$\kzz$ models with the nominal models (Figs.~\ref{fig_chem_acepac}, \ref{fig_chem_kinetics}), we find that strong vertical mixing greatly increases the methane and ammonia abundances to 1-10~ppm. The HCN fraction is likewise increased, because of the higher availability of parent species that participate in HCN photochemistry. Although methane and ammonia are quenched to the same abundance in the model with solar metallicity (Fig.~\ref{fig_kzz}, left), the high metallicity model produces ten times more ammonia than methane (Fig.~\ref{fig_kzz}, right).
A comparison with the retrieval models above (see Figs.~\ref{fig_CH4_metallicities}, \ref{fig_NH3_metallicities}) shows that the high $\kzz$ model with solar metallicity is disfavored, as it disagrees with the observed upper limits of methane (around 1~ppm on the night side). The high-metallicity model, however, shows a reduced night-side methane abundance, in agreement with upper limits from retrievals, as well as an increased ammonia concentration of several ppm. This combination would bring the model more in line with the retrieval results of \citet{Yang_simultaneous_2024}. It is clear, however, that more detected species -- or more stringent upper limits -- are needed to further pin down the atmospheric chemistry of WASP-43~b and enable meaningful constraints to $\kzz$. 

Finally, we note that the models with high $\kzz$ show larger compositional variations with longitude. The reason is that the timescale for vertical mixing changes from $\tau_\textrm{vert}$$\sim$$10^7$~s in the nominal case to $\tau_\textrm{vert}$$ \sim$$ 1000$~s in the case where $\kzz = 10^{11}$~cm$^{2}$~s$^{-1}$. As such, it becomes significantly shorter than the time it takes to advect gas zonally across the hemisphere ($\tau_\textrm{zonal}$$ \sim$$ 10^5$~s), and the atmospheric composition homogenizes vertically rather than horizontally \citep{baeyens_grid_2021}. This situation can lead to a significant build-up of quenched species on the night side of the planet. At the same time, a reduction in \chem{SO_2} can be seen, as the molecule no longer gets a chance to build up on the night side of the planet \citep{tsai_day-night_2023}. Only at elevated metallicity, moderate amounts of \chem{SO_2} are produced at the day side and the limbs (0.1~ppm), while the night-side abundance remains low ($< 10^{-10}$) (see Fig.~\ref{fig_kzz}, right). This result highlights that zonal advection is an important aspect in understanding the global distribution of photochemical species. On the flip side, it reveals how phase-dependent chemical observations are able to trace the underlying atmosphere dynamics, especially for photochemical species such as \ce{SO2}.

\subsection{Coupled three-dimensional transport}\label{sec_3Dchemistry}

\begin{figure*}
    \centering
    \includegraphics[width=0.49\linewidth]{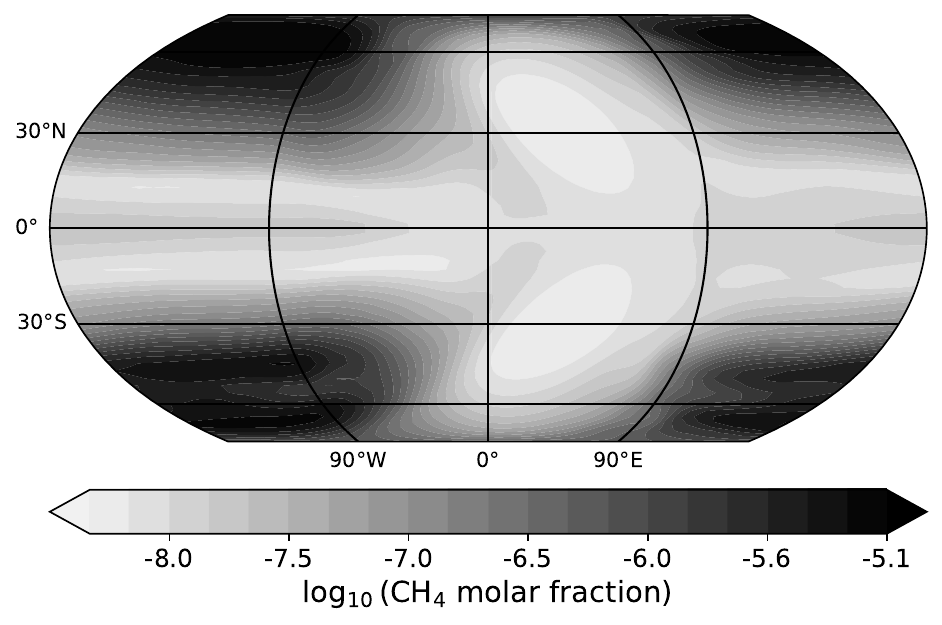}
    \includegraphics[width=0.49\linewidth]{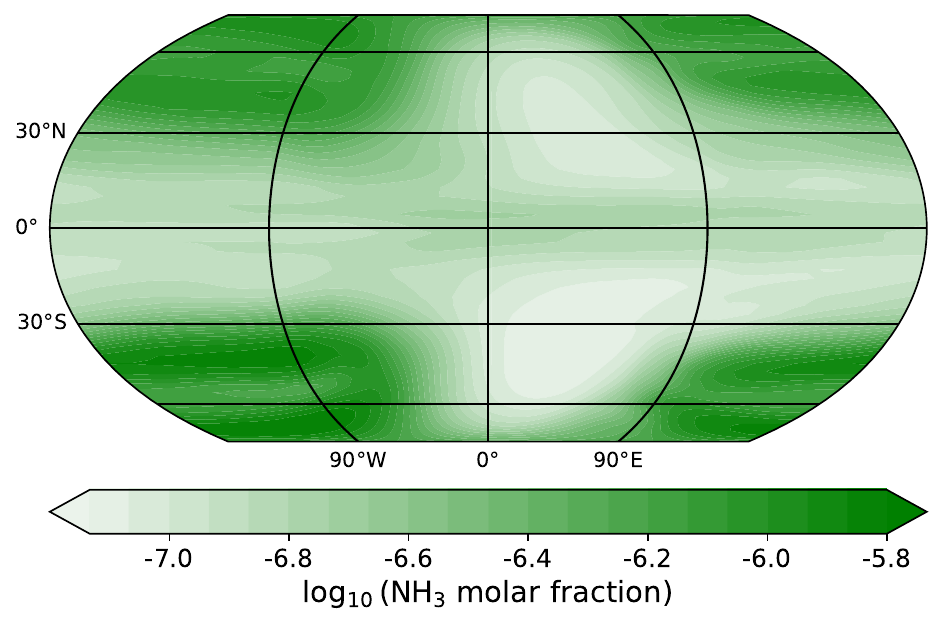}
    \includegraphics[width=0.49\linewidth]{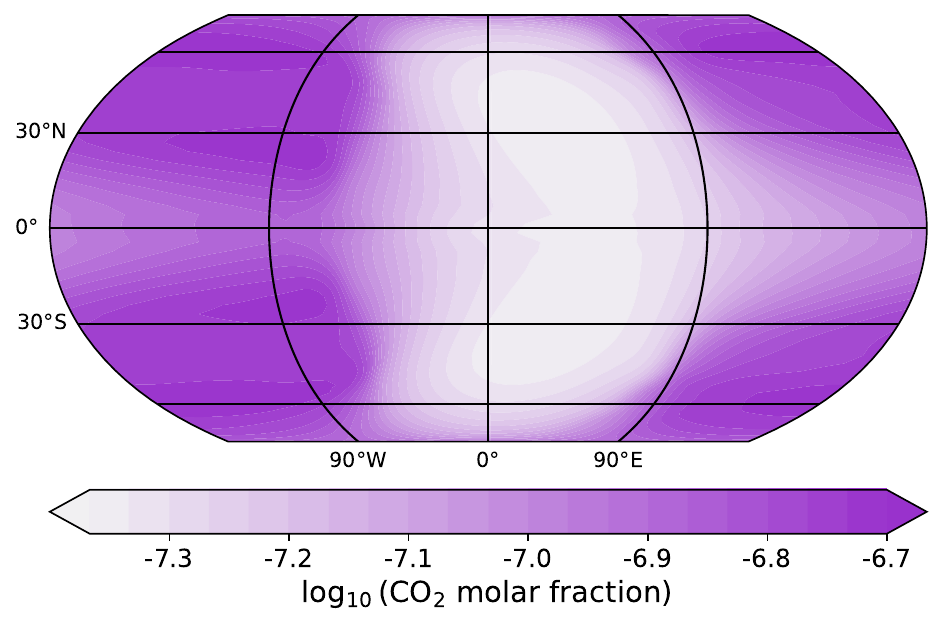}
    \includegraphics[width=0.49\linewidth]{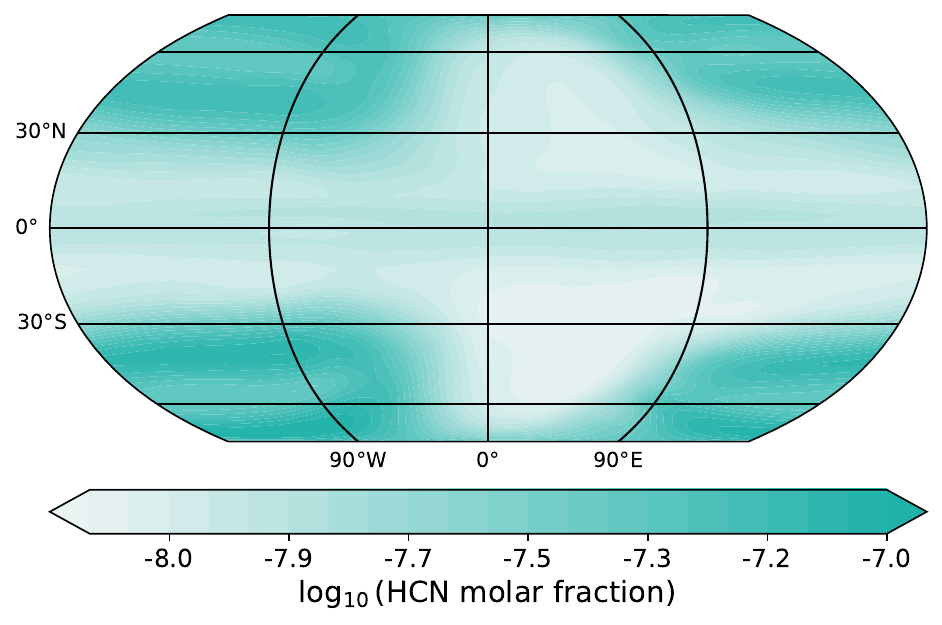}    
    \caption{Longitude-latitude maps of the \ce{CH4} (\textit{top left}), \ce{NH3} (\textit{top right}), \ce{CO2} (\textit{bottom left}) and \ce{HCN} (\textit{bottom right}) concentrations at 0.1~bar from the 3D \textit{Exo-FMS} GCM + \textit{mini-chem} simulation. The sub-stellar point is at the coordinates (0$^{\circ}$,0$^{\circ}$). Please note the differently scaled color bars.}
    \label{fig:maps_GCM}
\end{figure*}

\begin{figure}
    \centering
    \includegraphics[width=\linewidth]{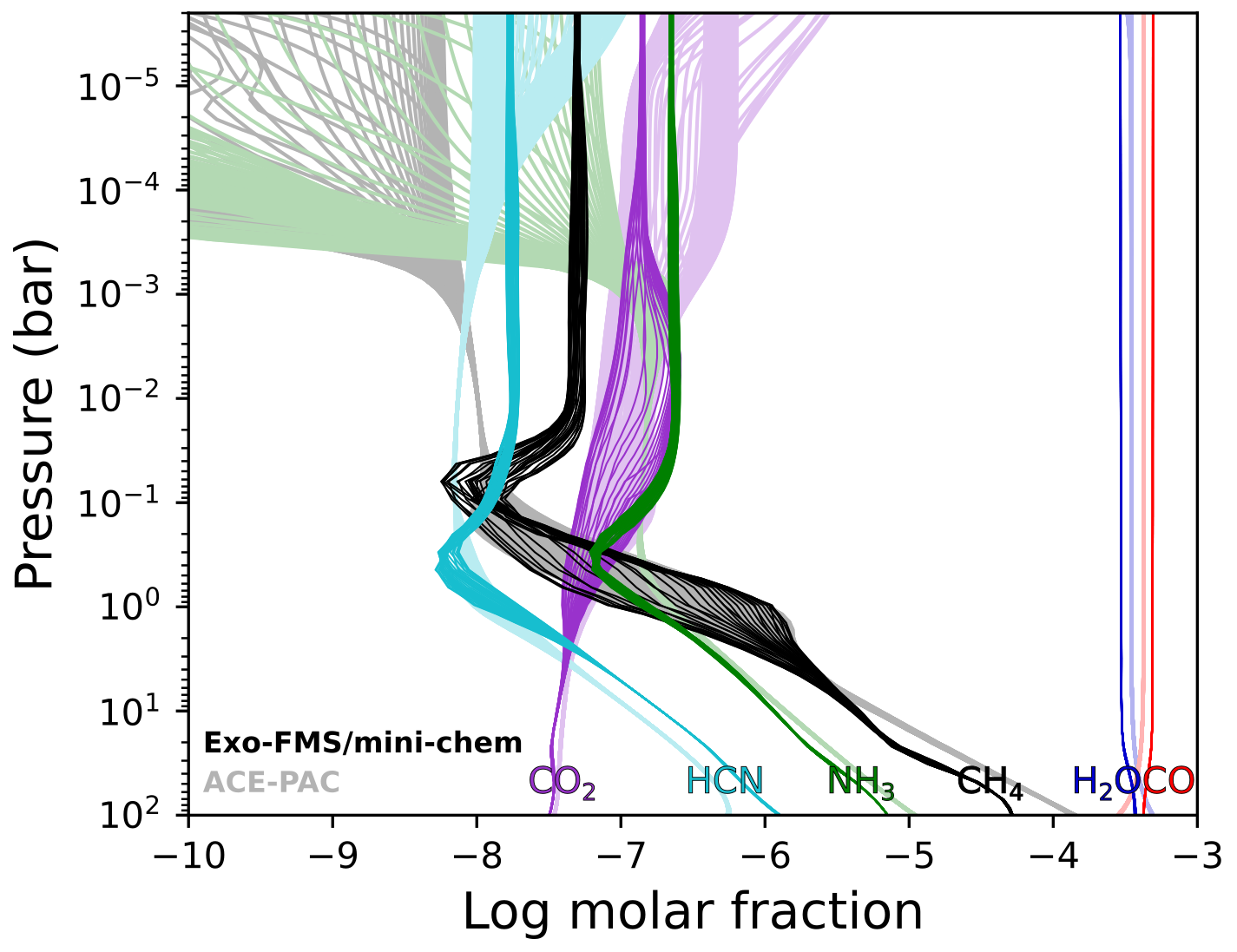}
    \caption{Equatorial abundances as a function of pressure of several species, computed using the 3D \textit{Exo-FMS} + mini-chem simulation. The abundances are cos(latitude)-weighted and averaged between $\pm20\degrees$ latitude. As a comparison, the pseudo-2D \textit{ACE-PAC} results are shown in muted colors.}
    \label{fig:equatorial_GCM}
\end{figure}

To complete the interpretation of atmospheric chemistry on WASP-43~b, we employed a coupled three-dimensional climate-chemistry model (\textit{Exo-FMS} with \textit{mini-chem}, see Sec.~\ref{sec_exo-FMS}), which is capable to simulate the time-dependent chemical reactions for the dominant atmospheric gases, as well as their thermal and radiative climate feedback. Isobaric maps at 100~mbar of \chem{CH_4}, \ce{NH3}, \chem{CO_2}, and \chem{HCN} gas distributions (Fig.~\ref{fig:maps_GCM}) are in good qualitative agreement with photochemical predictions above. They reveal a high degree of zonal homogenization for methane and ammonia, especially at equatorial latitudes. Likewise, species such as \chem{CO}, \chem{H_2O} (not shown), and \chem{CO_2} show negligible day-night contrasts. Quantitative differences between the pseudo-2D and 3D models are likewise within an order magnitude for all key species, except the photochemically produced \ce{SO2}. As such, our pseudo-2D approach of the equatorial region is proven to be well justified. At the same time, \ce{CH4}, \ce{NH3}, and HCN show small-scale local enhancements of 1--2 orders of magnitude at higher night-side latitudes. These regions correspond to off-equatorial gyres with cold temperatures, resulting in local maxima and potential meridional mixing for some species \citep{drummond_3D_2018, zamyatina_quenching-driven_2024}. Mid- and high-latitude flow is not captured in pseudo-2D photochemistry models, so it is important to quantify their effect on the overall atmospheric composition.

A detailed comparison between vertical profiles sampled at equatorial latitudes from our 3D climate-chemistry model and our nominal pseudo-2D photochemical model reveals excellent agreement (within $15\%$ for \ce{H2O}, CO, within $60\%$ for \ce{CO2}, HCN, and within $100\%$ for \ce{CH4}) between both models despite using different underlying thermal structure, chemistry, and wind geometry (Fig.~\ref{fig:equatorial_GCM}).  
Importantly, chemical equilibrium trends in the deep atmosphere ($p>1$~bar) are in agreement for all major species, and the abundances of \chem{H_2O}, CO, \chem{CO_2} correspond well to those described in the pseudo-2D photochemical models above. The quenching pressures, where divergence from chemical equilibrium is seen, are consistently between $\sim$1~bar and $\sim$100~mbar in both the 3D and pseudo-2D models. Both models likewise show partial zonal mixing of methane in that pressure range, followed by vertical quenching at lower pressures (unlike the pseudo-2D \textit{KINETICS} model that does not show vertical methane quenching, see Fig.~\ref{fig_chem_kinetics}). The correspondence highlights that both models do a good job at capturing the essential dynamical and chemical time scales. The agreement is also noteworthy because the implementation of vertical mixing in both models is fundamentally different: the nominal pseudo-2D model uses the eddy diffusion parameter ($\kzz$), while the 3D climate-chemistry model also computes advective mixing via atmospheric circulation. Although the 3D \textit{Exo-FMS} model does not include sulfur chemistry, the impact on non-sulfur-bearing species remains limited at $p>1$~mbar, because coupled carbon-sulfur and oxygen-sulfur chemistry mostly occur in the upper atmosphere (see Sec.~\ref{sec_S}).
On the other hand, an important discrepancy between the two models is the low-pressure composition ($p<1$~mbar). Here, photochemistry causes molecular dissociation and an increasingly contrasting day- and night-side compositions (see also Fig.~\ref{fig_H}). As the mini-chem scheme currently does not take photochemistry into account \citep{Tsai2022,Lee2023}, the 3D model maintains a homogeneous composition at low pressures. 

Finally, we observe the signature of meridional quenching of \chem{CH_4}, \chem{NH_3}, and HCN in the 100~mbar-regime (Fig.~\ref{fig:equatorial_GCM}). A slight increase in the equatorial mixing ratio of less than one order of magnitude can be observed, as these species are quenched from their chemical equilibrium and before they attain a vertically constant distribution at low pressure ($p< 10$~mbar). The same switchback in the vertical profile was found for the methane abundance of other hot Jupiters in earlier models coupling 3D dynamics and chemistry \citep{drummond_observable_2018, drummond_3D_2018}. This effect can be attributed to meridional transport of gas from the cold night-side gyres to the equator, enriching the low-latitude regions with methane, ammonia, and hydrogen cyanide.  

In summary, the emerging picture of the atmospheric chemistry of WASP-43~b -- and many hot Jupiters by extension -- is one that combines the insights of the pseudo-2D and 3D models. The former models inform us about photochemistry and coupled sulfur-chemistry, while the latter provides a more realistic picture of atmospheric circulation and the global distribution of species.

\section{Model emission spectra}\label{sec_spectra}


\begin{figure*}
    \centering
    \includegraphics[width=\textwidth]{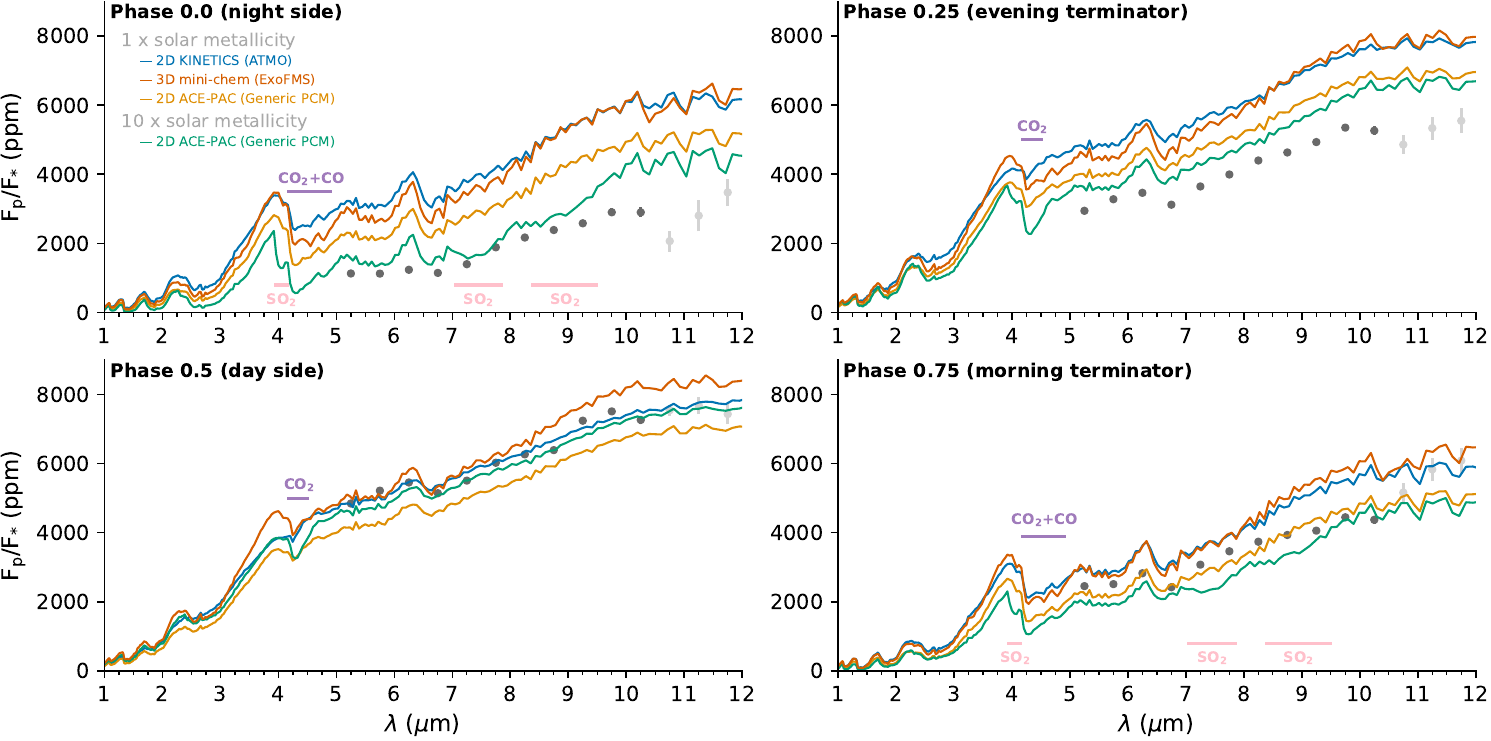}
    \caption{Synthetic cloud-free emission spectra of WASP-43~b at phases 0.00 (night), 0.25 (evening), 0.50 (day), and 0.75 (morning), generated using the 3D radiative transfer code \textit{gCMCRT} \citep{Lee_3D_2022}. Observational data by \cite{Bell_nightside_2024} (\textit{gray markers}) are shown as reference, but since the photochemical models do not include clouds, no accurate match to the observational data is pursued. Data points in the `shadowed region' ($\lambda>10.6$~$\mu$m) are shown in light gray and have high systematic errors \citep{Bell_nightside_2024}. The main absorption bands of \ce{SO2}, \ce{CO2}, and \ce{CO} are highlighted. Absorption by \ce{H2O} is also present throughout the spectrum.}
    \label{fig_emi_spec}
\end{figure*}

\subsection{Spectral code setup}

To quantify the potential effects of (disequilibrium) chemistry on the observed planetary emission, we compute synthetic emission spectra at different phases. To this end, we use the 3D Monte Carlo radiative-transfer model \textit{gCMCRT} \citep{Lee_3D_2022}.
\textit{gCMCRT} takes input from the pseudo-2D photochemical and 3D climate-chemistry simulations, and self-consistently calculates the spectral emission properties for a given phase angle, taking into account the three-dimensional spherical geometry of the atmosphere.
The 3D geometry calculation capabilities of \textit{gCMCRT} make it an ideal model to calculate emission spectra of the pseudo-2D and 3D model output in a consistent manner. As in \citet{Bell_nightside_2024}, we consider emission spectra at four phases: 0.0, 0.25, 0.5 and 0.75 (i.e.~the quadrature points of the orbit).

To calculate the emission spectra from the \textit{ACE-PAC} pseudo-2D model, we adopt the same thermal structure that was used in the \textit{ACE-PAC} model, namely the one resulting from the \textit{Generic PCM} climate model (see Sec.~\ref{sec_gpcm}). To accommodate the geometric leap from pseudo-2D to 3D, we assume uniform chemical abundances along all latitudes, equal to those calculated for the equator in the pseudo-2D photochemistry model.
In effect, we obtain the 3D temperature-pressure structure, but assume as a global chemical profile one that is reflected by the calculated chemical composition of the equator.
We process both the $1\times$ and $10\times$ solar metallicity \textit{ACE-PAC} simulations.
To calculate emission spectra for the equivalent \textit{KINETICS} pseudo-2D model, which adopted a two-dimensional temperature structure (see Sec.~\ref{sec_atmo}), we adopt an `orange slice' configuration. Both the temperature-pressure profiles and the photochemical abundances for all latitudes are assumed to be the same as the equatorial region. This approach has been shown to be a good approximation for the phase-dependent emission of the planet, which is dominated by the equatorial region \citep{cowan_inverting_2008}.
Finally, for the \textit{Exo-FMS} GCM + \textit{mini-chem} results, the full three-dimensional temperature and chemical structure are provided directly by the simulation output. Thus, we do a fully consistent 3D calculation of the emitted flux using \textit{gCMCRT} without any further approximation.

\subsection{Phase-dependent spectral signatures}

Model emission spectra at each phase are plotted in terms of flux contrast (F$_{\rm p}$/F$_{*}$), along with the phase dependent MIRI/LRS spectra from \citet{Bell_nightside_2024} as a guide (Fig.~\ref{fig_emi_spec}). We note that our forward models are not specifically designed to fit the observed spectra. Indeed, our models do not include the effects of cloud opacity, which are needed to explain the muted night-side flux \citep{Bell_nightside_2024}. Nonetheless, the comparison between the observed and model spectra may serve as a reference guide for spectral features.

Several species are spectrally active in the MIRI/LRS wavelength range (5~$\mu$m--12~$\mu$m). In the model emission spectra, absorption by \chem{H_2O} dominates the spectral range at all phases and for all chemical models. It likewise causes the distinct spectral feature near $\sim$$6.5$~$\mu$m. Besides water, the only spectral features that can be made out by eye are the broad 7.5~$\mu$m- and 8.8~$\mu$m-centered absorption bands of \chem{SO_2}, namely on the night side and morning terminator (phases 0.0 and 0.75) in the $10\times$ solar metallicity model. Based on Fig.~\ref{fig_SO2_metallicities}, we indeed expect to see \chem{SO_2} exclusively at higher metallicity. However, the appearance of spectral features also depends on the thermal structure, as no \chem{SO_2} absorption is visible at the day-side and evening phases, despite having a similar abundance as on the morning limb. It should be noted that the \chem{SO_2} concentration peaks high up in the atmosphere ($p<1$~mbar), where its contribution to the emission spectrum may be minimal. As such, abundance measurements of sulfur dioxide should always go hand in hand with spectral retrievals of the temperature profile. Finally, an absorption band of ammonia sits near $\sim$$11$~$\mu$m, but none of our models predict sufficient \chem{NH_3} to produce the feature.\footnote{Although the MIRI/LRS data appears to show absorption near 11~$\mu$m, systematics plague this part of the spectrum, the so-called shadowed region \citep[10.6--11.8~$\mu$m,][]{Bell_nightside_2024}. A retrieval model by \citet{Yang_simultaneous_2024} finds evidence for ammonia in the data, despite excluding this region from their analysis.}

\begin{figure}
    \centering
    \includegraphics[width=\columnwidth]{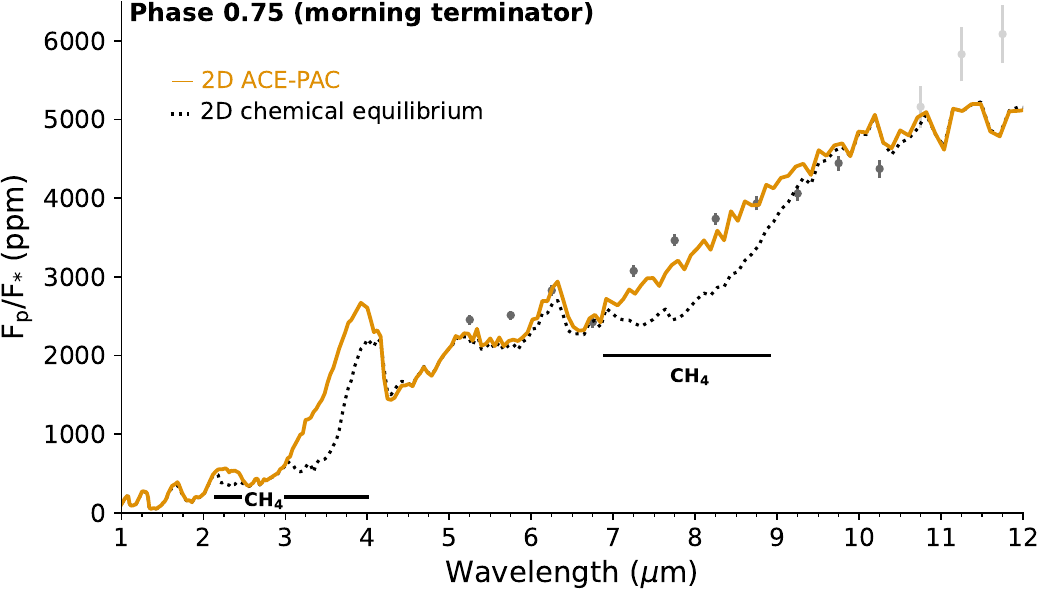}
    \caption{Comparison between the morning terminator emission spectrum of a model with horizontal transport (\textit{orange}) and a model with equilibrium chemistry (\textit{black dotted}). The MIRI/LRS observations \citep{Bell_nightside_2024} are shown as reference. The main difference between the models is due to methane absorption, highlighted in black.}
    \label{fig_emi_spec_equilibrium}
\end{figure}

Methane likewise has a strong and broad absorption band in the MIRI/LRS wavelength range, peaking near 8~$\mu$m. However, it is not present in any of the 2D/3D photochemical models shown in Fig.~\ref{fig_emi_spec}, because the methane abundance is kept below detectable limits through vigourous horizontal advection of hot day-side gas. To illustrate the impact of disequilibrium chemistry on the spectral emission of WASP-43~b, we computed an emission spectrum under the assumption of chemical equilibrium for the morning terminator of the planet (Fig.~\ref{fig_emi_spec_equilibrium}). The equilibrium model shows clear methane absorption in the wavelength range between 7~$\mu$m and 9~$\mu$m. The difference of more than 600~ppm between the two models is much larger than the data uncertainty, clearly disfavouring the chemical equilibrium model for this planet. Hence, we demonstrate with self-consistent photochemical models that disequilibrium chemistry -- and horizontal advection in particular -- leads to the reported non-detection of methane in \citet{Bell_nightside_2024}. This result does not depend on the assumed atmospheric metallicity, as we have shown in Fig.~\ref{fig_CH4_metallicities}.

Looking outside of the MIRI/LRS wavelength range, water still dominates the emission spectra, but features of \chem{CO_2} and CO become apparent as well. A broad combined \chem{CO_2}/CO absorption band, centered on 4.5~$\mu$m, is apparent in all of our model spectra. The feature increases with metallicity, as we would expect based on the increased \chem{CO_2} abundance (Fig.~\ref{fig_CO2_metallicities}). In addition, the $10 \times$ solar metallicity model exhibits \chem{SO_2} absorption near 4~$\mu$m. This is again most notable for the night-side and morning phases, in accordance with the strength of the mid-infrared \chem{SO_2} features. We conclude that \chem{CO_2} and \chem{SO_2} are likely detectable by NIRSpec/G395H, and their presence or absence can provide a strong constraint on the metallicity and thermal structure of WASP-43~b. Methane absorption, again, would not be detectable, despite having absorption bands in the NIRSpec wavelength range. A non-detection of \ce{CH4} with NIRSpec/G395H is expected to be significant, due to the $>800$~ppm difference (at 3.5~$\mu$m) between our self-consistent photochemistry models and the chemical equilibrium model.


\section{Quantifying implicit photochemical model uncertainties}\label{sec_comparison}

Different photochemical models have their own specific numerical implementations of the underlying physics and chemistry, eq.~\eqref{eq_chem}, and rely on equivalent but different chemical reaction networks. Additionally, reasonable choices made by modellers during model setup, such as the adopted stellar flux, can introduce further variations in the resulting composition. As such, photochemistry models can be expected to produce differing results even when they are set up in equivalent ways. The uncertainties associated with the photochemical model predictions often remain obscured, since the model output consisting of chemical abundance profiles does not allow for ambiguity. Here, we aim to quantify and highlight these implicit uncertainties for the benefit of the exoplanet community. We note, however, that various methods to quantify photochemical model uncertainties have been presented in the Solar System and astronomical literature \citep[e.g.][]{thompson_1991, dobrijevic_coupling_2014, byrne_2024}. 

In order to illustrate some of the intermodel uncertainties and garner a rough idea of the reliability of the abundance predictions, we compare the output of several photochemical codes. Comparing the results of different photochemical models provides a clear first diagnostic for the robustness of species predictions, as well as a starting point for future studies that aim to investigate underlying discrepancies. 

\subsection{Experiment setup}\label{sec_intercomp_setup}

Four photochemistry models commonly used in the exoplanet literature are compared: \textit{ACE-PAC}, \textit{KINETICS}, \textit{EPACRIS}, and \textit{VULCAN}. We use each model in a one-dimensional implementation, with agreed-upon input parameters. \textit{ACE-PAC} and \textit{KINETICS} are detailed in Secs.~\ref{sec_ace-pac} and \ref{sec_kinetics}. \textit{EPACRIS} and \textit{VULCAN} are discussed next.

The chemistry module of \textit{EPACRIS} can implement the chemical kinetics equation \eqref{eq_chem} in a variety of planetary environments \citep{hu_photochemistry_2012, hu_photochemistry_2013, hu_photochemistry_2014, yang2024automated}. Given WASP-43~b's equilibrium temperature \citep[$T_\mathrm{eq}$ $\approx 1400$~K,][]{lesjak_retrieval_2023}, we adopted a chemical network developed for modelling \ce{H2}-dominated atmospheres of warm and hot Jupiters within the $T_\mathrm{eq}$ range of 800~K -- 1500~K \citep{yang2024automated}. This network was constructed via Reaction Mechanism Generator \citep{Gao_2016,RMG-database}) and comprises a total of 126 species and 2578 reactions, including sulfur molecules \citep{yang2024automated}.

As a fourth model, we employ the one-dimensional photochemical model \textit{VULCAN} for model intercomparison. \textit{VULCAN} has been widely used to model the atmospheric composition of gas giants \citep[e.g.][]{tsai23wasp39b} as well as rocky planets \citep[e.g.][]{tsai_inferring_2021, tsai_biogenic_2024}.
We used a C-H-O-N-S chemical network\footnote{\url{https://github.com/exoclime/VULCAN/blob/master/thermo/SNCHO_photo_network_2025.txt}} updated from that in \cite{tsai23wasp39b}, which now includes \ce{H2CS} to provide a more complete carbon-sulfur coupling. 

With each of these models, we ran one-dimensional steady-state calculations representing four phases of WASP-43~b (night, day, morning, evening) at solar elemental composition \citep{lodders_solar_2019}. The models have been set up with the same system parameters and with equatorial temperature-pressure profiles drawn from the solar-metallicity \textit{Generic PCM} simulation. All models and all phases used identical vertical mixing ($\kzz$) profiles, namely the parametrized description of \citet{moses21pseudo2d} as described in Sec.~\ref{sec_ace-pac}. A zenith angle of incident irradiation of 0$\degrees$ was used for the day-side models, and 87$\degrees$ for the limb models. For the night-side models no photochemistry was computed.

\begin{figure}
    \centering
    \includegraphics[width=\linewidth]{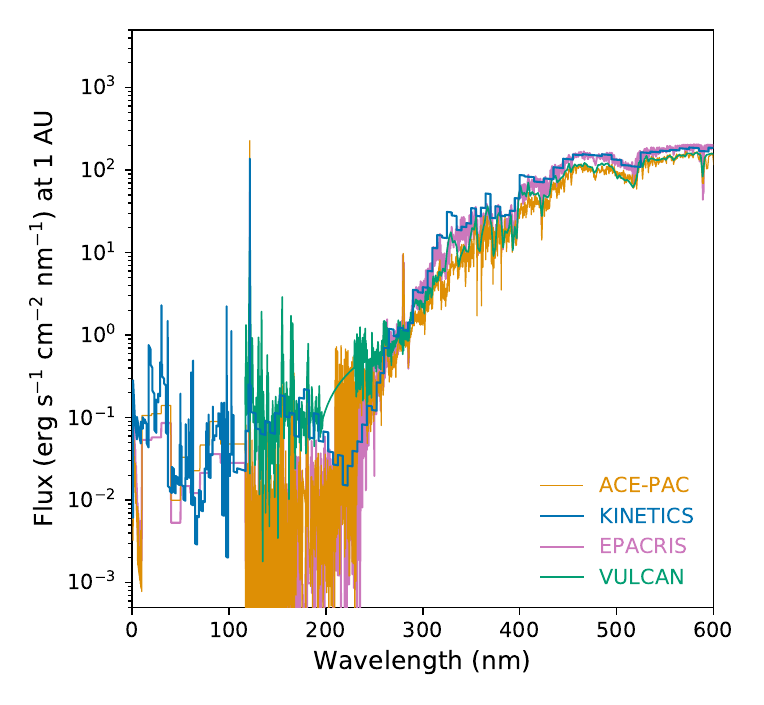}
    \caption{The four different stellar fluxes that are used by each photochemical model in the intercomparison. For details regarding each spectral energy distribution, see Sec.~\ref{sec_intercomp_setup}.}
    \label{fig_stellarspectra}
\end{figure}

Given the relative scarcity of (X)UV spectral observations of exoplanet host stars and their importance to photochemistry, atmospheric models often need to rely on observed proxy stars and flux reconstructions for certain spectral ranges \citep{teal_effects_2022, peacock_accurate_2022}. These best-practices introduce additional uncertainty into photochemical models. Hence, we conduct two experiments. First, we run the four photochemical models with identical stellar spectra. For this, we use the panchromatic WASP-43~b flux from the MUSCLES Extension program \citep{behr_muscles_2023}. Second, in order to test the additional uncertainty associated with proxied stellar fluxes, we take four different approaches for each model. 
\begin{itemize}
    \item \textit{KINETICS} adopts the composite spectrum from \citet{Venot_global_2020}, which is based on \textit{International Ultraviolet Explorer} observations of GL~15B and the solar spectrum, scaled to match the high-energy flux of WASP-43~b \citep{czesla_X-ray_2013}. 
    \item \textit{EPACRIS} adopts the panchromatic flux of HD~85512, a K6V ($T_\mathrm{eff}=4305$~K) proxy star from the MUSCLES survey III \citep{loyd_muscles_2016}. 
    \item \textit{VULCAN} adopts the K7V-type ($T_\mathrm{eff}=4250$~K) flux from the grid of \citet{Rugheimer2013}, a semi-empirical reconstruction based on \textit{International Ultraviolet Explorer} observations of the proxy star BY~Dra.
    \item \textit{ACE-PAC} continues to use the panchromatic WASP-43~b spectrum of \citet{behr_muscles_2023}.
\end{itemize}
Each of the four spectral energy distributions is shown in Fig.~\ref{fig_stellarspectra}. We compare the four photochemical models when their stellar input spectra are uniformized, and when each of their stellar fluxes is differently sourced.

\subsection{Photochemical model intercomparison}

\begin{figure*}
    \centering
    \includegraphics[width=\linewidth]{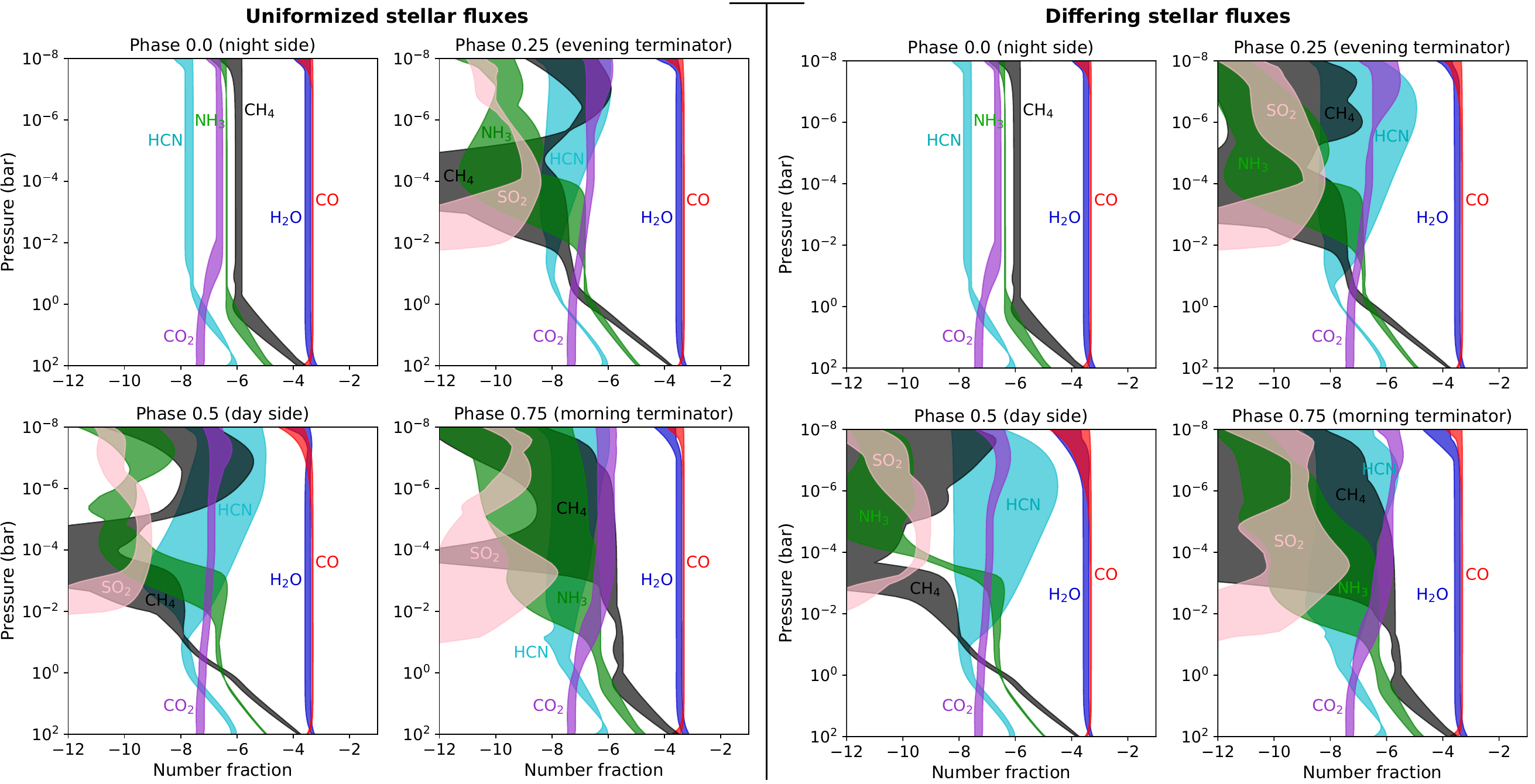}
    \caption{Illustration of the chemical species variation between four different but equivalent photochemical models: \textit{ACE-PAC}, \textit{KINETICS}, \textit{EPACRIS}, and \textit{VULCAN}. Left: All models use the same stellar spectrum as input. Right: All models use different stellar fluxes (see text for details). The composition is solar elemental composition. Colored areas indicate the maximum and minimum values resulting from all models.} 
    \label{fig_compare}
\end{figure*}

The solution space of several chemical species, consisting of the maximal and minimal values among all four models, is shown in Fig.~\ref{fig_compare} for the two cases of uniformized and differing stellar fluxes. For both cases, while model results are very consistent on the planetary night side, differences of up to multiple orders of magnitude can be seen at the other phases for trace species. Moreover, model differences in the deeper part of the atmospheres remain relatively small, while the model agreement becomes noticeably worse in the higher regions ($p < 100$~mbar). Out of all species shown, only \chem{H_2O}, CO, and \chem{CO_2} are relatively consistent between models. It should be noted that no single model is producing results that are clearly distinct from the other models, i.e.~the variations are driven by different models for different species and pressures. 

Given that the deep atmosphere is in chemical equilibrium, we can only attribute the small discrepancies in those layers to differences in the thermodynamic coefficients that are used in each model. This data determines the exact chemical composition at a given temperature and pressure and is used to reverse the rate coefficients in the chemical network. Despite these small differences, the agreement between models is good. Since all models use the same vertical mixing strength, this further leads to a consistent prediction of the quench points, i.e.~the pressures at which species diverge from their equilibrium concentrations, resulting in a consistent atmospheric composition throughout the night-side phase. 

Only when photochemistry is taken into account (phases 0.25, 0.5, 0.75), strong model divergence can be seen. These differences can be attributed to the different chemical reactions and corresponding rates used by each model, as well as the adopted photo-absorption and -dissociation cross-sections. The agreement between models remains good for the deepest layers. Indeed, each reaction network is specifically tuned to reproduce the expected chemical equilibrium composition under these conditions. Since there is no such guiding principle for the higher layers where photochemistry drives the composition out of equilibrium, however, discrepancies between the models grow. 

By comparing the set of phases in which all models use the same stellar input flux to those where they all use a different flux (respectively left and right in Fig.~\ref{fig_compare}), the effect of stellar irradiation becomes clearer. When all models use the same stellar spectrum, the uncertainties in modeled abundances are smaller. For instance, this trend is noticeable for HCN, where the typical 1~mbar-uncertainty is $\sim$1 order of magnitude for uniformized fluxes, but $\sim$2 orders of magnitude otherwise. The model uncertainties for \ce{CH4} and \ce{NH3} are already considerable for the uniformized fluxes ($\sim$4 and $\sim$3 orders of magnitude respectively at 1~mbar), but if models use differing stellar irradiation, these balloon to up to $\sim$6~orders of magnitude, due to the combined effect of different photodissociation rates with different chemical pathways. Remarkably, the abundance of \ce{SO2} does not depend strongly on the exact stellar input spectrum that is used, as its uncertainties ($<1$~order of magnitude at the pressure of maximal abundance, except on phase~0.75) are comparable across the two experiments.

These experiments show that, although the stellar spectra are roughly similar (Fig.~\ref{fig_stellarspectra}), small differences related to the adopted observations, activity levels, and (X)UV reconstructions in these spectra can propagate into the photodissociation rates. As such, the right side of Fig.~\ref{fig_compare} paints a more realistic view of the systematic uncertainty in photochemical models of exoplanets.

Overall, the intermodel differences are expected given the different implementations of each model. We highlight that it is not the purpose of this comparison to fully pinpoint all model discrepancies, but merely to illustrate the robustness of photochemical model results to the underlying implementation and the choice of chemical data and stellar spectrum. While we demonstrate that the adopted stellar flux introduces variations among the photochemical models, even the case of uniformized stellar fluxes shows strong model discrepancies. Follow-up work detailing the key photochemical pathways for each network \citep[e.g.][]{tsai_comparative_2021} and fully benchmarking the radiative transfer and photolysis rates \citep[e.g.][]{baudino_toward_2017} is needed to identify the remaining causes of model variations.

An additional source of uncertainty, not explored here, lies in the reaction rate error. Sensitivity tests have shown that uncertain reaction rates further increase the variance observed in the chemical abundances, especially in the upper atmosphere \citep{lira-barria_DARWEN_2024, agundez_quantification_2025}. Our experiment thus highlights the importance of rigorous methodology in kinetic modelling, including validation with experimental work \citep{veillet_extensively_2024, veillet_inclusion_2025}, to improve our predictions of the atmospheric compositions of exoplanets. 


\section{Discussion}\label{sec_discussion}


\subsection{Night-side methane depletion}\label{sec_disc_methane}

Although spectroscopically active and expected to be abundant at warm ($\lesssim 1000$~K) temperatures, methane has eluded detection on many exoplanets \citep[e.g.][]{morley_forward_2017, kreidberg_water_2018, carone_indications_2021, barat_metal-poor_2024}, until it was finally found with JWST in a handful of warm transiting giants and sub-Neptunes \citep[e.g.][]{bell_methane_2023, madhusudhan_carbon-bearing_2023, beatty_sulfur_2024, benneke_JWST_2024, welbanks_high_2024,hu2025water}. Nonetheless, it remains unclear what determines the presence and observability of methane. Several explanations have been brought forward, including interior or tidal heating \citep{agundez_puzzling_2014, fortney_beyond_2020, welbanks_high_2024}, a low carbon-to-oxygen ratio \citep{moses13gj436b, molliere_exoplanet_2022}, atmospheric circulation supplying gas from hotter regions \citep{Agundez_pseudo_2014, baeyens_grid_2021, zamyatina_quenching-driven_2024}, or condensation into clouds and photochemical haze, for which methane is an important precursor molecule \citep{gao_aerosol_2020}. An extensive parameter study can be found in \citet{mukherjee_effects_2025}. All of the above mechanisms can result in chemical disequilibrium, and a combination of them might be responsible for the observed methane depletion in \citet{Bell_nightside_2024}. 


We show that zonal mixing, such as day-night transport via the equatorial jet stream, is an important mechanism that naturally explains the methane depletion on the night side of WASP-43~b, in accordance with upper limits from retrieval models (Fig.~\ref{fig_CH4_metallicities}). For this mechanism to occur, only moderately fast wind speeds of $\gtrsim 500$~m/s are required (Fig.~\ref{fig_CH4_windspeeds}), well below jet speed predictions from the 3D GCMs in this paper, hot Jupiters in general \citep[2 -- 4~km~s$^{-1}$,][]{showman_atmospheric_2020}, and the measured jet speed of WASP-43~b from Doppler-shifted molecular absorption lines \citep[$5.4^{+2.8}_{-2.1}$~km~s$^{-1}$,][]{lesjak_retrieval_2023}. As such, we can reasonably assume it to occur on WASP-43~b. In fact, this type of transport-induced disequilibrium chemistry on WASP-43~b has been investigated before, with numerous studies highlighting the potential depleting effect on \ce{CH4} \citep{mendonca_three-dimensional_2018, Venot_global_2020}. Moreover, recent observations of the similarly hot exoplanet NGTS-10A~b also found evidence of a night side depleted in methane \citep{parmentier_horizontal_2026}. The authors pin horizontal transport as the cause, ruling out any alternative explantions such as vertical mixing or non-solar elemental abundances.

Given the prevalence of fast ($\sim$km~s$^{-1}$) superrotating jet streams on hot transiting exoplanets across a large parameter space \citep{Roth_hot_2024}, strong equatorial advection extends to other hot gaseous exoplanets. Atmospheric circulation is expected to result in chemically homogeneous day and night sides for planets with effective temperatures below $\sim$$1400$~K \citep{baeyens_grid_2021}. As such, it appears that not the night side temperature, but the global mean effective temperature could be a better indicator for methane presence, regardless of the observed phase. Indeed, one of the hottest transiting planet on which \ce{CH4} has yet been detected\footnote{Here we don't consider tentative \ce{CH4} detections made via high-resolution cross-correlation spectroscopy of a few warmer planets \citep{giacobbe_five_2021, guilluy_GAPS_2022}. Outstanding tension with low-resolution space-based spectroscopy calls into question some of these results \citep{xue_JWST_2024}.}$^{,}$\footnote{Surprisingly, evidence for night-side methane was also found on the ultra-hot Jupiter WASP-121~b, despite a hot night-side temperature of $\sim$1200~K \citep{evans-soma_SiO_2025}. Given that temperature, its presence is potentially linked to anomalous elemental ratios, e.g.~super-stellar C/O, and a dynamical decoupling from the day side composition.} is WASP-80~b, with equilibrium temperature $\Teq=850$~K \citep{bell_methane_2023}. This temperature is sufficiently low to support high amounts of methane not only on the night-side hemisphere, but also the day side. As such, even in the presence of strong day-night mixing, methane remains abundant.  Future observations of planets in the $\Teq = 800-1000$~K regime will further uncover the important role of day-night transport for the detectability of molecular absorption.

At the same time, our 3D \textit{Exo-FMS} model shows that the mid-latitudes on the planetary night side are local maxima for \ce{CH4} and \ce{NH3}. Meridional winds carry these species from their production sites toward the equator, resulting in global enrichment. This effect has been thought significant for methane on HD~189733~b \citep{drummond_3D_2018, drummond_implications_2020}, a 1200~K planet that is in a sweet spot to observe transport-induced quenching \citep{zamyatina_observability_2023}. Even in these 3D coupled-chemistry studies, however, meridional quenching is a secondary effect and the large-scale composition is primarily determined by zonal (and vertical) mixing. For the hotter planet WASP-43~b, our models show that meridional quenching indeed plays a minor role for the disequilibrium chemistry of \ce{CH4}, \ce{NH3}, and HCN.  


An alternative scenario to explain the low abundance of methane on some exoplanets is strong vertical mixing, in combination with a high intrinsic temperature ($\Tint$), such as caused by tidal heating \citep{agundez_puzzling_2014, fortney_beyond_2020}. In that scenario, hot methane-depleted gas is dredged up from the deep interior to observable pressures at a fast rate, thereby preventing the chemical reactions to equilibrate. Provided that the strength of vertical eddy diffusion ($\kzz$) is well-understood, it is in principle possible to use this relation as a \textit{thermometer} and constrain the interior temperature based on the observed methane abundance \citep{sing_warm_2024, welbanks_high_2024}. Due to observational degeneracies and model limitations, however, such analyses may lead to conflicting results \citep{konings_reliability_2025}.

We investigated the effect of a high-temperature interior on the chemical composition, but found no significant differences in the observable atmosphere ($p < 1$~bar) between a hot or a cold interior (Fig.~\ref{fig_chem_kinetics}). Although the equilibrium composition at depth is indeed affected, with \ce{CH4} depleted by an order of magnitude in the case of a hot interior, the assumed $\kzz$ value ($10^{7}$~cm$^2$~s$^{-1}$ at 1~bar) is too low for that composition to propagate upward through vertical mixing.

In Sec.~\ref{sec_kzz}, the possibility of strong vertical mixing ($\kzz=10^{11}$~cm$^2$~s$^{-1}$) was explored as well. We found that increased vertical mixing would lead to an increase in tropospheric \ce{CH4}, as well as \ce{NH3} and HCN. While a high \ce{CH4} abundance is disfavored by \citet{Bell_nightside_2024}, a combination of strong vertical mixing and 10$\times$ solar metallicity results in an increased \ce{NH3} abundance while keeping \ce{CH4} low (Fig.~\ref{fig_kzz}), bringing the result more in line with the retrieval analysis of \citet{Yang_simultaneous_2024} \citep[see also][for a 3D perspective of this effect]{zamyatina_quenching-driven_2024}. This example shows that pinning down the atmospheric metallicity is essential to understand the full picture of atmosphere dynamics, and vice-versa, as these aspects are linked through disequilibrium chemistry.

It should be noted that WASP-43~b, with a mass of 2.0~M$_\textrm{J}$ and 1.0~R$_\textrm{J}$ \citep{gillon_TRAPPIST_2012}, is one of the rare non-inflated hot Jupiters. As such, it is unlikely that the planet has a hot interior \citep{thorngren_intrinsic_2019, sarkis_evidence_2021}. Additionally, dynamical models have shown that high-gravity planets such as WASP-43~b tend to have slower vertical winds and a low $\kzz$ \citep{baeyens_grid_2021}. Therefore, based on first principles, we do not necessarily expect $\kzz$ to attain very high values on this planet.


Finally we discuss the effect of photochemistry, in particular sulfur chemistry, on the methane concentration. Methane is strongly affected by photochemistry, because its CH$_3$-H bond is easily broken, either by direct photolysis of by reaction with atomic H (see Fig.~\ref{fig_H}). This effect, however, typically only manifests in the upper atmosphere ($p<1$~mbar), whereas the JWST/MIRI observations probe deeper layers (1~bar - 1~mbar). Even if the stellar irradiation is increased by several orders of magnitude, for example following a stellar flare, the chemical impact is largely confined to the upper atmosphere \citep{Konings_impact_2022, barat_metal-poor_2024}. As a result, photodissociation plays a role in the observed methane depletion, but is unlikely to be the key factor.

The detection of sulfur-bearing molecules on WASP-39~b and several other exoplanets by the JWST \citep{tsai23wasp39b, dyrek_SO2_2024, thao_featherweight_2024, fu_hydrogen_2024} has provided new chemical constraints, highlighting the complexity of sulfur photochemistry and its importance in exoplanet atmospheres. The inclusion of sulfur-bearing molecules and reactions forms one of the key differences in our study of WASP-43~b compared to \citet{Venot_global_2020}. We found that coupled carbon-sulfur chemistry has a big impact on the concentration of methane seen in our chemical models (Sec.~\ref{sec_S}). Specifically, the reaction S + \ce{CH3} $\rightarrow$ \ce{H2CS} + H acts as a sink for the methyl radical and inhibits the (re)formation of methane.\footnote{On cooler planets, other reactions such as \ce{CH3} + S $\rightarrow$ \ce{CH2SH} will lead to the formation of a C-S bond \citep{hu2025water}.} Thus, the inclusion of sulfur in the chemical network results in a reduced \ce{CH4} abundance compared to pure C-H-O-N chemistry \citep[see also Fig.~18 in][]{tsai_comparative_2021}. New observations of species like \ce{CS2} \citep{benneke_JWST_2024} and OCS/COS \citep{kirk_BOWIE-ALIGN_2025}, in tandem with photochemical modelling \citep[][Moses et al., \textit{in prep.}]{zahnle_photolytic_2016, veillet_inclusion_2025}, will shed more light on the role of sulfur in the atmospheric chemistry of exoplanets in the future.

\subsection{Night-side clouds}


In this paper, we have opted to omit cloud modelling in order to discuss the phase-dependent chemistry of WASP-43~b in detail. Nonetheless, phase curves have consistently yielded day-night brightness contrasts larger than predicted by 3D GCMs, indicating the presence of night-side clouds \citep{stevenson_spitzer_2017, mendonca_revisiting_2018, keating_uniformly_2019, Bell_nightside_2024}. Here we discuss the effect night-side clouds may have on the chemical composition of WASP-43~b.

Clouds affect the gas-phase chemistry through depletion and fractionation of certain species. Condensation of mineral clouds (e.g.~SiO$_2$, MgSiO$_3$) on hot Jupiters can deplete the available oxygen reservoir, resulting in an increased C/O of the gas \citep{molliere_model_2015}. As a consequence, the composition may contain more CH$_4$ than CO. Given that methane appears depleted rather than enhanced, we conclude that this effect is not important for WASP-43~b.

\citet{helling_cloud_2021} performed microphysical cloud modelling for WASP-43~b and found that silicate clouds could form between 100~mbar and 0.1~mbar, with metal-oxide clouds taking over around 1~bar. Cloud modelling by \citet{Venot_global_2020} and \citet{chubb_dark_2024} likewise found that silicates take up large volume fractions of the night-side and even (patchy) day-side clouds. As a result, the gas C/O content may be enhanced from solar value (C/O = 0.55) to 0.75, which is nonetheless unlikely to impact the overall chemistry strongly \citep{moses_chemical_2013}. 
Additionally, sulfur-bearing clouds with a lower condensation temperature (ZnS, MnS, Na$_2$S) may form on the night side at pressures lower than $p<0.1$~bar \citep{Venot_global_2020}, especially in the presence of cloud-condensation nuclei \citep{gao_microphysics_2018}. Such clouds would deplete the gas-phase sulfur reservoir. Similar to the silicate clouds, however, the cloud-chemistry feedback is throttled by the available metals. If their abundances scale with the solar abundances \citep{lodders_solar_2019}, even the extreme case in which all zinc, manganese, and sodium is condensed in sulfur clouds would only deplete about $8\%$ of available sulfur. As a result, direct cloud-chemistry feedback is unlikely to drastically change the results in this paper.

Finally, night-side clouds may also affect the phase-dependent chemistry in indirect ways. \citet{Bell_nightside_2024} propose that the wind jet speed on WASP-43~b is reduced below typical values to about 2.0 -- 2.5~km~s$^{-1}$. This inefficient circulation, potentially caused by night-side clouds \citep{mendonca_revisiting_2018, roman_modeled_2019}, would be more in line with the observed phase curve hot spot offset ($7.34\pm0.38\degrees$). We show here that methane is zonally quenched by wind jet speeds faster than $\gtrsim 500$~m~s$^{-1}$ (Fig.~\ref{fig_CH4_windspeeds}). Hence, even a strongly reduced jet stream that is suppressed through night-side clouds would be sufficient to deplete methane on the night side.


\subsection{WASP-43~b's metallicity}

Using photochemical kinetics models of varying atmospheric metallicity, we have demonstrated that \ce{CO2} and \ce{SO2} are among the clearest metallicity indicators (Figs.~\ref{fig_CO2_metallicities}, \ref{fig_SO2_metallicities}), as known from literature \citep{lodders_atmospheric_2002, moses13gj436b, konings_reliability_2025}. Unfortunately, these species have not been detected in the MIRI/LRS phase curve data \citep{Bell_nightside_2024}, so we can only compare to upper limits, as well as weak (2.5$\sigma$) evidence for \ce{CO2} in \citet{Yang_simultaneous_2024}.

Previous investigations of WASP-43~b, mostly based on \textit{Hubble} and \textit{Spitzer Space Telescope} data, place its atmospheric metallicity at near-solar \citep{kreidberg_precise_2014, stevenson_spitzer_2017} or (slightly) super-solar values \citep{changeat_exploration_2021, chubb_exoplanet_2022}. High-resolution spectroscopy of the planet's day side leads to comparable constraints, with a metallicity of 0.4 to 8 $\times$ solar based on abundance measurements of CO and \ce{H2O} \citep{lesjak_retrieval_2023}. Finally, a recent analysis of the JWST/MIRI phase curve aligns with this result, finding a metallicity of 0.6 to 6.5$\times$ solar \citep{Yang_simultaneous_2024}. In summary, despite leaving an order of magnitude uncertainty margin, various studies tend to converge on a roughly solar metallicity.

Complementary observations of WASP-43~b using NIRSpec/G395H (GTO 1224, PI: Stephan Birkmann) are best suited to observe signatures of \ce{CO2} (4.3~$\mu$m) and \ce{SO2} (4.0~$\mu$m). As a result, they will likely lead to a more stringent metallicity constraint than currently available. Our models predict that \ce{CO2} and CO should be visible at all phases, as well as \ce{SO2} depending on the phase and metallicity. We conclude that a clear spectroscopic detection of \ce{CO2} and potential upper limit to \ce{SO2} could finally pin down the elemental composition of WASP-43~b.

\section{Conclusions and outlook}\label{sec_conclusions}

In this paper, we conducted a theoretical study of the phase-dependent chemistry of WASP-43~b using a suite of chemical kinetics models that take into account day-night transport. We investigated how gas-phase disequilibrium chemistry shapes the global chemical inventory of the planet, and how these changes can be probed via spectroscopic phase curves.

We find that horizontal quenching is the most likely scenario to explain the observed methane depletion on the planet's night side \citep{Bell_nightside_2024}. This result is robust across the different models we tested, does not depend on the atmospheric metallicity, and requires only moderately strong wind speeds of $\gtrsim 500$~m~s$^{-1}$. As a result, even if the presence of night-side clouds suppresses the atmospheric circulation and generation of a strong wind jet \citep{mendonca_revisiting_2018}, the zonal advection is still sufficient to result in low night-side methane concentrations. In addition, we identified a secondary mechanism causing a reduction in methane, namely coupled carbon-sulfur chemistry. We show that atomic sulfur can sweep up methyl radicals, leading to the formation of C-S-bonded molecules that can compete with the formation of CH$_4$. While fast atmospheric circulation is expected to be ubiquitous on hot transiting exoplanets, future observations of species such as \ce{CS2} and OCS may shed new light on the role that sulfur plays in the atmospheric chemistry of different exoplanets \citep[cf.][]{benneke_JWST_2024, kirk_BOWIE-ALIGN_2025, veillet_inclusion_2025}.


By varying the metallicity in our models and comparing the results with atmospheric retrievals, we have attempted to further constrain the metallicity of WASP-43~b. We confirm that both \ce{CO2} and \ce{SO2} are good probes of atmospheric metallicity. Notably, we show that sulfur dioxide would have been observable at 10 $\times$ solar metallicity in the MIRI/LRS dataset, thereby providing a clear-atmosphere upper limit. The presence of clouds, however, may mute the spectral signal depending on the altitude and opacity of the cloud deck. Future observations, in particular the NIRSpec/G395H phase curve (GTO 1224), will be better suited to constrain the atmospheric metallicity, by measuring the absorption features of \ce{CO2}, CO, and \ce{SO2}. 

Finally, upon comparing the results of four different 1D photochemical models, we find that the abundances of \ce{H2O}, CO, and \ce{CO2} are robustly modeled. Additionally, the quenching pressures, where atmosphere dynamics become quicker than chemical reactions, are robustly predicted for most main species. By contrast, the abundances of photochemically active species such as \ce{CH4}, \ce{NH3}, \ce{SO2}, and HCN show variations of more than an order of magnitude from model to model. The underlying causes lie in the different chemical pathways and radical species used by each model, as well the differences in the adopted stellar high-energy spectra, which are typically uncertain for most exoplanets. Given the increasing importance of photochemistry in hot exoplanet atmospheres \citep[e.g.][]{tsai23wasp39b}, we advocate for the development of a photochemistry benchmark to quantify differences between chemical kinetics models, and provide a sense of robustness -- a \textit{model error bar} -- for photochemical predictions. In parallel, there is an opportunity for rate sensitivity tests \citep{lira-barria_DARWEN_2024, agundez_quantification_2025} and experimental validation \citep{veillet_extensively_2024, veillet_inclusion_2025} to improve photochemical data and reaction schemes to further reduce model uncertainties.


\begin{acknowledgements}
      RB acknowledges support from the \textit{Origins} investment incentive, the research program Vidi \textit{New Frontiers in Exoplanetary Climatology} with project number 614.001.601, and the Veni ENW research program with project number VI.Veni.242.190, which are (partly) financed by the Dutch Research Council (NWO). 
      JM acknowledges support from the NASA Exoplanet Research Program 80NSSC23K0281. 
      JB acknowledges the support received in part from the NYUAD IT High Performance Computing resources, services, and staff expertise.
      EL acknowledges funding from the CSH through the Bernoulli Fellowship.
      SMT is supported by the National Science and Technology Council (grants 114-2112-M-001-065-MY3).
      SK, LC, and CH acknowledge funding from the European Union H2020-MSCA-ITN-2019 under grant agreement no.~860470 (CHAMELEON).
      AP acknowledges funding from a UK Science and Technology Facilities Council (STFC) Small Award, grant number UKRI/ST/B001171/1.
      OV acknowledges funding from the ANR project ‘EXACT’ (ANR-21-CE49-0008-01) and from the Centre National d’Etudes Spatiales (CNES).
      The authors acknowledge the use of the Lisa and Snellius Computing clusters hosted by SURFsara. Part of the simulations (project id 72245 and 71773) were performed on the Austrian Scientific Computing (ASC) infrastructure, in particular the Vienna Science Cluster (VSC).
      For this work we have made use of the GCM post-processing library \textit{gcm\_toolkit} \citep{Schneider_gcm_2022}.
      This work is based on observations made with the NASA/ESA/CSA JWST. The data were obtained from the Mikulski Archive for Space Telescopes at the Space Telescope Science Institute, which is operated by the Association of Universities for Research in Astronomy, Inc., under NASA contract NAS 5-03127 for JWST. These observations are associated with program JWST-ERS-01366. Support for program JWST-ERS-01366 was provided by NASA through a grant from the Space Telescope Science Institute.
      This research has made use of NASA's Astrophysics Data System Bibliographic Services
\end{acknowledgements}

%
%

  \bibliographystyle{aa} 
  \bibliography{my_bib} 

\begin{appendix}

\section{Atmospheric retrievals}\label{sec_retrievals}

\subsection{Additional atmospheric retrieval models}

In \citet{Bell_nightside_2024}, we conducted a comprehensive retrieval analysis of the individual molecular abundances in the MIRI/LRS data at four orbital phases using multiple retrieval frameworks, including free chemistry retrievals with \textit{PyratBay} \citep{cubillos_PYRAT_2021}, \textit{HyDRA} \citep{gandhi_retrieval_2018}, and \textit{NEMESIS} \citep{irwin_NEMESIS_2008}. 
A variety of models were generated under different assumptions in order to test the robustness of the results. We tested the impact of including:
\begin{itemize}
    \vspace{-1pt}
    \item different species in the final model, removing molecules for which the retrieval showed less constrained posteriors,
    \item a \textit{dilution} parameter, which uniformly scales the emission spectrum by a factor $<1$ in order to account for temperature inhomogeneities \citep{taylor_understanding_2020},
    \item a wavelength-independent \textit{error inflation} parameter to account for model uncertainty \citep{Bell_nightside_2024}.
    \vspace{-1pt}
\end{itemize} 

The latter two parameters act to increase the data uncertainties in order to avoid any bias that can be introduced by the three-dimensional (3D) nature of planetary emission and to prevent overfitting. Indeed, their inclusion was statistically preferred in the retrievals, suggesting that models excluding these parameters may fit the data slightly less well, though with the statistical evidence remaining below the threshold for strong support \citep{kass1995bayes}. It should be noted that a better treatment of 3D inhomogeneities should also lead to a better fit without including these parameters. 

In \citet{Bell_nightside_2024}, we reported retrievals which include free parameters for the abundances of H$_2$O, CH$_4$, NH$_3$, CO, and CO$_2$ or HCN (depending on the phase), as well as an error inflation parameter for each phase and a dilution parameter for each phase except for the night side (for which the dilution parameter was not statistically preferred; see Bell et al. Extended Data Table~2). Retrieval models with this setup have been color coded \textbf{black} throughout this paper. These assumptions provided the most consistent results across all frameworks and were statistically favored. However, the abundances of trace species could not be robustly determined, likely due to insufficient modeling of 3D effects, so we reported the statistically significant constraints only for H$_2$O and CH$_4$. Conversely, \citealt{Yang_simultaneous_2024} (color-coded \textcolor{YangBlue}{\bf blue} throughout) include a more comprehensive treatment of day-night thermal inhomogeneities in 2D, including opacity from H$_2$O, CH$_4$, NH$_3$, CO, and CO$_2$.

\begin{figure}
    \centering
    \includegraphics[width=\linewidth]{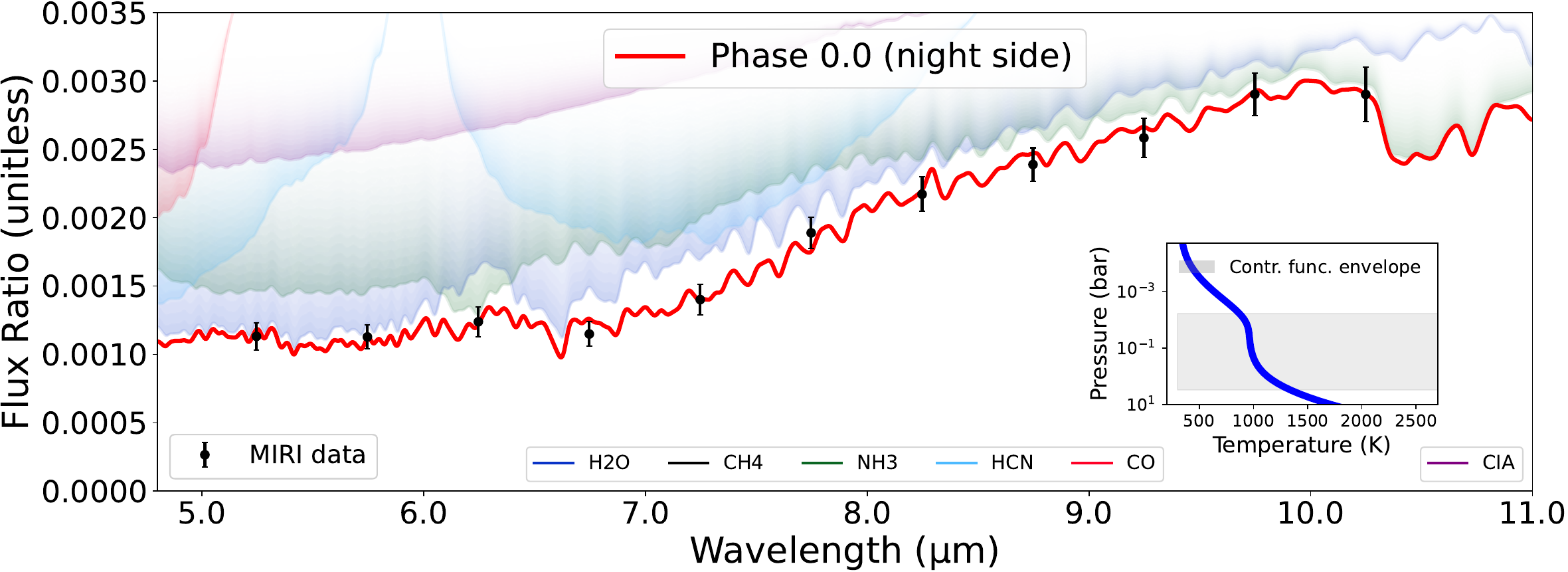}
    \includegraphics[width=\linewidth]{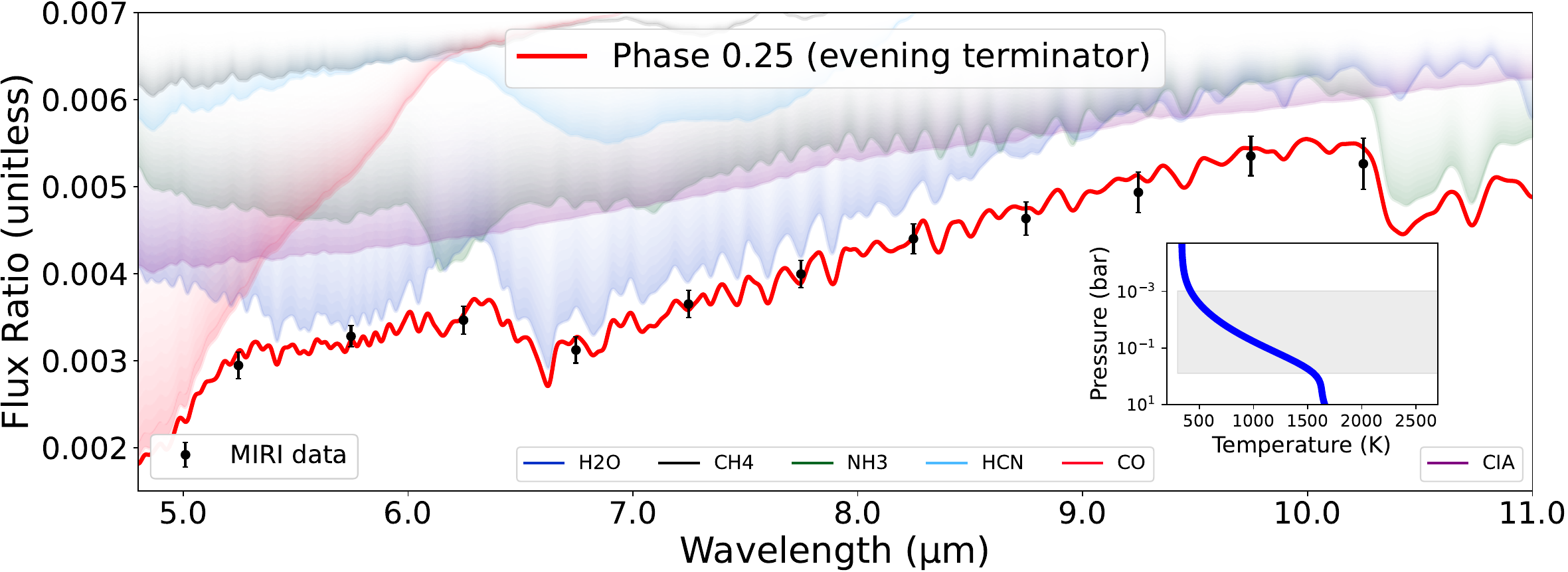}
    \includegraphics[width=\linewidth]{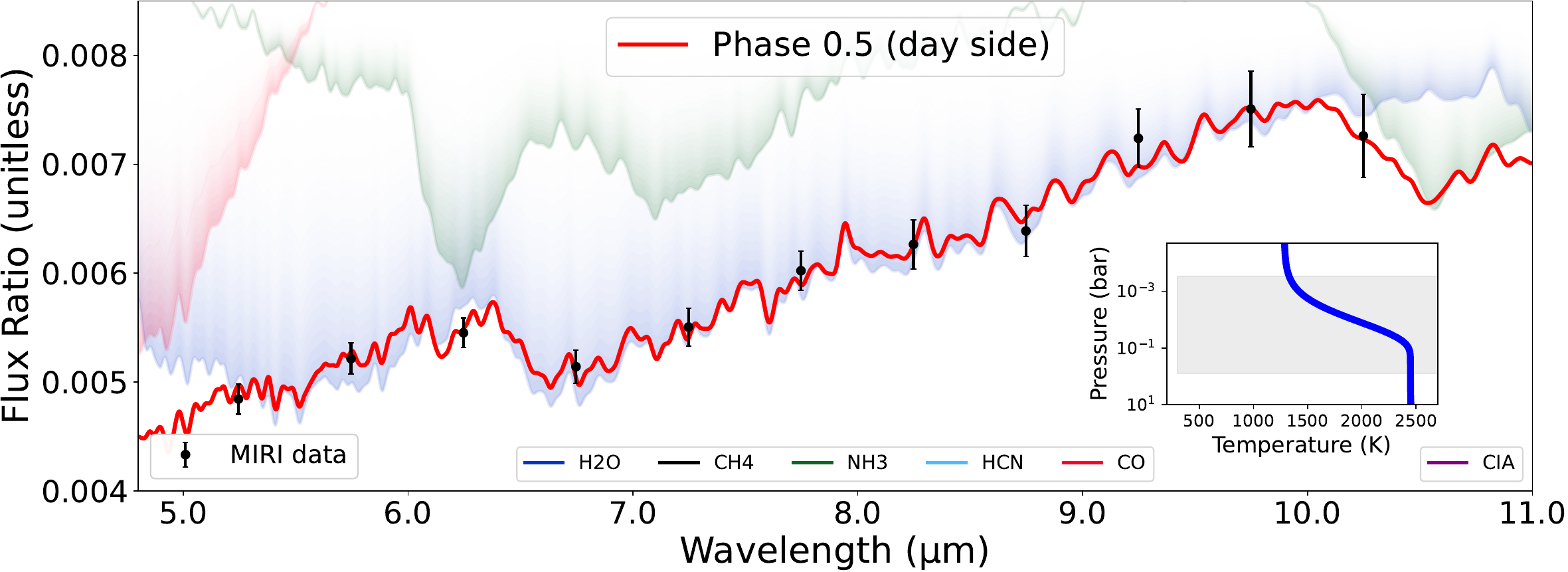}
    \includegraphics[width=\linewidth]{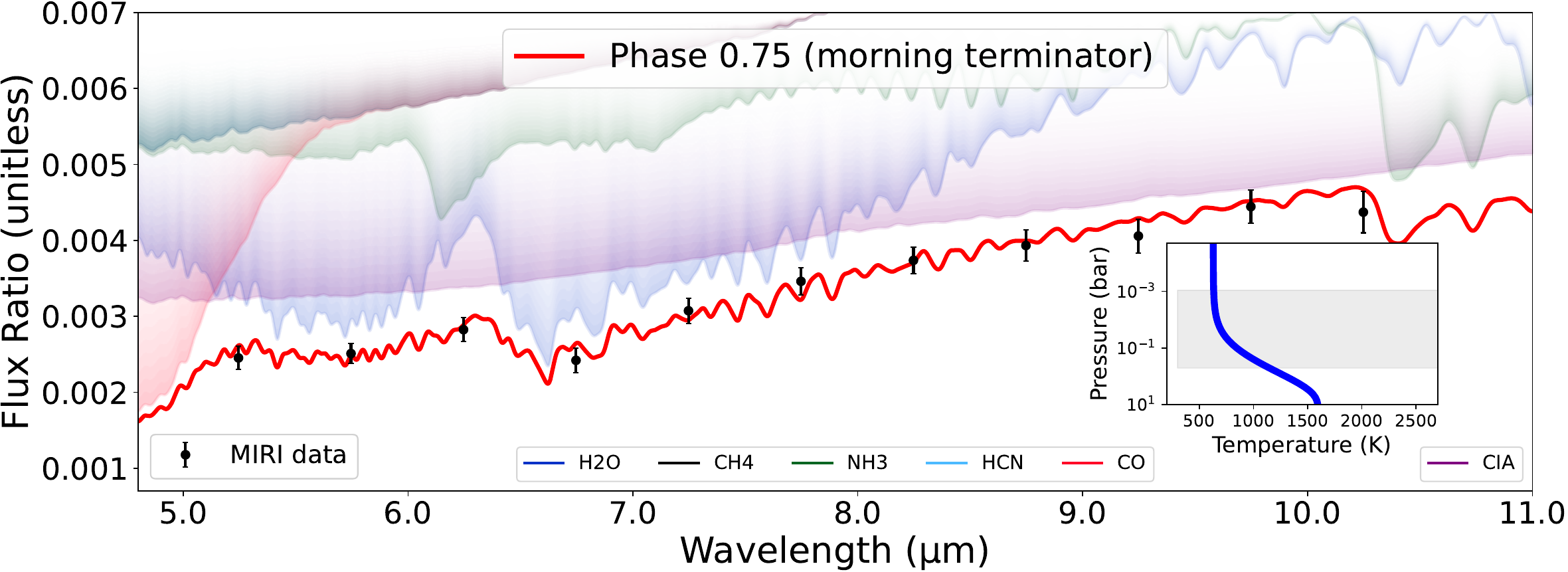}
    \caption{Contributions from individual opacity sources for the best-fit spectrum at each orbital phase (\textit{red}), as reported in \citet{Bell_nightside_2024} for \textit{\textbf{PyratBay}}. The underlying model has the most stringent assumptions, using only five molecular species, error inflation, and a dilution parameter for all phases except 0.5. 
    The color transparency scales with the optical depth of each molecular absorber. The inset panels display the corresponding best-fit temperature–pressure profiles and the extent of the contribution functions.} 
    \label{fig:Jasmina_contributions}
\end{figure}

\begin{table*}
\centering
\caption{Summary of chemical species constraints for the different retrieval models presented in this paper.} 
\label{tab:BellEtal}
\footnotesize
\begin{tabular}{|c|c|c|c|c|c|p{0.4cm}|c|c|}
\cline{1-6}\cline{8-9}
{\bf Species} &
{\bf Phase} &
{\shortstack[c]{\textcolor{YangBlue}{\bf Yang et al. (2D)}}} &
{\shortstack[c]{{\bf PyratBay} (1$\sigma$)}} &
{\shortstack[c]{{\bf HyDRA}}} &
{\shortstack[c]{{\bf NEMESIS}}} &
& 
{\shortstack[c]{\textcolor{PyratRed}{\bf PyratBay} (1$\sigma$)}} &
{\shortstack[c]{{\bf Species}}} \\
\cline{1-6}\cline{8-9}
\multirow{4}{*}{\bf H$_2$O}
 & 0.00 & \multirow{4}{*}{$-3.86^{+0.69}_{-0.36}\,\,(1\sigma)$} & $-1.9^{+1.1}_{-1.5}$ & $-3.5^{+1.4}_{-0.8}\,\,(1\sigma)$ & $-3.6^{+1.6}_{-0.5}\,\,(1\sigma)$ & & $-1.6\pm1.2$ & \multirow{4}{*}{\bf H$_2$O} \\
 & 0.25   &  & $-1.9^{+1.1}_{-1.6}$ & $-3.2\pm1.1\,\,(1\sigma)$  & $-2.1^{+0.7}_{-1.5}$ \,\,(1$\sigma$) & & $-3.3\pm2.0$ &  \\
 & 0.50   &  & $-1.9^{+1.0}_{-1.5}$ & $-3.4^{+1.2}_{-1.1}\,\,(1\sigma)$  & $-1.8^{+0.5}_{-0.8}$ \,\,(1$\sigma$) & & $-3.8\pm0.8$ &  \\
 & 0.75   &  & $-2.2^{+1.3}_{-2.0}$ & $-3.3^{+1.4}_{-1.5}\,\,(1\sigma)$  & $-4.4^{+2.2}_{-4.9}$ \,\,(1$\sigma$) & & $-5.4\pm0.2$ &  \\
\cline{1-6}\cline{8-9}
\multirow{4}{*}{\bf NH$_3$}
 & 0.00 & \multirow{4}{*}{$-5.27^{+0.46}_{-0.30}\,\,(1\sigma)$} & $-6.6^{+2.5}_{-3.3}$ & $<-3.05\,\,(99\%)$ & $-7.3^{+2.9}_{-3.0}\,\,(1\sigma)$ & & $-3.7\pm1.2$ & \multirow{4}{*}{\bf NH$_3$} \\
 & 0.25   &  & $-7.8\pm2.6$ & $<-4.03\,\,(99\%)$           & $-7.8\pm2.5$ \,\,(1$\sigma$) & & $-7.6\pm2.9$ &  \\
 & 0.50   &  & $-7.9^{+2.6}_{-2.5}$ & $<-4.23\,\,(99\%)$   & $-8.1^{+2.7}_{-2.4}$ \,\,(1$\sigma$) &  & $-8.8\pm2.0$ &  \\
 & 0.75   & & $-7.6\pm2.7$ & $<-2.29\,\,(99\%)$            & $-6.4^{+3.5}_{-3.4}$ \,\,(1$\sigma$) &  &  $-7.3\pm1.7$ &  \\
\cline{1-6}\cline{8-9}
\multirow{4}{*}{\bf CH$_4$}
 & 0.00 & \multirow{4}{*}{$<-6.93\,(95\%)$} & $-8.6^{+2.2}_{-2.1}$ ; $<-5.2\,\,(95\%)$ & $<-5.8\,\,(95\%)$ & $<-5.9\,\,(95\%)$ & &  $-8.5\pm2.1$ & \multirow{4}{*}{\bf SO$_2$} \\
 & 0.25   &  & $-8.3^{+2.4}_{-2.3}$ ; $<-4.4\,\,(95\%)$ & $<-5.3\,\,(95\%)$ & $<-4.9\,\,(95\%)$ & &  $-6.4\pm2.2$ &  \\
 & 0.50   &  & $-7.9^{+2.6}_{-2.5}$ ; $<-3.5\,\,(95\%)$ & $<-4.6\,\,(95\%)$ & $<-4.9\,\,(95\%)$ & &  $-8.9\pm2.0$ &  \\
 & 0.75   &  & $-8.4^{+2.3}_{-2.2}$ ; $<-4.1\,\,(95\%)$ & $<-5.6\,\,(95\%)$ & $<-1.9\,\,(95\%)$ & &  $-10.3\pm1.1$ &  \\
\cline{1-6}\cline{8-9}
\multirow{4}{*}{\bf CO}
 & 0.00 & \multirow{4}{*}{$-3.25^{+0.58}_{-0.46}\,\,(1\sigma)$} & $-6.6\pm3.3$ & N/A & $-7.6^{+3.0}_{-2.8}\,\,(1\sigma)$ & & $-6.7\pm3.2$  & \multirow{4}{*}{\bf H$_2$S} \\
 & 0.25   &  & $-6.0^{+3.3}_{-3.7}$ & FLAT & $-6.0^{+2.9}_{-3.5}\,\,(1\sigma)$ & &  $-7.3\pm3.2$ &  \\
 & 0.50   &  & $-6.0^{+3.2}_{-3.6}$ & FLAT & $-5.6^{+2.9}_{-4.0}\,\,(1\sigma)$ & &  $-8.3\pm2.5$ &  \\
 & 0.75   &  & $-6.6\pm3.3$ & N/A & $-6.8^{+3.3}_{-3.2}\,\,(1\sigma)$ & & $-8.9\pm1.9$  &  \\
\cline{1-6}\cline{8-9}
\multirow{4}{*}{\bf CO$_2$}
 & 0.00 & \multirow{4}{*}{$-4.60^{+0.55}_{-0.50}\,\,(1\sigma)$} & \multirow{4}{*}{N/A}  & N/A & FLAT & &  $-6.6\pm3.2$& \multirow{4}{*}{\bf CO$_2$} \\
 & 0.25   &  &                    & N/A  & $<-2.2\,\,(99\%)$ & &  $-8.1\pm2.6$ &  \\
 & 0.50   &  &                    & N/A  & $<-3.9\,\,(99\%)$ & & $-8.6\pm2.3$  &  \\
 & 0.75   &  &                    & $<-2.68\,\,(99\%)$  & FLAT & &  $-7.7\pm2.5$ &  \\
\cline{1-6}\cline{8-9}
\multirow{4}{*}{\bf HCN}
 & 0.00
    & \multirow{4}{*}{} 
    & $-7.6\pm-2.7$
    & \multicolumn{1}{|c}{\multirow{4}{*}{}}
    & \multicolumn{1}{c}{\multirow{4}{*}{}} \\
 & 0.25
    & 
    & $-7.3^{+3.0}_{-2.9}$ \\
 & 0.50
    & 
    & $-7.1^{+3.2}_{-3.0}$ \\
 & 0.75
    & 
    & $-7.7^{+2.8}_{-2.6}$ \\
\cline{1-2}\cline{4-4}
\noalign{\smallskip}
\end{tabular}
\begin{minipage}{\linewidth} 
    {\footnotesize \textbf{Notes.} The first part of the table shows species constraints obtained from the 2D retrieval of \citet{Yang_simultaneous_2024} (\textcolor{YangBlue}{\bf blue}), as well as results from the most conservative models reported in \citet{Bell_nightside_2024} (\textbf{black}). The second part of the table provides an extended list of species constraints obtained using 
    a more permissive \textit{PyratBay} model (\textcolor{PyratRed}{\bf red}), which also includes additional species beyond the basic five, but omits both dilution and error inflation parameters. The column header colors correspond to those used in the figures in Secs.~\ref{sec:metallicity} and~\ref{sec:JB_retr_compar}, which show the comparison between \textit{ACE–PAC} pseudo-2D kinetics, chemical equilibrium, and retrieval models. Note that the first and second sections of the table list different molecular species, each corresponding to the models listed within the same section.
    }
\end{minipage}
\end{table*}

Although the abundances of several trace chemical species could not be robustly constrained by the retrievals reported in \citet{Bell_nightside_2024}, it can be useful to compare upper limits on these abundances to photochemical models. In this paper, we discuss retrievals that include additional molecular species. Beyond the models with the molecules mentioned above, we also generated models that included SO$_2$, H$_2$S, and C$_2$H$_2$. These species were chosen because they can be expected to influence the WASP-43~b spectra given the wavelength range probed by the MIRI observations and the planet’s temperature regime in all orbital phases. 
We explored models in which the error inflation and dilution parameters were omitted entirely. This \textit{PyratBay} model is color coded \textbf{\color{PyratRed}{red}}. An overview of the numeric constraints obtained with the two-dimensional \textbf{\color{YangBlue}{Yang et al.}}, more conservative \textbf{Bell et al.}, and a more permissive \textbf{\color{PyratRed}{PyratBay}} retrieval model is given in Table~\ref{tab:BellEtal}.


Before we turn to the results of these additional retrievals, we note that 
the analyses performed for \citet{Bell_nightside_2024} were not fully uniform in their implementation. The differences arise not only from variations in modelling assumptions, such as the inclusion/exclusion of dilution and error inflation parameters and/or different temperature profile parameterizations, but also from the use of different opacity databases, underlying opacity treatments (e.g., $k$-correlated versus line-by-line opacities), the selection of molecular species included in individual models, and, ultimately, the manner in which results were reported (e.g., 1$\sigma$ confidence intervals versus 95\% or 99\% upper limits). Since some of these factors could not be controlled or reconciled, we explicitly acknowledge them and carefully document all relevant differences. In particular, we clearly indicate the confidence levels associated with each reported constraint, even when these are not uniform across the models, to ensure transparency. In this context, we report all the confidence intervals obtained from these analyses to allow future studies using the same or additional instruments to compare against our results. In Table~\ref{tab:BellEtal}, we list confidence intervals for all models discussed in this section. We encourage the reader to carefully consider the associated uncertainties and confidence intervals when interpreting the results.

\subsection{Molecular contributions in the MIRI/LRS range}

We first illustrate how individual chemical species contribute to the best-fit spectrum reported in Bell et al.~under the most conservative modelling assumptions (Fig.~\ref{fig:Jasmina_contributions}). Other molecular species, not reported in Bell et al., may show tentative signatures that can arise under alternative model assumptions. We present only the \textit{\textbf{PyratBay}} results, but they are representative given the strong agreement among all retrieval models \citep[][their Figure~4, Extended Data Table~2, and Extended Data Figs.~3, 5, and~7]{Bell_nightside_2024}. 
 Under these strict modelling assumptions
, H$_2$O is the dominant contributor to the spectra across all orbital phases and most wavelengths, except at wavelengths longer than 10~$\mu$m, where NH$_3$ becomes increasingly important (constraints not reported in Bell et al.). Unfortunately, this spectral region overlaps largely with the shadowed region at $> 10.6$~$\mu$m, in which strong systematics were reported.

Carbon monoxide exhibits spectral features below 5~$\mu$m and contributes opacity in all phases except the night side, although its influence on the first wavelength bin in the MIRI/LRS data remains marginal. 
As expected by the retrieved pressure-temperature profiles of the night side and morning terminator, and discussed in Sec.~\ref{sec_discussion}, carbon should be primarily found in CH$_4$ rather than CO. However, this expectation is not fully realized for this planet, since the chemical timescales required for CH$_4$ formation are significantly longer than the atmospheric transport timescale, leading to horizontal quenching. Consistent with this picture, the retrievals show only limited contributions of methane, despite the cool temperatures.

Finally, the retrieval model shows a moderate contribution from HCN, most prominently on the night side (not reported in Bell et al.). Smaller HCN contributions are inferred on the evening and morning sides, although not at levels sufficient to significantly influence the spectra. 
On the day side of the planet, no HCN is inferred, and the observed spectrum is primarily shaped by CO, NH$_3$ and H$_2$O.


\subsection{Retrieval comparison to photochemical models}
\label{sec:JB_retr_compar}

To illustrate the full extent of the retrieval model results obtained across all analyses performed for \citet{Bell_nightside_2024}, we compare the constraints for retrieved molecules to photochemical model predictions at each orbital phase. In particular, we present results from the \textit{ACE–PAC} pseudo-2D kinetics and equilibrium models at 1--10$\times$ solar metallicities, and plot them with constraints from the different retrieval models, some of which were unreported in \citet{Bell_nightside_2024}.

\begin{figure*}
    \centering
    \includegraphics[width=0.499\textwidth]{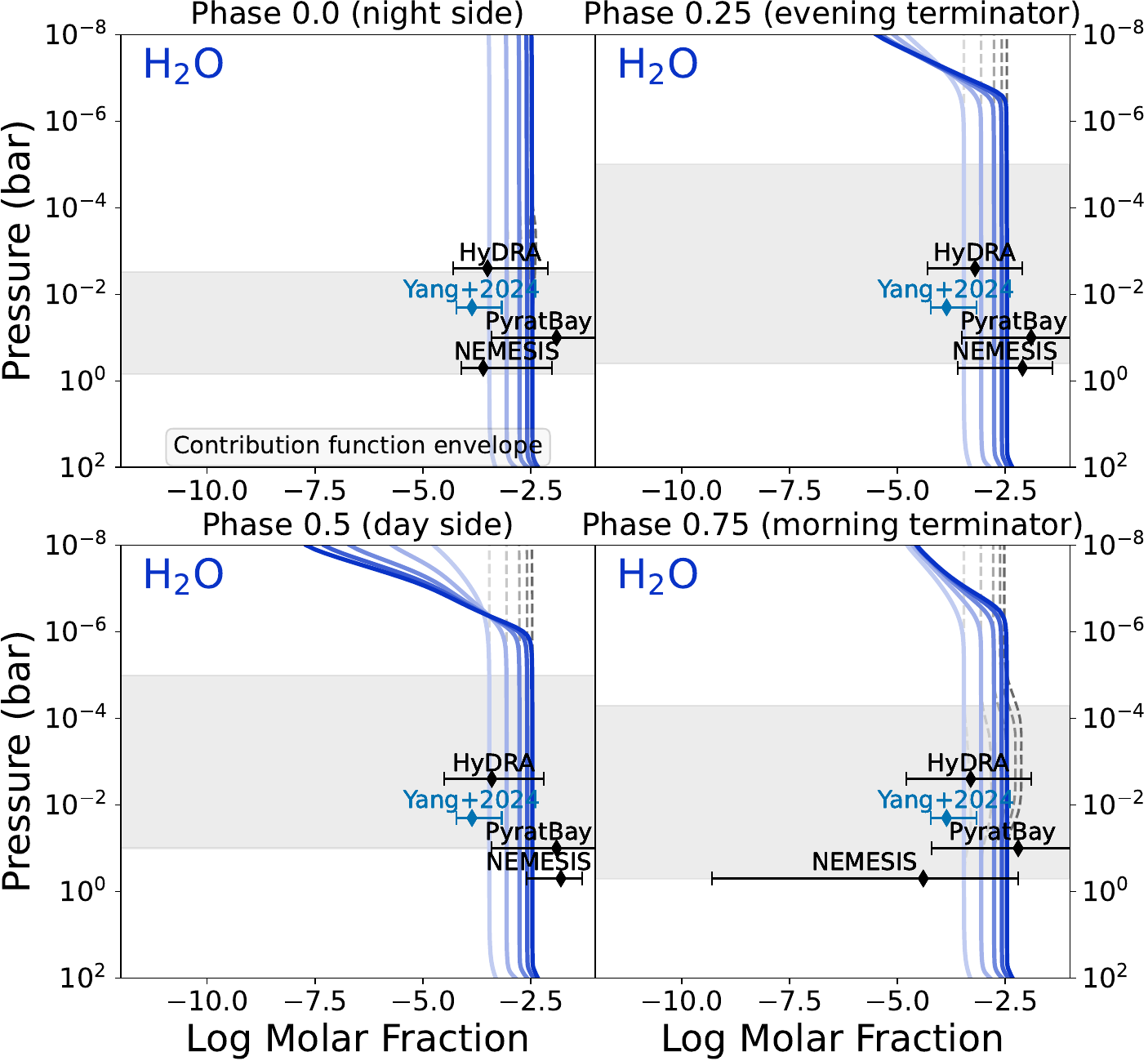}%
    \includegraphics[width=0.499\textwidth]{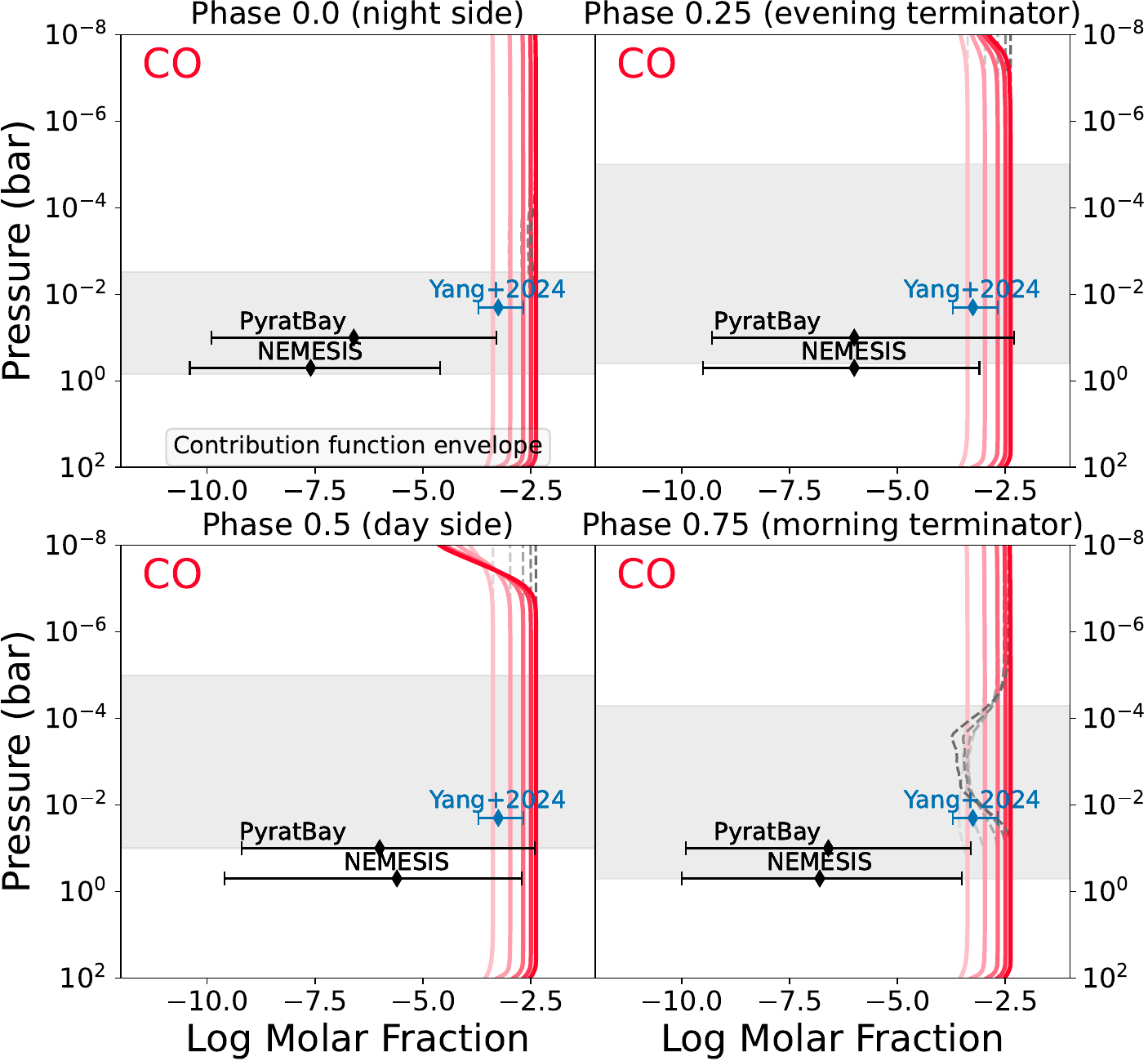}\\%
    \includegraphics[width=0.499\textwidth]{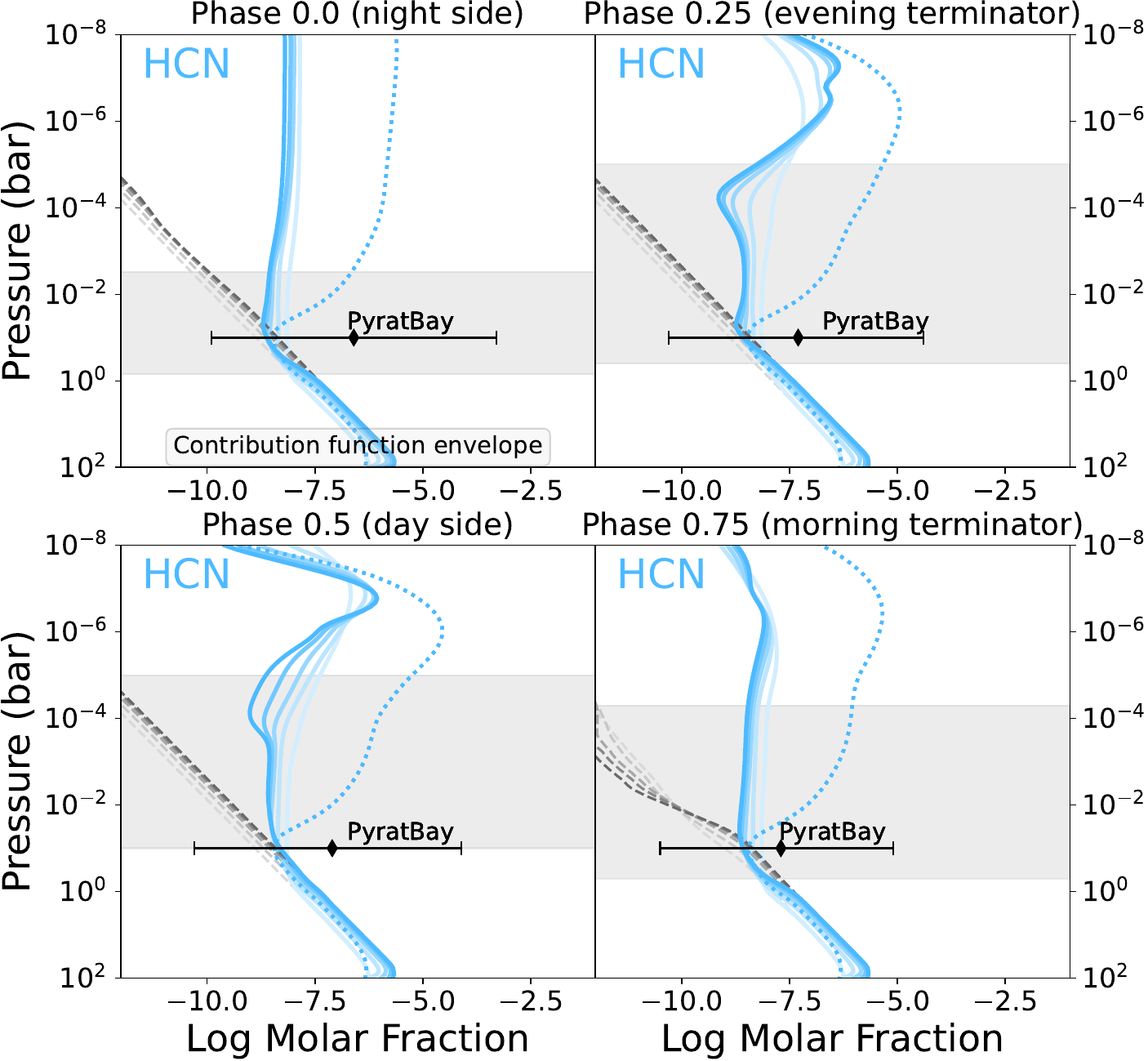}%
    \includegraphics[width=0.499\textwidth]{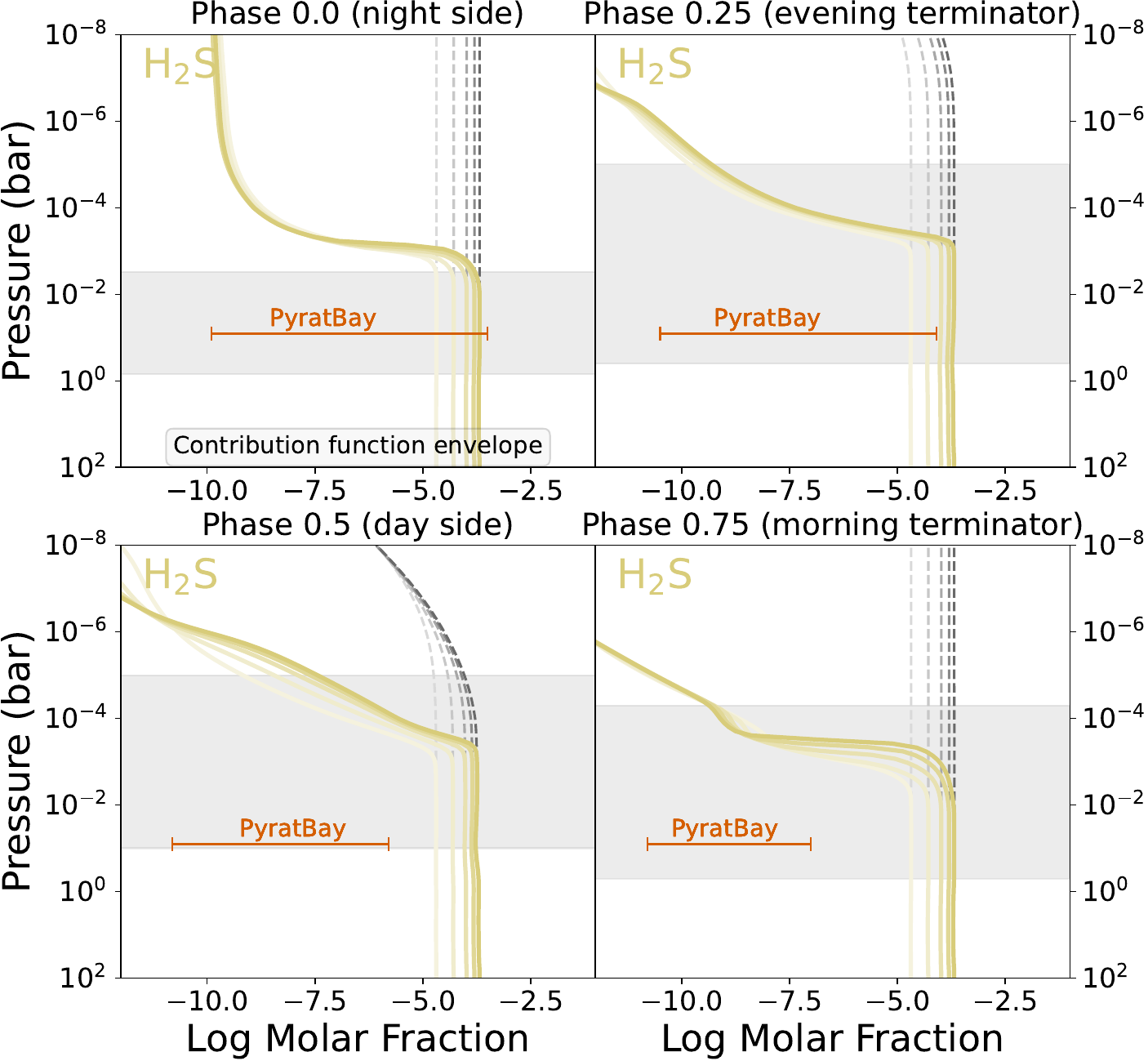}%
    \caption{Abundances of \ce{H2O}, CO, HCN, and \ce{H2S} from the \textit{ACE–PAC} pseudo-2D kinetics (\textit{solid}) and chemical equilibrium (\textit{dashed}) calculations, spanning metallicities from 1$\times$ (\textit{lightest}) to 10$\times$ (\textit{darkest}) solar. Retrieval constraints under the most conservative modelling setup are shown \citep[][\textbf{black}]{Bell_nightside_2024}, as well as from a 2D retrieval \citep[][\textbf{\color{YangBlue}{blue}}]{Yang_simultaneous_2024}. These are supplemented with 
    the most permissive \textit{PyratBay} model (\textbf{\color{PyratRed}{red}}). The lines corresponding to different retrieval models are vertically displaced for clarity. The contribution envelope from the best-fit model in \citet{Bell_nightside_2024} is indicated for each phase (\textit{gray}). Given the reported HCN abundance difference between \textit{ACE-PAC} and \textit{KINETICS} (see Sec.~\ref{sec_N}), we additionally plot the 1$\times$ solar HCN abundance from the \textit{KINETICS} model (\textit{dotted}).}
    \label{fig:JB_all_species}
\end{figure*}

Figure~\ref{fig:JB_all_species} shows the abundance constraints of \ce{H2O}, CO, HCN, and \ce{H2S} obtained from the various retrieval frameworks detailed above and in Table~\ref{tab:BellEtal}. Analogous figures for \ce{CH4}, \ce{CO2}, \ce{SO2}, and \ce{NH3} are presented in the main text (Figs.~\ref{fig_CH4_metallicities}--\ref{fig_SO2_metallicities}). Retrievals of \ce{H2O} are consistent along all four phases, and agree well with photochemical predictions, hence, why water absorption was reported in \citet{Bell_nightside_2024}. Spectral retrievals of the other molecules, however, show large uncertainties, consistent with a very low abundance or non-detection in the atmosphere of WASP-43~b.

Overall, the results shown in Fig.~\ref{fig:JB_all_species} indicate that, across the range of retrieval frameworks and modelling assumptions explored, the inferred abundances are generally compatible with both predictions from the \textit{ACE–PAC} pseudo-2D kinetics and from the chemical equilibrium models within the pressure ranges probed by the observations, as illustrated by the overlap with the gray regions marking the contribution-function envelopes for each phase. This level of agreement does not imply a detection, but it is consistent with scenarios in which multiple molecular species may contribute to the observed spectra of WASP-43~b. 

Still, some trends may be highlighted in order to guide future research. It is clear that the 2D retrieval of \citet{Yang_simultaneous_2024} has more precise constraints than the other retrieval models, which are one-dimensional retrievals of each phase separately. Increasing the dimensionality may thus provide a path toward better retrievals of spectroscopic phase curves, as long as the existing discrepancies between 1D and 2D retrieval models can be understood (e.g.~for ammonia, Fig.~\ref{fig_NH3_metallicities}). Additionally, further observations of WASP-43~b will provide a much needed broader spectral baseline, beyond the MIRI/LRS wavelength range (Crouzet et al., \textit{in prep.}). A spectral retrieval analysis of the combined dataset will likely yield not only multiple molecular detections, but also smaller uncertainties. This will prove to be essential in distinguishing between equilibrium or disequilibrium chemistry (\ce{CH4}, \ce{SO2}), atmospheric metallicity (\ce{CO2}, \ce{SO2}), and even the accuracy of underlying photochemical models (HCN, Fig.~\ref{fig:JB_all_species}).



\section{Thermal structure models}\label{sec_temperature}

This appendix provides a more detailed description of the models that were used to generate thermal profiles. The equatorial pressure-temperature profiles from these models serve as input parameters for the photochemical models in this paper. Specifically, we use \textit{Generic PCM} models (Sec.~\ref{sec_gpcm}) at various atmospheric metallicities in conjunction with \textit{ACE-PAC}, and the \textit{2D-ATMO} radiative-advective equilibrium model (Sec.~\ref{sec_atmo}) in conjunction with \textit{KINETICS}. These models provide the necessary flexibility for our study, as well as allow for a detailed comparison with \citet{Venot_global_2020}. For further information on how different GCMs fit the MIRI phase curve, we refer to \citet{Bell_nightside_2024, hammond_two-dimensional_2024}.

\subsection{Three-dimensional climate model: Generic PCM}\label{sec_gpcm}

In this work, we use the \textit{Generic Planetary Climate Model} (PCM) to model the atmosphere of WASP-43~b. This model has been developed for the study of giant planets of the Solar System \citep{boissinot_global_2024, spiga_global_2020,milcareck_radiative-convective_2024}, as well as mini-Neptunes \citep{charnay_3d_2015,charnay_3d_2015-1,charnay_formation_2021} and recently also  hot Jupiters \citep{teinturier_radiative_2024,teinturier_warm_2024}. The model is a coupling between a dynamical core that solves the primitive hydrostatic equations of meteorology on a longitude-latitude-pressure grid \citep{Hourdin_LMDZ6A_2020} and a physics package. The physics package contains parametrizations for the radiative transfer, based on the k-correlated method, the convection, using a convective adjustment scheme, and a vertical turbulent mixing scheme. Although the model has the ability to simulate the formation and evaporation of any type of cloud condensates, taking into account their radiative effect on the local thermodynamical structure, we only simulate cloud-free atmospheres in this work. 

Variations to the atmospheric metallicity are introduced using the \textit{Exo-REM} model \citep{charnay_self-consistent_2018}. We computed one-dimensional atmospheric columns for WASP-43~b with metallicities scaled by $1\times$, $2.5\times$, $5\times$, $7.5\times$, and $10\times$ the solar composition \citep{lodders_solar_2019}. For each metallicity, \textit{Exo-REM} yields a vertical model of the temperature and chemical composition, taking into account opacities of the following species: H$_{\rm 2}$O, CO, CH$_{\rm 4}$, CO$_{\rm 2}$, FeH, HCN, H$_{\rm 2}$S, TiO, VO, Na, K, PH$_{\rm 3}$, and NH$_{\rm 3}$. Then, these profiles were used to compute k-tables depending on temperature and pressure, using the \textit{exo\_k} package \citep{leconte_spectral_2021}. The different k-tables are used when running the 3D GCM at different atmospheric metallicities. For each simulation, the mean molecular weight of the atmosphere and the heat capacity $c_{p}$ were varied accordingly. 

\begin{figure}
    \centering
    \includegraphics[width=\linewidth]{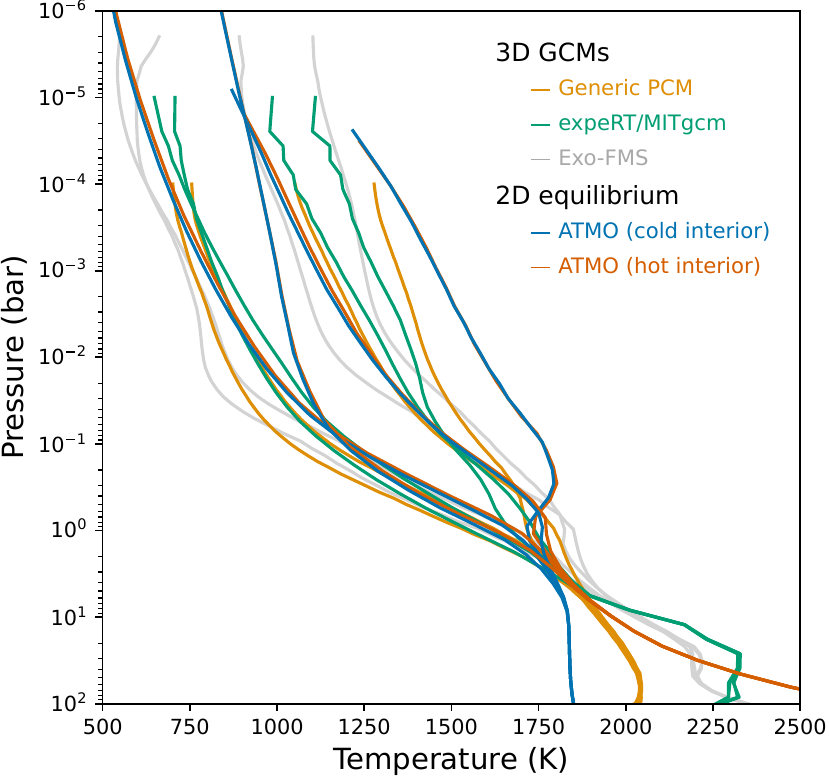}
    \caption{Comparison of pressure-temperature profiles used for the photochemical calculations in this study. Temperatures are sampled at four longitudes (0$\degrees$, 90$\degrees$, 180$\degrees$, -90$\degrees$) on the planetary equator (cosine-weighted average $\pm20\degrees$ in latitude), and computed for WASP-43~b at solar metallicity. 
    }
    \label{fig_temperature}
\end{figure}

All simulations were run for 2000 planetary years (with exception of the solar metallicity simulation, which ran for 9000 planetary years) so that a steady state is reached. The final results are averaged over the last 100 years. The solar and $10\times$ solar metallicity simulations are the same as in \cite{Bell_nightside_2024} and \cite{teinturier_radiative_2024}. 

Figure~\ref{fig_temperature} shows a comparison of the temperature profiles of the $1\times$ solar metallicity model with those of an equivalent model that was also used in \citet{Bell_nightside_2024} \citep[\textit{expeRT/MITgcm},][]{carone_equatorial_2020, Schneider_exploring_2022}, as well as with a chemistry-coupled GCM (\textit{Exo-FMS}, Sec.~\ref{sec_exo-FMS}), and a 2D radiative-advective equilibrium code (\textit{ATMO}, Sec.~\ref{sec_atmo}). There is generally good agreement between the predicted day-side and night-side temperatures. At 100~mbar, \textit{Generic PCM} and \textit{Exo-FMS} show temperatures ranging between 1100--1600~K. The temperature distribution in the \textit{expeRT/MITgcm} is somewhat more uniform (1200--1500~K). The 2D radiative-advective model shows agreement with the GCMs for the coolest profiles, but slightly hotter day-side temperatures up to $\sim$1750~K. Clear divergence among models can be seen in the deep atmosphere ($p> 5$~bar). The disagreement arises because GCMs often struggle in capturing deep dynamics, resulting in extremely long convergence times \citep{mendonca_angular_2020, Wang_extremely_2020, carone_equatorial_2020}. The \textit{2D-ATMO} models do not suffer from this problem, but a free parameter is used to control the deep temperature (see Sec.~\ref{sec_atmo}). Although the thermal and dynamical state of the deep atmosphere can to a certain degree affect the atmospheric circulation \citep{carone_equatorial_2020} and cloud microphysics \citep{helling_cloud_2021}, we show in Sec.~\ref{sec_chemistry}, Fig.~\ref{fig_chem_kinetics} that the deep temperature has a minimal effect on the resulting observable atmospheric composition of WASP-43~b.

\subsection{Two-dimensional radiative-advective equilibrium: ATMO}\label{sec_atmo}

In order to facilitate a direct comparison to the chemical models of \citet{Venot_global_2020}, we draw upon the same thermal structure model, namely that computed using the radiative transfer code \textit{2D-ATMO} \citep{tremblin_advection_2017}. Besides the heating of the atmosphere through stellar irradiation and propagation, the code takes into account vertical and horizontal energy advection to compute a two-dimensional thermal profile as a function of pressure and longitude in the equatorial plane. The free parameter $\alpha^{-1}$ determines the vertical to meridional mass flux, with high values of $\alpha$ corresponding to weak vertical heat advection and a relatively cold interior temperature, and low $\alpha$ causing a hot interior. As in \citet{Venot_global_2020}, we adopt two scenarios for WASP-43~b: a cold interior ($\alpha = 10^4$, $T_\textrm{100~bar}\approx 1800$~K) and a hot interior ($\alpha = 10$, $T_\textrm{100~bar}\approx 2800$~K). Calibration tests with GCMs have shown that $\alpha=1$--100 is representative for inflated hot Jupiters \citep{tremblin_advection_2017}. However, WASP-43~b has a high gravity and is not inflated, so $\alpha\approx10^4$ should be more realistic for this planet. Both thermal structure models are shown in comparison to the GCM temperatures in Fig.~\ref{fig_temperature}. For further details about the \textit{2D-ATMO} setup, we refer to \citet{Venot_global_2020}.

\end{appendix}

\end{document}